\documentclass{article}
\pdfoutput=1
\usepackage{amsmath, amssymb, amsthm, graphicx, textcomp}
\usepackage{setspace, float, url, hyperref, enumitem, microtype}
\usepackage[titletoc]{appendix}
\usepackage[english]{babel}
\usepackage{scalerel}
\usepackage[dvipsnames]{xcolor}
\usepackage{caption}[2008/04/01]
\usepackage{wrapfig}
\usepackage{subfig}
\usepackage{lipsum}  
\usepackage{tikz}
\usetikzlibrary{svg.path}
\usepackage[autostyle, english = american]{csquotes}
\MakeOuterQuote{"}

\vbadness=\maxdimen

\definecolor{orcidlogocol}{HTML}{A6CE39}
\tikzset{
  orcidlogo/.pic={
    \fill[orcidlogocol] svg{M256,128c0,70.7-57.3,128-128,128C57.3,256,0,198.7,0,128C0,57.3,57.3,0,128,0C198.7,0,256,57.3,256,128z};
    \fill[white] svg{M86.3,186.2H70.9V79.1h15.4v48.4V186.2z}
                 svg{M108.9,79.1h41.6c39.6,0,57,28.3,57,53.6c0,27.5-21.5,53.6-56.8,53.6h-41.8V79.1z M124.3,172.4h24.5c34.9,0,42.9-26.5,42.9-39.7c0-21.5-13.7-39.7-43.7-39.7h-23.7V172.4z}
                 svg{M88.7,56.8c0,5.5-4.5,10.1-10.1,10.1c-5.6,0-10.1-4.6-10.1-10.1c0-5.6,4.5-10.1,10.1-10.1C84.2,46.7,88.7,51.3,88.7,56.8z};
  }
}

\newcommand\orcidicon[1]{\href{https://orcid.org/#1}{\mbox{\scalerel*{
\begin{tikzpicture}[yscale=-1,transform shape]
\pic{orcidlogo};
\end{tikzpicture}
}{|}}}}

\newcommand{\mathsym}[1]{{}}
\newcommand{\unicode}[1]{{}}

\begin{document}
\newtheorem{Theorem}{Theorem}
\newtheorem{Definition}{Definition}
\newtheorem{Note}{Note}
\newtheorem{Lemma}{Lemma}
\newtheorem{Corollary}{Corollary}
\newtheorem{Example}{Example}

\renewcommand{\abstractname}{}

\title{The role of (non)contextuality in Bell's theorems from the perspective of an operational modeling framework}
\author{Michael L. Ulrey  
\footnote{\url{https://www.researchgate.net/profile/Michael_Ulrey/research}}\footnote{\url{https://www.linkedin.com/in/mike-ulrey-9bb9047b/}}
\footnote{\url{https://turqlink.com/interests-and-goals/}}
\orcidicon{00000-0002-5143-973X}\,, {IEEE, SIAM}\\
\textbf{e-mail:} \textit{mulrey@hotmail.com }}
\maketitle

\abstract{%
A novel approach for analyzing "classical" alternatives to quantum mechanics for explaining the statistical results of an EPRB-like experiment is proposed. This perspective is top-down instead of bottom-up. Rather than beginning with an inequality derivation, a hierarchy of model types is constructed, each distinguished by appropriately parameterized conditional probabilities. This hierarchy ranks the "classical" model types in terms of their ability to reproduce QM statistics or not. The analysis goes beyond the usual consideration of model types that "fall short" (i.e., satisfy all of the CHSH inequalities) to ones that are "excessive" (i.e., not only violate CHSH but even exceed a Tsirelson bound). This approach clearly shows that noncontextuality is the most general property of an operational model that blocks replication of at least some QM statistical predictions. Factorizability is naturally revealed to be a special case of noncontextuality. The same is true for the combination of remote context independence and outcome determinism (RCI+OD). It is noncontextuality that determines the dividing line between "classical" model instances that satisfy the CHSH inequalities and those that don't. Outcome deterministic operational models are revealed to be the "building blocks" of all the rest, including quantum mechanical, noncontextual, and contextual ones. The set of noncontextual model instances is exactly the convex hull of all 16 RCI+OD model instances, and furthermore, the set of all model instances, including all QM ones, is equal to the convex hull of the 256 OD model instances. It is shown that, under a mild assumption, the construction of convex hulls of finite ensembles of OD model instances is (mathematically) equivalent to the traditional hidden variables approach. Via the introduction of operational models that possess outcome and measurement "predictability", a new perspective is gained on the impossibility of faster-than-light transfer of information in an EPRB experiment. Finally, many plots and figures, some of which appear to be new, provide visual affirmation of many of the results.

\textbf{Keywords}: Bell's theorem(s); Bell-Kochen-Specker theorem; hidden variables; convex hull; outcome determinism; locality; non-contextuality; predictability; correlation plots; modeling hierarchy
}

\section{Introduction}
\label{sec:introduction}

\begin{figure}[H]
\centering
\includegraphics[width=0.75\linewidth]{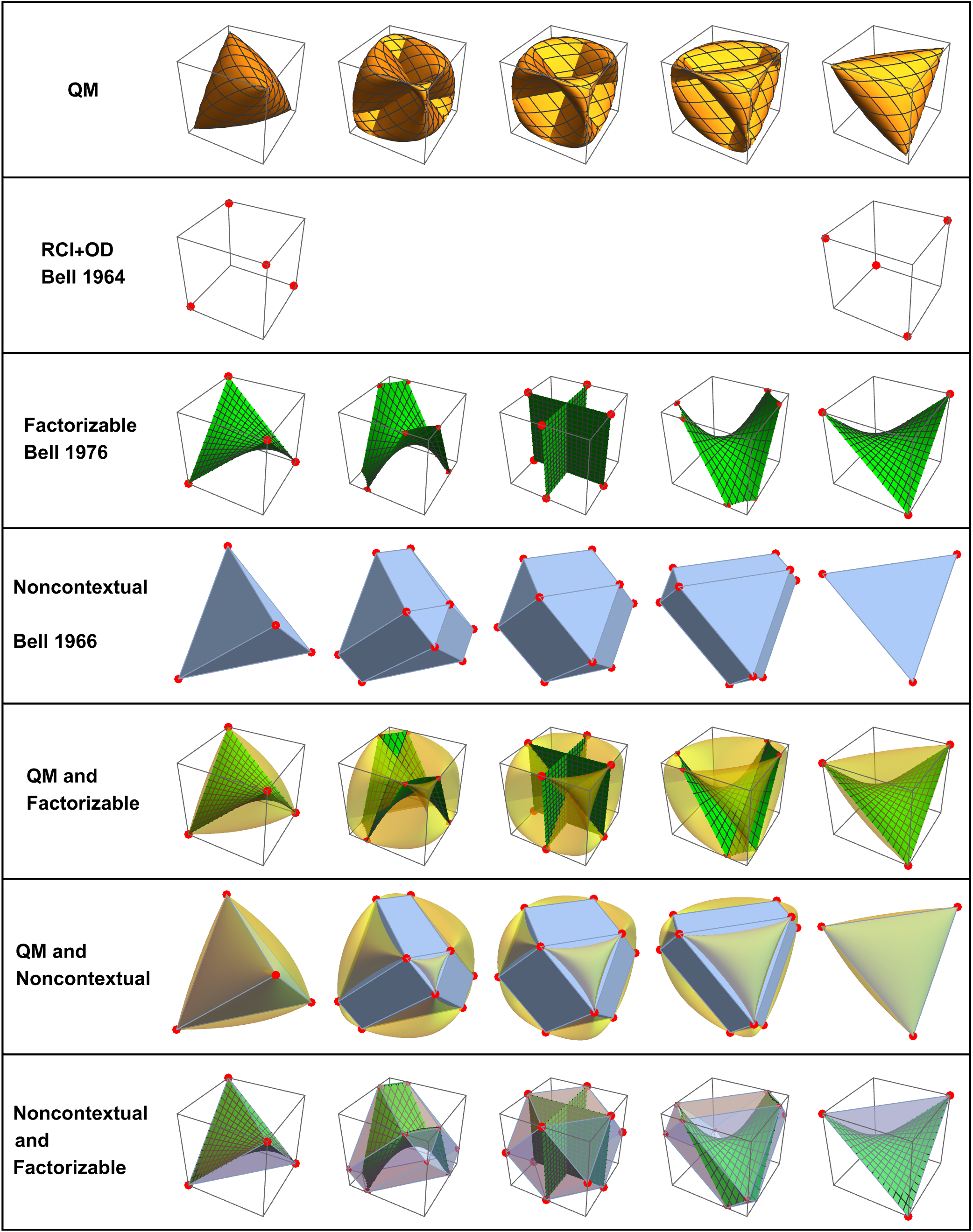}
\caption{3D slices through sets of 4D correlation vectors corresponding to various probabilistic assumptions, and combinations thereof. In each row, exactly one correlation is fixed in each of the five plots at the values $-1,-\frac{1}{2},0,\frac{1}{2},\text{ and } 1$, from left to right. The plots in the noncontextual row (4) correspond to the polyhedra defined by the CHSH inequalities (see Corollary~\ref{cor:ncPolytopes}). The frame for each plot is the "correlation cube" $[-1,1] \times [-1,1] \times [-1,1]$. See Appendix~\ref{sec:corrPlots} for more details.}
\label{fig:models5SelectedIntro}
\end{figure}

Consider Fig.~\ref{fig:models5SelectedIntro}, which shows sets of 3-tuples of correlations attached to different probabilistic models of an EPRB experiment (the fourth correlation is fixed at $-1,-\frac{1}{2},0,\frac{1}{2},\text{ or } 1$, from left to right in each row). The first four rows show correlations corresponding to the assumptions, respectively, of Hilbert-space quantum mechanics (QM), remote context independence together with outcome determinism (RCI+OD), factorizability (or Bell locality), and noncontextuality. The remaining rows show pairwise comparisons of selected plots. 

Rows 2-4 and the last row strongly suggest that RCI+OD and factorizability are simply stronger notions of noncontextuality. Notice how the factorizable correlations (green surfaces) fit entirely inside the noncontextual correlations (blue solids) and the RCI+OD correlations (red dots) are "vertices" of both of them (in the first and last columns only, where the correlation in the 4th dimension is fixed at $-1$ and $+1$, respectively).  In other words, in terms of sets of correlations,
\begin{equation}
RCI+OD \subset Factorizable \subset Noncontextual.
\label{eq:rciFCNC}
\end{equation}
Furthermore, all of the correlations in each of these categories satisfy all of the CHSH inequalities.\footnote{Clauser, Horne, Shimony, and Holt. See~\cite{CHSH1969} and~\cite{Aspect2002}.} 

Wiseman~\cite{Wiseman2014} points out that many of the debates surrounding "Bell's theorem" stem from the fact that there are really two Bell's theorems. In~\cite{Bell1964}, Bell showed that RCI+OD were sufficient to derive his inequality, and in~\cite{Bell1976}, he showed that factorizability is also sufficient to derive a similar inequality. Both inequalities are inconsistent with at least some statistical predictions of QM. 

Mermin~\cite{Mermin1993} compared Bell's 1964 paper~\cite{Bell1964} to the earlier so-called "Bell-Kochen-Specker" (BKS) paper~\cite{Bell1966}\footnote{Due to a publication delay, this paper did not appear until 1966. Kochen and Specker independently proved the same result by different means in~\cite{KochenSpecker1967}.}. In the BKS paper, Bell dismantled von Neumann's "silly" argument against a hidden variables theory and replaced it with his own argument, which he then self-criticized because of the tacit assumption that the measurement of a given observable should yield the same value, regardless of what other measurements are made simultaneously. Although he did not name it at the time, this assumption later came to be called "noncontextuality" -- again see~\cite{Mermin1993} and also the noncontextuality discussion in~\cite{StanfordBell2021}.  

The increasing level of generality of Eq.~\ref{eq:rciFCNC} thus reflects the content of Bell's three papers~\cite{Bell1964,Bell1976,Bell1966}, mathematically if not chronologically (also noted in Fig.~\ref{fig:models5SelectedIntro} in the descriptions to the left of rows 2, 3, and 4). They each indicate Bell's ideas at the time about what kind of hidden variables theory is blocked from explaining at least some QM statistical predictions. This culminates in the BKS paper~\cite{Bell1966}, which effectively shows there can be no noncontextual (NC) hidden variables theory that is consistent with all QM statistical predictions.

The central importance of noncontexuality was already evident in the pioneering work of Fine~\cite{Fine1982-1,Fine1982-2}. For example in~\cite{Fine1982-1} he said: "\textit{Finally, I believe that Proposition (1) -- conjoined with the other two -- shows what hidden variables and the Bell inequalities are all about; namely, imposing requirements to make well-defined precisely those probability distributions for noncommuting observables whose rejection is the very essence of quantum mechanics.}"

Abramsky and Bradenburger~\cite{AbramskyBrandenburger2011}, exploring nonlocality and contextuality in the very general framework of sheaf theory, echo this conclusion by saying: "\textit{We show that contextuality, and non-locality as a special case, correspond exactly to obstructions to global sections.}"

In Held~\cite{StanfordKS2018}, the Kochen-Specker theorem and noncontextuality are linked by the statement: "\textit{The second important no-go theorem against HV theories is the theorem of Kochen and Specker (KS) which states that, given a premise of noncontextuality (to be explained presently) certain sets of QM observables cannot consistently be assigned values at all (even before the question of their statistical distributions arises).}"

Klyachko, et. al.~\cite{Klyachko2008} emphasize the important implications for quantum computing, given that the contextuality of QM is more fundamental than nonlocality: "\textit{However, for quantum computation the magic ability of entanglement to bypass constraints imposed by the so-called classical realism is far more important. The latter is understood here as the existence of hidden parameters, or equivalently a joint probability distribution of all involved quantum observables.}" See also~\cite{Spekkens2009} for a specific example of better-than-classical performance of quantum protocols using polarization qubits, enabled by preparation contextuality.

For deep connections of noncontextuality to graph theoretic concepts, see for example~\cite{Winter2010,Bharti2019}, and for a surprising connection to relational database theory, see~\cite{Abramsky2013}.

In this paper, within the context of a Bell/Aspect-like experiment, many of the concepts surrounding Bell's theorems are organized into a hierarchy of parameterized operational model types, together with sets of their instantiations, that represent various concepts, such as remote context independence, remote outcome independence, factorizability, outcome determinism, noncontextuality, and combinations thereof. In this parameterized modeling framework, it is possible to "rank" the different model types in their ability to replicate QM predictions while retaining certain "classical" features (or not). This top-down approach provides a broader perspective than mere manipulation of inequalities. It lends  intuition into why certain known results are true, reveals a few underappreciated facts, renders many proofs almost trivial, leads to a more intuitive, simpler formulation of "hidden variables", provides a different insight into why the experiment cannot be used for faster-than-light communication, and provides many relevant plots and figures, some of which appear to be new.

In particular, this parameterized operational modeling paradigm can be used to confirm the suggestiveness of Fig.~\ref{fig:models5SelectedIntro}, namely that the hierarchy of noncontextual correlations in Eq.~\ref{eq:rciFCNC} reflects a similar relationship  among the model types representing these properties. Both RCI+OD and factorizability are merely special cases of noncontextuality. If one assumes "locality", either as RCI in the combination RCI+OD, or as factorizability, one is implicitly assuming noncontextuality. Noncontextuality is the most general "classical" notion among these three that blocks complete duplication of QM statistical behaviors. In fact, the 8 CHSH inequalities become a "witness" for noncontextuality. That is, if noncontextuality is assumed, all 8 CHSH inequalities hold and (almost) conversely, if a model instance (not necessarily noncontextual) has correlations that satisfy all of the CHSH inequalities, \emph{there exists} a (possibly different) noncontextual model instance with those same correlations. 

\section{The operational model types}
\label{sec:definitionsAndOperModel}

The basic motivation behind a "hidden variables" theory is to posit an alternative explanation of quantum mechanics that explains its behaviors in a "classical" way. Some "hidden variables" theories can be eliminated straight away if they lead to mathematical models that produce statistical predictions that are inconsistent with QM. The strategy in this paper is to consider a specific class of statistical models, which will be called "operational" models, and determine if certain properties that they possess are consistent (or not) with QM.

An operational model type is characterized by a set of 16 conditional probabilities 
\[P(A=s,B=t|M_A=u,M_B=v)\]
for Alice and Bob's outcomes $A$ and $B$ given their measurement choices (context) $M_A$ and $M_B$, where $s,t=\pm1$ and  $u,v=1 \text{ or } 2$. The novel approach in this paper is to represent these conditional probabilities with symbolic parameters 
\[\pmb{\gamma}=(\gamma_1,\gamma_2,...,\gamma_{16}),\]
and then suitably "re-parameterize" this \emph{generic} representation to characterize various \emph{specialized} model types.

\subsection{Generic}
\label{sec:genericType}

The generic representation of an \emph{operational} (that is, statistical) model of an EPRB experiment is shown in Table~\ref{tab:genericTypeIntro}.
\begin{table}[H]
\caption{The \textbf{generic} model type, where the $\gamma_k$'s represent the 16 conditional probabilities $P(A=s,B=t|M_A=u,M_B=v),$ where $s,t=\pm1$ denote Alice and Bob's measurement outcomes and $u,v=1 \text{ or } 2$ indicate their measurement choices (context), respectively. Each row consists of nonnegative real numbers summing to 1.}
\label{tab:genericTypeIntro}
\[
\begin{array}{|c|c|c|c|c|}
\hline
\text{Context} & \multicolumn{4}{c|}{\text{Outcome }(s,t)} \\
\hline
(u,v) & (-1,-1)  & (-1,1)   & (1,-1)   & (1,1) \\
\hline
(1,1) & \gamma_1 & \gamma_2 & \gamma_3 & \gamma_4 \\
\hline
(1,2) & \gamma_5 & \gamma_6 & \gamma_7 & \gamma_8 \\
\hline
(2,1) & \gamma_9 & \gamma_{10} & \gamma_{11} & \gamma_{12} \\
\hline
(2,2) & \gamma_{13} & \gamma_{14} & \gamma_{15} & \gamma_{16} \\
\hline
\end{array}
\]
\end{table}
In this table, each $\gamma_k$ represents a conditional probability, that is,
\[\gamma_k=P(A=s,B=t|M_A=u,M_B=v)\]
for appropriate $s,t=\pm1$, $u,v=1,2$, and $k=1,2,...,16$. For example,
\[\gamma_7=P(A=1,B=-1|M_A=1,M_B=2),\] 
and similarly for the other 15 entries. Table~\ref{tab:genericTypeIntro} corresponds to what Abramsky and Bradenburger~\cite{AbramskyBrandenburger2011} call an \emph{empirical} model, or in this case, a $(2,2,2)$ \emph{Bell-type scenario}\footnote{$2$ observers, $2$ measurements per observer, $2$ outcomes per measurement.}. This is just one example in their comprehensive analysis of non-locality and contextuality in the language of sheaf theory. Whereas~\cite{AbramskyBrandenburger2011} encompasses general $(n,k,l)$ Bell-type scenarios (and more), the focus here is on the $(2,2,2)$ case only.

This model type is called "generic" because the $\gamma_k$'s are unspecified, except for the following restrictions, which apply to each and every model type in this paper, whether generic or specialized.

\begin{Definition}
\label{def:contextualIntro}
The \textbf{measurement context constraints} say that
\begin{align}
\label{eq:contextualIntro}
&\gamma_1+\gamma_2+\gamma_3+\gamma_4=1, \nonumber \\
&\gamma_5+\gamma_6+\gamma_7+\gamma_8=1,  \\
&\gamma_9+\gamma_{10}+\gamma_{11}+\gamma_{12}=1, \nonumber \\
&\gamma_{13}+\gamma_{14}+\gamma_{15}+\gamma_{16}=1. \nonumber
\end{align}
\end{Definition}

\begin{Definition}
The term \textbf{model type} refers to any table like Table~\ref{tab:genericTypeIntro}, which includes a specification of the outcomes, context, and conditional probabilities  $\pmb{\gamma}=(\gamma_1,\gamma_2,...,\gamma_{16})$. These conditional probabilities might be written in terms of a different set of symbolic parameters. 
\end{Definition}

\begin{Definition}
A \textbf{model instance} is a model type in which the symbolic parameters have been assigned specific numeric values.
\end{Definition}

\begin{Definition}
\label{def:cpvIntro}
A \textbf{conditional probability vector}, or \textbf{cpv}, is an ordered set of 16 nonnegative real numbers $\pmb{\gamma}=(\gamma_1,\gamma_2,...,\gamma_{16})$ that satisfy the measurement context constraints of Def.~\ref{def:contextualIntro}. 
\end{Definition}

In this paper, the terms "model type" and "cpv" will be used interchangeably, when they involve symbolic parameters, and "model instance" and "cpv" used interchangeably, when they involve numeric parameters.

\subsection{Correlations, s-functions, and the CHSH inequalities}
\label{sec:defineCHSH}

Before proceeding with the definitions of the more specialized model types, it is convenient to define the correlations, $s$-functions, and CHSH inequalities in terms of the generic model type parameters. These definitions will then apply \emph{mutatis mutandis} to any other specialized model type by a suitable change of parameters according to the table that defines the relevant model type. The results for all model types are summarized in Appendix~\ref{sec:correlations}.
\begin{Definition}
Define the four \textbf{correlations} $(w,x,y,z)$ of any generic model type (and hence any other model type in this paper) in terms of its parameters $\pmb{\gamma}=(\gamma_1,\gamma_2,...,\gamma_{16})$:

\begin{align}
\label{eq:defineCorrIntro}
&w=\gamma_1-\gamma_2-\gamma_3+\gamma_4, \nonumber \\
&x=\gamma_5-\gamma_6-\gamma_7+\gamma_8,  \\
&y=\gamma_9-\gamma_{10}-\gamma_{11}+\gamma_{12}, \nonumber \\
&z=\gamma_{13}-\gamma_{14}-\gamma_{15}+\gamma_{16}. \nonumber
\end{align}

Now define the four $\pmb{s}$-\textbf{functions}\footnote{The use of the letter $s$ for these expressions is taken from Aspect~\cite{Aspect2002}.} of the correlations $(w,x,y,z)$:
\begin{align}
\label{eq:defineSFunIntro}
&s_1=-w+x+y+z,  \nonumber \\
&s_2=w-x+y+z,   \\
&s_3=w+x-y+z,  \nonumber \\
&s_4=w+x+y-z.  \nonumber
\end{align}

Finally the 8 \textbf{CHSH inequalities} are:
\begin{equation}
-2 \leq s_1,s_2,s_3,s_4 \leq 2. \label{eq:chshIntro}
\end{equation}
\end{Definition}
The following two matrices will come in handy for reducing lengthy lists of equations/inequalities to a  more compact matrix format. 

\begin{Definition}
\label{def:cMatrix}
The \textbf{correlation matrix} $C$ turns a set of generic parameters $\pmb{\gamma}=(\gamma_1,\gamma_2,...,\gamma_{16})$ into a set of correlations $(w,x,y,z)^T=C \pmb{\gamma}^T$. See Eq.~\ref{eq:defineCorrIntro}.
\begin{equation}
C =
\left( \begin{array}{rrrrrrrrrrrrrrrr}
1 & -1 & -1 & 1 & 0 & 0 & 0 & 0 & 0 & 0 & 0 & 0 & 0 & 0 & 0 & 0\\
0 & 0 & 0 & 0 & 1 & -1 & -1 & 1 & 0 & 0 & 0 & 0 & 0 & 0 & 0 & 0 \\
0 & 0 & 0 & 0 & 0 & 0 & 0 & 0 & 1 & -1 & -1 & 1 & 0 & 0 & 0 & 0\\
0 & 0 & 0 & 0 & 0 & 0 & 0 & 0 & 0 & 0 & 0 & 0 & 1 & -1 & -1 & 1
\end{array} \right)
\end{equation}
\end{Definition}

\begin{Definition}
The matrix $S$ turns a set of correlations $(w,x,y,z)$ into the corresponding $s$-functions  $(s_1,s_2,s_3,s_4)^T=S(w,x,y,z)^T$. See Eq.~\ref{eq:defineSFunIntro}.
\begin{equation}
S =
\left( \begin{array}{cccc}
-1 & 1 & 1 & 1 \\
1 & -1 & 1 & 1 \\
1 & 1 & -1 & 1\\
1 & 1 & 1 & -1
\end{array} \right).
\end{equation}
\end{Definition}

The CHSH inequalities (Eq.~\ref{eq:chshIntro}) can now be written in terms of the generic model type parameters $\pmb{\gamma}=(\gamma_1,\gamma_2,...,\gamma_{16})$ as
\begin{equation}
\label{eq:genericCHSH}
-2 \leq SC \pmb{\gamma}^T \leq 2.
\end{equation}
Inequalities like these are meant to be interpreted as applying componentwise to the vector in the middle. One of the main tasks is to characterize the sets of generic parameters $\pmb{\gamma}=(\gamma_1,\gamma_2,...,\gamma_{16})$ that either satisfy all or violate at least one of the inequalities represented by Eq.~\ref{eq:genericCHSH}. 

\begin{table}[H]
\caption{The correlations and $s$-functions for the generic type. Not only is it possible to violate a CHSH inequality, it is possible to exceed a Tsirelson bound ($\pm2 \sqrt{2}$).}
\label{tab:genericCandSTableIntro}
\[
\begin{array}{|c||l|}
\hline
w & \gamma_1-\gamma_2-\gamma_3+\gamma_4 \\
\hline
x & \gamma_5-\gamma_6-\gamma_7+\gamma_8 \\
\hline
y & \gamma_9-\gamma_{10}-\gamma_{11}+\gamma_{12} \\
\hline
z & \gamma_{13}-\gamma_{14}-\gamma_{15}+\gamma_{16} \\
\hline
\hline
s_1 & -\gamma_1+\gamma_2+\gamma_3-\gamma_4+\gamma_5-\gamma_6-\gamma_7+\gamma_8+ \\
& \gamma_9-\gamma_{10}-\gamma_{11}+\gamma_{12}+\gamma_{13}-\gamma_{14}-\gamma_{15}+\gamma_{16} \\
\hline
s_2 & \gamma_1-\gamma_2-\gamma_3+\gamma_4-\gamma_5+\gamma_6+\gamma_7-\gamma_8+ \\
& \gamma_9-\gamma_{10}-\gamma_{11}+\gamma_{12}+\gamma_{13}-\gamma_{14}-\gamma_{15}+\gamma_{16} \\
\hline
s_3 & \gamma_1-\gamma_2-\gamma_3+\gamma_4+\gamma_5-\gamma_6-\gamma_7+\gamma_8- \\
& \gamma_9+\gamma_{10}+\gamma_{11}-\gamma_{12}+\gamma_{13}-\gamma_{14}-\gamma_{15}+\gamma_{16} \\
\hline
s_4 & \gamma_1-\gamma_2-\gamma_3+\gamma_4+\gamma_5-\gamma_6-\gamma_7+\gamma_8+ \\
& \gamma_9-\gamma_{10}-\gamma_{11}+\gamma_{12}-\gamma_{13}+\gamma_{14}+\gamma_{15}-\gamma_{16} \\
\hline
\hline
\text{Min and Max} & \multicolumn{1}{c|}{-4 \text{ and } 4} \\
\hline
\end{array}
\]
\end{table}
It should be clear from Table~\ref{tab:genericCandSTableIntro} that, since the $\gamma_k$'s are nonnegative real numbers that sum to 4, the $s$-functions of the generic model type can be as high as +4 or as low as -4. 

\subsection{QM}
\label{sec:qmType}

Table~\ref{tab:qmTypeIntro} shows the model type derived from standard Hilbert-space quantum mechanics (QM)\footnote{These probabilities correspond to the photon-polarization version of the experiment, \emph{not} the spin-$\frac{1}{2}$ singlet pair version. This assumption holds throughout this paper.}. It is clear that this is an example of the generic model type, since the sum of row entries is 1 for all four rows.

\begin{figure}[H]
\centering
\includegraphics[width=0.3\linewidth]{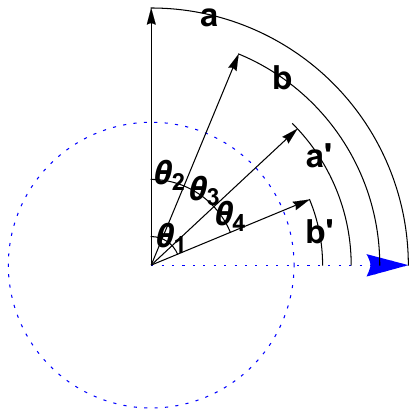}
\caption{The relationships among the measurement difference angles $\theta_k$.}
\label{fig:thetaConstraintsIntro}
\end{figure}

\begin{table}[H]
\caption{The \textbf{QM} model type is defined by the following 16 expressions in terms of the parameters $\theta_k$ for $k=1,2,3,4$. They represent the conditional probabilities $P(A=s,B=t|M_A=u,M_B=v)$ of Alice and Bob's outcomes $s,t=\pm1$ given their measurement choices (context) $u,v=1 \text{ or } 2$, respectively. Note the sum of values in each of the four rows is 1.}
\label{tab:qmTypeIntro}
\[
\begin{array}{|c|c|c|c|c|}
\hline
\text{Context} & \multicolumn{4}{c|}{\text{Outcome }(s,t)} \\
\hline
(u,v) & (-1,-1)  & (-1,1)   & (1,-1)   & (1,1) \\
\hline
(1,1) & \frac{1}{2} \cos^2 \theta_1 & \frac{1}{2} \sin^2 \theta_1 & \frac{1}{2} \sin^2 \theta_1 & \frac{1}{2} \cos^2 \theta_1 \\
\hline
(1,2) & \frac{1}{2} \cos^2 \theta_2 & \frac{1}{2} \sin^2 \theta_2 & \frac{1}{2} \sin^2 \theta_2 & \frac{1}{2} \cos^2 \theta_2 \\
\hline
(2,1) & \frac{1}{2} \cos^2 \theta_3 & \frac{1}{2} \sin^2 \theta_3 & \frac{1}{2} \sin^2 \theta_3 & \frac{1}{2} \cos^2 \theta_3 \\
\hline
(2,2) & \frac{1}{2} \cos^2 \theta_4 & \frac{1}{2} \sin^2 \theta_4 & \frac{1}{2} \sin^2 \theta_4 & \frac{1}{2} \cos^2 \theta_4 \\
\hline
\end{array}
\]
\end{table}
Comparing to Table~\ref{tab:genericTypeIntro}, for example, the entry in row 2, column 3 is
\[\gamma_7=P(A=1,B=-1|M_A=1,M_B=2)=\frac{1}{2}\sin^2 \theta_2.\] 
The $\theta_k$'s represent the measurement \emph{difference} angles as shown in Eq.~\ref{eq:measDiffIntro} and Fig.~\ref{fig:thetaConstraintsIntro}, where $a,a'$ and $b,b'$ represent Alice and Bob's photon polarization analyzer orientation choices, respectively. This notation is based on (not completely equivalent to) Aspect's account~\cite{Aspect2002}. 

\begin{equation}
\theta_1=a-b', \;  \theta_2=a-b, \; \theta_3=b-a', \; \theta_4=a'-b'. \label{eq:measDiffIntro}
\end{equation}

It is clear from this figure that the measurement difference angles satisfy the constraint in Eq.~\ref{eq:thetaConstraintsIntro}. If the names or orientations are different, it will be assumed that they can always be renamed so as to conform to Fig.~\ref{fig:thetaConstraintsIntro}.

\begin{equation}
\theta_1=\theta_2+\theta_3+\theta_4.
\label{eq:thetaConstraintsIntro}
\end{equation} 
The correlations and $s$-functions for the QM model type are shown in Table~\ref{tab:qmCandSTableIntro}. These can be calculated from Eq.'s~\ref{eq:defineCorrIntro} and~\ref{eq:defineSFunIntro} by substituting the QM conditional probabilities from Table~\ref{tab:qmTypeIntro} for the corresponding $\gamma_k$'s in Table~\ref{tab:genericTypeIntro}.

\begin{table}[H]
\caption{The correlations and $s$-functions for the QM model type. The proof that the $s$-function bounds are $\pm 2 \sqrt{2}$ can be found in Appendix~\ref{sec:qmMaxMin}.}
\label{tab:qmCandSTableIntro}
\[
\begin{array}{|c||l|}
\hline
w & \cos 2\theta_1 \\
\hline
x & \cos 2\theta_2 \\
\hline
y & \cos 2\theta_3 \\
\hline
z & \cos 2\theta_4 \\
\hline
\hline
s_1 & -\cos 2\theta_1+\cos 2\theta_2+\cos 2\theta_3+\cos 2\theta_4 \\
\hline
s_2 & \cos 2\theta_1-\cos 2\theta_2+\cos 2\theta_3+\cos 2\theta_4 \\
\hline
s_3 & \cos 2\theta_1+\cos 2\theta_2-\cos 2\theta_3+\cos 2\theta_4 \\
\hline
s_4 & \cos 2\theta_1+\cos 2\theta_2+\cos 2\theta_3-\cos 2\theta_4 \\
\hline
\hline
\text{Min and Max} &  \multicolumn{1}{c|}{-2\sqrt{2} \text{ and } 2\sqrt{2}} \\
\hline
\end{array}
\]
\end{table}

\subsection{Noncontextual}
\label{sec:ncType}

It is important to note that the probabilistic structure postulated in Tables~\ref{tab:genericTypeIntro} or~\ref{tab:qmTypeIntro} consists of four probability mass functions (expressed as conditional probabilities) on four distinct probability spaces, each associated with a different measurement context. 
\begin{Definition}
A \textbf{probability mass function} (\textbf{pmf}) is a finite ordered set of real numbers $\pmb{\pi}=(\pi_1,\pi_2,...,\pi_n)$ where all $\pi_k \geq 0$ and $\sum_k \pi_k=1.$
\end{Definition}
So far there are no assumptions concerning an overall joint pmf from which all 16 marginal pmfs can be derived. This should be evident from the fact that the QM probabilities from Table~\ref{tab:qmTypeIntro} are a special case of the ones in Table~\ref{tab:genericTypeIntro}, and it is well known that there does not always exist an overall joint distribution from which all QM probabilities can be derived as marginals.

If one \emph{does} make such an assumption, however, then is is easy to show, by marginalization, that the resulting four sets of marginal probabilities must be of the \emph{form} shown in Table~\ref{tab:predeterminedTypeIntro}. If these are viewed as double conditional probabilities, they define the noncontextual (NC) model type. It is assumed that $\pmb{\rho}=(\rho_1,\rho_2,...,\rho_{16})$ is a pmf, that is, $0 \leq \rho_k \leq 1$ for all $k=1,2,..,16$, and $\sum_{k=1}^{16} \rho_k=1$. Note that the $\rho_k$'s do not \emph{individually} denote conditional probabilities, only appropriate sums of four as indicated. In terms of generic parameters, for example, 
\[\gamma_7=P(A=1,B=-1|M_A=1,M_B=2)=\rho_3+\rho_4+\rho_{11}+\rho_{12}\] 
is the conditional probability in row 2, column 3.

\begin{table}[H]
\setlength\arraycolsep{2pt}
\caption{The \textbf{noncontextual (NC)} model type. Each entry represents a conditional probability $P(A=s,B=t|M_A=u,M_B=v)$ where $s,t=\pm1$ denote Alice and Bob's measurement outcomes and $u,v=1 \text{ or } 2$ indicate their measurement choices (context), respectively. Since the $\rho_k$'s form a pmf, the conditional probabilities in each row sum to $\sum_{k=1}^{16} \rho_k=1$.} 
\label{tab:predeterminedTypeIntro}
\[
\begin{array}{|c|c|c|c|c|}
\hline
\text{Context} & \multicolumn{4}{c|}{\text{Outcome }(s,t)} \\
\hline
(u,v) & (-1,-1)  & (-1,1)   & (1,-1)   & (1,1) \\
\hline
(1,1) & \rho_1+\rho_2+\rho_5+\rho_6 & \rho_9+\rho_{10}+\rho_{13}+\rho_{14} & \rho_3+\rho_4+\rho_7+\rho_8 & \rho_{11}+\rho_{12}+\rho_{15}+\rho_{16} \\
\hline
(1,2) & \rho_1+\rho_2+\rho_9+\rho_{10} & \rho_5+\rho_{6}+\rho_{13}+\rho_{14} & \rho_3+\rho_4+\rho_{11}+\rho_{12} & \rho_{7}+\rho_{8}+\rho_{15}+\rho_{16} \\
\hline
(2,1) & \rho_1+\rho_3+\rho_5+\rho_7 & \rho_9+\rho_{11}+\rho_{13}+\rho_{15} & \rho_2+\rho_4+\rho_6+\rho_8 & \rho_{10}+\rho_{12}+\rho_{14}+\rho_{16} \\
\hline
(2,2) & \rho_1+\rho_3+\rho_9+\rho_{11} & \rho_5+\rho_{7}+\rho_{13}+\rho_{15} & \rho_2+\rho_4+\rho_{10}+\rho_{12} & \rho_{6}+\rho_{8}+\rho_{14}+\rho_{16} \\
\hline
\end{array}
\]
\end{table}
The correlations and $s$-functions for the noncontextual model type are shown in Table~\ref{tab:preDetCandSTable2Intro}. 

\begin{table}[H]
\caption{The correlations and $s$-functions for the noncontextual type. The fact that $\pmb{\rho}=(\rho_1,\rho_2,...,\rho_{16})$ is a pmf guarantees that all $s$-functions lie between -2 and +2. Hence all 8 CHSH inequalities are satisfied.}
\label{tab:preDetCandSTable2Intro}
\[
\begin{array}{|c||l|}
\hline
w & \rho_1+\rho_2-\rho_3-\rho_4+\rho_5+\rho_6-\rho_7-\rho_8- \\
& \rho_9-\rho_{10}+\rho_{11}+\rho_{12}-\rho_{13}-\rho_{14}+\rho_{15}+\rho_{16} \\
\hline
x & \rho_1+\rho_2-\rho_3-\rho_4-\rho_5-\rho_6+\rho_7+\rho_8+ \\
& \rho_9+\rho_{10}-\rho_{11}-\rho_{12}-\rho_{13}-\rho_{14}+\rho_{15}+\rho_{16} \\
\hline
y & \rho_1-\rho_2+\rho_3-\rho_4+\rho_5-\rho_6+\rho_7-\rho_8- \\
& \rho_9+\rho_{10}-\rho_{11}+\rho_{12}-\rho_{13}+\rho_{14}-\rho_{15}+\rho_{16} \\
\hline
z & \rho_1-\rho_2+\rho_3-\rho_4-\rho_5+\rho_6-\rho_7+\rho_8+ \\
& \rho_9-\rho_{10}+\rho_{11}-\rho_{12}-\rho_{13}+\rho_{14}-\rho_{15}+\rho_{16} \\
\hline
\hline
s_1 & 2(\rho_1-\rho_2+\rho_3-\rho_4-\rho_5-\rho_6+\rho_7+\rho_8+ \\ 
& \rho_9+\rho_{10}-\rho_{11}-\rho_{12}-\rho_{13}+\rho_{14}-\rho_{15}+\rho_{16}) \\
\hline
s_2 & 2(\rho_1-\rho_2+\rho_3-\rho_4+\rho_5+\rho_6-\rho_7-\rho_8- \\
& \rho_9-\rho_{10}+\rho_{11}+\rho_{12}-\rho_{13}+\rho_{14}-\rho_{15}+\rho_{16}) \\
\hline
s_3 & 2(\rho_1+\rho_2-\rho_3-\rho_4-\rho_5+\rho_6-\rho_7+\rho_8+ \\
& \rho_9-\rho_{10}+\rho_{11}-\rho_{12}-\rho_{13}-\rho_{14}+\rho_{15}+\rho_{16}) \\
\hline
s_4 & 2(\rho_1+\rho_2-\rho_3-\rho_4+\rho_5-\rho_6+\rho_7-\rho_8- \\
& \rho_9+\rho_{10}-\rho_{11}+\rho_{12}-\rho_{13}-\rho_{14}+\rho_{15}+\rho_{16}) \\
\hline
\hline
\text{Min and Max} & \multicolumn{1}{c|}{-2 \text{ and } 2}\\
\hline
\end{array}
\]
\end{table}
Note how the parameterized operational modeling approach makes the following statement obvious. 

\begin{Theorem}
All  8 CHSH inequalities are satisfied for \emph{every} NC model instance.
\label{thm:ncImpliesCHSH}
\end{Theorem}
\begin{proof}
The range of each $s$-function is restricted to $[-2,2]$, since $\pmb{\rho}=(\rho_1,\rho_2,...,\rho_{16})$ is a pmf by definition of the NC model type.
\end{proof}

\subsection{Factorizable}
\label{sec:fcType}

The model types RCI+OD and factorizable are relevant to Bell's original derivations of his inequalities~\cite{Bell1964,Bell1976}. It turns out that both of these are special cases of noncontextuality. RCI+OD is defined in Sec.~\ref{sec:OD}. Factorizability is defined here.

\begin{Definition} A model instance will be called \textbf{factorizable} or \textbf{Bell local} if 
\label{def:factorization}
\begin{align}
&P(A=s,B=t|M_{A}=u,M_{B}=v)= \nonumber \\
&P(A=s|M_{A}=u)\times P(B=t|M_{B}=v).
\end{align} 
for all $s,t=\pm1$ and $u,v=1,2$. It is sometimes said that the \textbf{factorization condition} (\textbf{FC}) is satisfied. Both~\cite{Wiseman2014} and~\cite{StanfordBell2021}  use the term \textit{Bell local} to describe models that satisfy the factorization condition.
\end{Definition}

Factorizability is closely related to local causality. Bell developed the idea of local causality in~\cite{Bell1976}. Also see~\cite{Wiseman2014} for a discussion of this topic. It was used by Bell to capture the idea that anything outside of Alice's past light cone, such as Bob's settings or outcomes, are statistically irrelevant to her outcome, and similarly for Bob's outcome with regard to Alice's actions or observations. 

\begin{Definition} \textbf{Local causality} or \textbf{LC}:
\begin{align}
&P(A=s|B=t,M_{A}=u,M_{B}=v)=P(A=s|M_{A}=u) \text{ and } \nonumber \\
&P(B=t|A=s,M_{A}=u,M_{B}=v)=P(B=t|M_{B}=v),
\end{align} 
for all $s,t=\pm1$ and $u,v=1,2$, but only if all of the conditional probabilities involved are well-defined. 
\label{def:LC}
\end{Definition}
The reason for caution in the definition of LC is that the conditional probabilities on the left may be undefined, because of possible zero conditioning probabilities. However, if all conditional probabilities are well-defined, it is equivalent to factorizability. See Sec.~\ref{sec:lcDefineCond} for details. Based on these definitions, the factorizable model type is defined by Table~\ref{tab:bellLocalTypeIntro}.

\begin{table}[H]
\caption{The conditional probabilities $P(A=s,B=t|M_A=u,M_B=v)$ (where $s,t=\pm1$ and $u,v=1,2$) for the \textbf{factorizable} (or \textbf{Bell local}) model type are characterized by 8 parameters $\alpha_k \text{ and } \beta_k$ for $k=1,2,3,4$. These conditional probabilities satisfy the measurement context constraints (each row sums to 1) because of the factorization constraints in Eq.~\ref{eq:alphabetaIntro}.}
\label{tab:bellLocalTypeIntro}
\[
\begin{array}{|c|c|c|c|c|}
\hline
\text{Context} & \multicolumn{4}{c|}{\text{Outcome }(s,t)} \\
\hline
(u,v) & (-1,-1)  & (-1,1)   & (1,-1)   & (1,1) \\
\hline
(1,1) & \alpha_1 \beta_1 & \alpha_1 \beta_3 & \alpha_3 \beta_1 & \alpha_3 \beta_3 \\
\hline
(1,2) & \alpha_1 \beta_2 & \alpha_1 \beta_4 & \alpha_3 \beta_2 & \alpha_3 \beta_4 \\
\hline
(2,1) & \alpha_2 \beta_1 & \alpha_2 \beta_3 & \alpha_4 \beta_1 & \alpha_4 \beta_3 \\
\hline
(2,2) & \alpha_2 \beta_2 & \alpha_2 \beta_4 & \alpha_4 \beta_2 & \alpha_4 \beta_4 \\
\hline
\end{array}
\]
\end{table}
As an example of the correspondence to the generic parameters, the entry in row 2, column 3 is 
\[\gamma_7=P(A=1,B=-1|M_A=1,M_B=2)=\alpha_3 \beta_2.\] 

\begin{Definition}
The \textbf{factorization parameters} are defined by
\begin{align}
\label{eq:blVarDef}
&\alpha_{1} = P(A=-1|M_A=1), &&\beta_{1} = P(B=-1|M_B=1), \nonumber \\
&\alpha_{2} = P(A=-1|M_A=2), && \beta_{2} = P(B=-1|M_B=2), \nonumber \\
&\alpha_{3} = P(A=+1|M_A=1),  && \beta_{3} = P(B=+1|M_B=1), \\
&\alpha_{4} = P(A=+1|M_A=2),  &&\beta_{4} = P(B=+1|M_B=2). \nonumber
\end{align}
\end{Definition}

\begin{Definition}
The factorization parameters satisfy the \textbf{factorization constraints} given by 
\begin{align}
\label{eq:alphabetaIntro}
\alpha_{1}+\alpha_{3}=\alpha_{2}+\alpha_{4}=1, \; \beta_{1}+\beta_{3}=\beta_{2}+\beta_{4}=1.
\end{align}
\end{Definition}

\subsection{Factorization implies noncontextuality}
\label{sec:fcImpliesNC}

\begin{Theorem}
The factorizable model type is a special case of the noncontextual model type.
\label{thm:fcImpliesPDIntro}
\end{Theorem}
\begin{proof}
Assume a set of factorizable conditional probabilities (Table~\ref{tab:bellLocalTypeIntro}) based on parameters 
$\alpha_1,\alpha_2,\alpha_3,\alpha_4$ and $\beta_1,\beta_2,\beta_3,\beta_4$. To construct the corresponding noncontextual conditional probabilities (Table~\ref{tab:predeterminedTypeIntro}), simply assign the parameters $\pmb{\rho}=(\rho_1,\rho_2,...,\rho_{16})$ according to Eq.~\ref{eq:fcImpliesPDIntro}. Then use the factorization constraints of Eq.~\ref{eq:alphabetaIntro} to show that $\pmb{\rho}$ is in fact a pmf, and that if one substitutes these values for the $\rho_k$'s into Table~\ref{tab:predeterminedTypeIntro} one gets Table~\ref{tab:bellLocalTypeIntro}.

\begin{equation}
\label{eq:fcImpliesPDIntro}
\left(
\begin{array}{l}
\rho_1 \\
\rho_2 \\
\rho_3 \\
\rho_4 \\

\rho_5 \\
\rho_6 \\
\rho_7 \\
\rho_8 \\

\rho_9 \\
\rho_{10} \\
\rho_{11} \\
\rho_{12} \\

\rho_{13} \\
\rho_{14} \\
\rho_{15} \\
\rho_{16} \\
\end{array}
\right)=
\left(
\begin{array}{l}
\alpha_1 \alpha_2 \beta_1 \beta_2 \\
\alpha_1 \alpha_4 \beta_1 \beta_2 \\
\alpha_2 \alpha_3 \beta_1 \beta_2 \\
\alpha_3 \alpha_4 \beta_1 \beta_2 \\

\alpha_1 \alpha_2 \beta_1 \beta_4 \\
\alpha_1 \alpha_4 \beta_1 \beta_4 \\
\alpha_2 \alpha_3 \beta_1 \beta_4 \\
\alpha_3 \alpha_4 \beta_1 \beta_4 \\

\alpha_1 \alpha_2 \beta_2 \beta_3 \\
\alpha_1 \alpha_4 \beta_2 \beta_3 \\
\alpha_2 \alpha_3 \beta_2 \beta_3 \\
\alpha_3 \alpha_4 \beta_2 \beta_3 \\

\alpha_1 \alpha_2 \beta_3 \beta_4 \\
\alpha_1 \alpha_4 \beta_3 \beta_4 \\
\alpha_2 \alpha_3 \beta_3 \beta_4 \\
\alpha_3 \alpha_4 \beta_3 \beta_4 \\
\end{array}
\right)
\end{equation}
\end{proof}
This statement can be strengthened to say that factorizability is a \emph{strictly stronger} classical notion than noncontextuality, meaning that there is at least one noncontextual cpv that is not also factorizable (in fact infinitely many). See Example~\ref{ex:pdNotFCIntro} in Sec.~\ref{sec:NCDoesNotImplyFC}.

The correlations and $s$-functions for the factorizable type are shown in Table~\ref{tab:blCandSTableIntro}. Being a special case of the noncontextual type, all of its $s$-functions also satisfy all 8 CHSH inequalities.
Note that since the single conditional expectations are given by
\begin{align}
&E[A|M_{A}=1]=\alpha_3-\alpha_1 \; & E[B|M_{B}=1]=\beta_3-\beta_1,  \nonumber \\
&E[A|M_{A}=2]=\alpha_4-\alpha_2 \; & E[B|M_{B}=2]=\beta_4-\beta_2,
\end{align}
it follows that each correlation (conditional expectation of the product) is the product of the corresponding single conditional expectations.
\begin{align}
w=E[AB|M_A=1,M_B=1]=E[A|M_A=1] \times E[B|M_B=1], \nonumber \\
x=E[AB|M_A=1,M_B=2]=E[A|M_A=1] \times E[B|M_B=2],  \label{eq:wxyzExpectationsIntro} \\
y=E[AB|M_A=2,M_B=1]=E[A|M_A=2] \times E[B|M_B=1], \nonumber \\
z=E[AB|M_A=2,M_B=2]=E[A|M_A=2] \times E[B|M_B=2]. \nonumber
\end{align}
\begin{table}[H]
\caption{The correlations and $s$-functions for the factorizable (Bell local) type. As a special case of the NC model type (Thm.~\ref{thm:fcImpliesPDIntro}), all factorizable model instances satisfy all CHSH inequalities.}
\label{tab:blCandSTableIntro}
\[
\begin{array}{|c||l|}
\hline
w & (\alpha_3-\alpha_1)(\beta_3-\beta_1) \\
\hline
x & (\alpha_3-\alpha_1)(\beta_4-\beta_2) \\
\hline
y & (\alpha_4-\alpha_2)(\beta_3-\beta_1) \\
\hline
z & (\alpha_4-\alpha_2)(\beta_4-\beta_2) \\
\hline
\hline
s_1 & -(\alpha_3-\alpha_1)(\beta_3-\beta_1)+(\alpha_3-\alpha_1)(\beta_4-\beta_2)+ \\& (\alpha_4-\alpha_2)(\beta_3-\beta_1)+(\alpha_4-\alpha_2)(\beta_4-\beta_2) \\
\hline
s_2 & (\alpha_3-\alpha_1)(\beta_3-\beta_1)-(\alpha_3-\alpha_1)(\beta_4-\beta_2)+ \\
& \alpha_4-\alpha_2)(\beta_3-\beta_1)+(\alpha_4-\alpha_2)(\beta_4-\beta_2) \\
\hline
s_3 & (\alpha_3-\alpha_1)(\beta_3-\beta_1)+(\alpha_3-\alpha_1)(\beta_4-\beta_2)- \\
& (\alpha_4-\alpha_2)(\beta_3-\beta_1)+(\alpha_4-\alpha_2)(\beta_4-\beta_2) \\
\hline
s_4 & (\alpha_3-\alpha_1)(\beta_3-\beta_1)+(\alpha_3-\alpha_1)(\beta_4-\beta_2)+ \\
& (\alpha_4-\alpha_2)(\beta_3-\beta_1)-(\alpha_4-\alpha_2)(\beta_4-\beta_2) \\
\hline
\hline
\text{Min and Max} & \multicolumn{1}{c|}{-2 \text{ and } 2}\\
\hline
\end{array}
\]
\end{table}

From either Eq.~\ref{eq:wxyzExpectationsIntro} or Table~\ref{tab:blCandSTableIntro}, it is easy to see that for the factorizable correlations $(w,x,y,z)$,
\[wz=xy,\]
and therefore
\begin{equation}
\left\{ \begin{array}{ll}
               z=w^{-1}{xy}  &  \mbox{if $w \neq 0$} \\
               xy=0            & \mbox{if $w=0$.}
        \end{array}
\right.
\end{equation}
This explains the shapes of the correlation plots in the factorizable row 3 in Fig.~\ref{fig:models5SelectedIntro}. At the far left ($w=-1$) and far right ($w=1$), the surfaces are hyperbolic paraboloids. In the middle ($w=0$), $xy=0$ determines two intersecting planes.

\subsection{Outcome determinism}
\label{sec:OD}

\begin{Definition}
\label{def:ODIntro}
A generic model instance will be called \textbf{outcome deterministic} (\textbf{OD}) if $\gamma_k=0 \text{ or } 1$ for all $k$. 
\end{Definition}

An example is shown in Table~\ref{tab:odExampleIntro}. There are 256 model instances like this (4 ways to place exactly one 1 in each of the four rows, hence a total of $4^4=256$ such arrangements of 1's).
\begin{table}[H]
\caption{Example of an \textbf{outcome deterministic (OD)} model instance. There are exactly 256 such model instances (including this one). Obviously each row sums to 1 in all cases.}
\label{tab:odExampleIntro}
\[
\begin{array}{|c|c|c|c|c|}
\hline
\text{Context} & \multicolumn{4}{c|}{\text{Outcome }(s,t)} \\
\hline
(u,v) & (-1,-1)  & (-1,1)   & (1,-1)   & (1,1) \\
\hline
(1,1) & 1 & 0 & 0 & 0 \\
\hline
(1,2) & 1 & 0 & 0 & 0 \\
\hline
(2,1) & 1 & 0 & 0 & 0 \\
\hline
(2,2) & 1 & 0 & 0 & 0 \\
\hline
\end{array}
\]
\end{table}
These turn out to be the fundamental building blocks for all other model types (including QM), via convex combinations. In Sec.~\ref{sec:hvCVHEquivalent}, the traditional hidden variables approach is reformulated in terms of convex combinations of finite ensembles of OD model instances.

It is fairly obvious that all of the correlations of any of the 256 OD model instances must be $\pm1$. See Def.~\ref{def:ODIntro} and Eq.~\ref{eq:defineCorrIntro}. This property defines an even larger (in fact infinite) class of model instances as follows.

\begin{Definition}
A model instance will be called \textbf{perfectly correlated} (\textbf{PC}) if its correlations are all $\pm1$. In other words, in terms of generic parameters,
\begin{align}
\label{eq:perfectCorr}
 &w=\gamma_{1}-\gamma_{2}-\gamma_{3}+\gamma_{4}=\pm1, \nonumber \\
 &x=\gamma_{5}-\gamma_{6}-\gamma_{7}+\gamma_{8}=\pm1, \nonumber \\
 &y=\gamma_{9}-\gamma_{10}-\gamma_{11}+\gamma_{12}=\pm1, \\
 &z=\gamma_{13}-\gamma_{14}-\gamma_{15}+\gamma_{16}=\pm1. \nonumber
\end{align}
\label{def:perfectCorr}
\end{Definition}
Since the subsets of $\gamma_{k}$'s that go into defining each of the correlations are mutually disjoint, it is clear that $(w,x,y,z)$ can be any of 16 possible sequences of $\pm1$'s. The $s$-functions are determined by the patterns of the $\pm1$'s according to Table~\ref{tab:perfectCorrTable}. Although the arrangement of numbers in any row of the last four columns might be different, the numbers that appear will be the same. 
Table~\ref{tab:perfectCorrTable} shows that perfectly correlated model instances, including the 256 OD model instances, are inconsistent with the predictions of QM, either by having $s$-functions that are "too small" or "too big". 

\begin{table}[H]
\caption{All possible values of $s$-functions for a perfectly correlated model instance. Note there are cases where the CHSH inequalities are satisfied and others where even a Tsirelson bound is exceeded.}
\label{tab:perfectCorrTable}
\[
\begin{array}{|c|c|r|r|r|r|}
\hline
\multicolumn{2}{|c|}{\text{Perfect correlations}} & \multicolumn{4}{c|}{\text{Typical s-functions }} \\
\hline
\text{No. -1's} & \text{No. +1's}  & s_1   & s_2   & s_3 & s_4 \\
\hline
4 & 0 & -2 & -2 & -2 & -2 \\
\hline
3 & 1 & 0 & 0 & 0 & -4 \\
\hline
2 & 2 & 2 & 2 & -2 & -2 \\
\hline
1 & 3 & 4 & 0 & 0 & 0 \\
\hline
0 & 4 & 2 & 2 & 2 & 2 \\
\hline
\end{array}
\]
\end{table}
The instances that have $s$-functions which exceed one of the Tsirelson bounds come from a set of correlations with either one +1 and three -1's or a set with one -1 and three +1's (table~\ref{tab:perfectCorrTable}). Consider the case with one -1 and three +1's. There are 4 ways to choose which of $w,x,y,z$ will be the -1. Within that correlation definition, there are just two $\gamma_k$ choices to set to 1 in order for that to happen. In the remaining three correlations, there are two choices for which $\gamma_k$ to set to 1 as well. This yields $4 \times 2 \times 2 \times 2 \times 2=64$ ways to get one -1 and three +1's. The same argument works for one +1 and three -1's. Hence a total of $64+64=128$ sets of correlations with either one -1 and three +1's or one +1 and three -1's, which are the ones with an $s$-function that violates a CHSH inequality. This leaves $256-128=128$ sets of correlations with either zero, two, or four +1's (equivalent to four, two, or zero -1's). These are the cases that satisfy CHSH.

\subsection{RCI+OD}
\label{sec:rciOD}

Only 16 of 256 OD model instances satisfy RCI. Their cpvs are shown as columns in Table~\ref{tab:rciODTabIntro}.\footnote{This format was chosen because 16 individual tables take up too much space.}
\begin{Definition} A model instance satisfies \textbf{remote context independence (RCI)} if:  
\begin{align}
&P(A=s|M_{A}=u,M_{B}=v)=P(A=s|M_{A}=u) \text{ and } \nonumber \\
&P(B=t|M_{A}=u,M_{B}=v)=P(B=t|M_{B}=v)
\end{align} 
for all $s,t=\pm1$ and $u,v=1,2$. That is, Alice's outcome probability conditioned on her measurement choice is independent of Bob's measurement choice, and vice-versa. RCI is sometimes called \textbf{parameter independence}.
\label{def:RCI}
\end{Definition}
\begin{table}[H]
\caption{The 16 \textbf{RCI+OD} model instances (shown as columns). They are also factorizable and noncontextual. Any NC (including factorizable) model instance can be written as a convex combination of these 16. The first column (1) corresponds to the model instance shown in Table~\ref{tab:odExampleIntro}.}
\label{tab:rciODTabIntro}
\[
\begin{array}{|r|r|r|r||c|c|c|c|c|c|c|c|c|c|c|c|c|c|c|c|}
\hline
s & t & u & v & 1 & 2 & 3 & 4 & 5 & 6 & 7 & 8 & 9 & 10 & 11 & 12 & 13 & 14 & 15 & 16 \\
\hline
\hline
-1 & -1 & 1 & 1 & 1 & 1 & 0 & 0 & 1 & 1 & 0 & 0 & 0 & 0 & 0 & 0 & 0 & 0 & 0 & 0 \\
\hline
-1 & 1 & 1 & 1 & 0 & 0 & 0 & 0 & 0 & 0 & 0 & 0 & 1 & 1 & 0 & 0 & 1 & 1 & 0 & 0 \\
\hline
1 & -1 & 1 & 1 & 0 & 0 & 1 & 1 & 0 & 0 & 1 & 1 & 0 & 0 & 0 & 0 & 0 & 0 & 0 & 0 \\
\hline
1 & 1 & 1 & 1 & 0 & 0 & 0 & 0 & 0 & 0 & 0 & 0 & 0 & 0 & 1 & 1 & 0 & 0 & 1 & 1 \\
\hline

-1 & -1 & 1 & 2 & 1 & 1 & 0 & 0 & 0 & 0 & 0 & 0 & 1 & 1 & 0 & 0 & 0 & 0 & 0 & 0 \\
\hline
-1 & 1 & 1 & 2 & 0 & 0 & 0 & 0 & 1 & 1 & 0 & 0 & 0 & 0 & 0 & 0 & 1 & 1 & 0 & 0 \\
\hline
1 & -1 & 1 & 2 & 0 & 0 & 1 & 1 & 0 & 0 & 0 & 0 & 0 & 0 & 1 & 1 & 0 & 0 & 0 & 0 \\
\hline
1 & 1 & 1 & 2 & 0 & 0 & 0 & 0 & 0 & 0 & 1 & 1 & 0 & 0 & 0 & 0 & 0 & 0 & 1 & 1 \\
\hline

-1 & -1 & 2 & 1 & 1 & 0 & 1 & 0 & 1 & 0 & 1 & 0 & 0 & 0 & 0 & 0 & 0 & 0 & 0 & 0 \\
\hline
-1 & 1 & 2 & 1 & 0 & 0 & 0 & 0 & 0 & 0 & 0 & 0 & 1 & 0 & 1 & 0 & 1 & 0 & 1 & 0 \\
\hline
1 & -1 & 2 & 1 & 0 & 1 & 0 & 1 & 0 & 1 & 0 & 1 & 0 & 0 & 0 & 0 & 0 & 0 & 0 & 0 \\
\hline
1 & 1 & 2 & 1 & 0 & 0 & 0 & 0 & 0 & 0 & 0 & 0 & 0 & 1 & 0 & 1 & 0 & 1 & 0 & 1 \\
\hline

-1 & -1 & 2 & 2 & 1 & 0 & 1 & 0 & 0 & 0 & 0 & 0 & 1 & 0 & 1 & 0 & 0 & 0 & 0 & 0 \\
\hline
-1 & 1 & 2 & 2 & 0 & 0 & 0 & 0 & 1 & 0 & 1 & 0 & 0 & 0 & 0 & 0 & 1 & 0 & 1 & 0 \\
\hline
1 & -1 & 2 & 2 & 0 & 1 & 0 & 1 & 0 & 0 & 0 & 0 & 0 & 1 & 0 & 1 & 0 & 0 & 0 & 0 \\
\hline
1 & 1 & 2 & 2 & 0 & 0 & 0 & 0 & 0 & 1 & 0 & 1 & 0 & 0 & 0 & 0 & 0 & 1 & 0 & 1 \\
\hline
\end{array}
\]
\end{table}
  
\begin{figure}[H]
\centering
\includegraphics[width=0.5\linewidth]{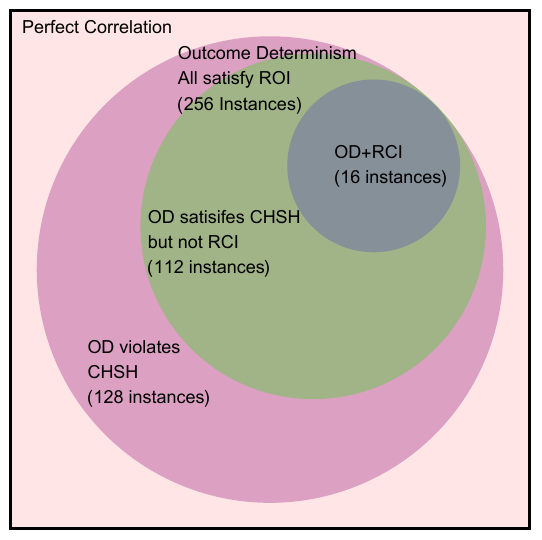}
\caption{Relationships among classes of OD model instances. All 256 of them satisfy ROI. Only 16 satisfy RCI. 112 others satisfy CHSH but dot not satisfy RCI, hence are neither noncontextual or factorizable. The remaining 128 not only violate CHSH but exceed a Tsirelson bound.}
\label{fig:vennDeterminism}
\end{figure}

\emph{Mathematica}\footnote{The computational workhorse behind everything in this paper is the Wolfram language platform \emph{Mathematica}, including the initial idea exploration and feasibility analysis, the heavy-duty matrix generation and calculations, function optimization and equation solving, the computation of expectations, marginal and conditional distributions, and especially the generation of the plots and visualizations.} was used to identify these 16 special model instances. In fact, \emph{Mathematica} was used to classify \emph{all} OD model instances according to certain  properties. Fig.~\ref{fig:vennDeterminism} shows the finite set of OD model instances embedded in the infinite set of perfectly correlated model instances. Subsets that satisfy various combinations of properties are indicated, together with their numbers, as follows:
\begin{itemize}
\item RCI+OD (16)
\item CHSH+OD (128)
\item CHSH+OD but do not satisfy RCI (112)
\item ROI+OD (256)
\item ROI+OD but do not satisfy CHSH (128)
\end{itemize}

\begin{Definition} A model instance satisfies \textbf{remote outcome independence (ROI)} if: 
\begin{align}
&P(A=s,B=t|M_{A}=u,M_{B}=v)= \nonumber \\
&P(A=s|M_{A}=u,M_{B}=v) \times P(B=t|M_{A}=u,M_{B}=v).
\end{align}
for all $s,t=\pm1$ and $u,v=1,2$. That is, the outcomes $A$ and $B$ are \textbf{conditionally independent} given the measurement choice pair $(M_{A},M_{B})$.
\label{def:ROI}
\end{Definition}
 
Here are examples from each of the four regions shown in Fig.~\ref{fig:vennDeterminism}.
\begin{enumerate}
\item PC but not OD. Choose generic parameters 
\[\gamma_1=\gamma_4=\gamma_6=\gamma_7=\gamma_{10}=\gamma_{11}=\gamma_{13}=\gamma_{16}=\frac{1}{2} \text{ and all other } \gamma_k=0.\]
Then the correlations are (see Eq.~\ref{eq:defineCorrIntro})
\[(w,x,y,z)=(1,-1,-1,1).\]

\item OD but does not satisfy CHSH. Choose generic parameters
\[\gamma_2=\gamma_6=\gamma_{10}=\gamma_{13}=1 \text{ and all other } \gamma_k=0.\]
Then the correlations are (see Eq.~\ref{eq:defineCorrIntro})
\[(w,x,y,z)=(-1,-1,-1,1)\] and the $s$-functions are (see Eq.~\ref{eq:defineSFunIntro})
\[(s_1,s_2,s_3,s_4)=(0,0,0,-4).\]

\item OD and satisfies CHSH but is not RCI. Choose generic parameters
\[\gamma_1=\gamma_7=\gamma_{9}=\gamma_{14}=1 \text{ and all other } \gamma_k=0.\]
Then the correlations are (see Eq.~\ref{eq:defineCorrIntro})
\[(w,x,y,z)=(1,-1,1,-1)\] and the $s$-functions are (see Eq.~\ref{eq:defineSFunIntro})
\[(s_1,s_2,s_3,s_4)=(-2,2,-2,2),\] but this is not one of the sixteen RCI+OD instances in Table~\ref{tab:rciODTabIntro}.
\label{ex:chshNotPD}
\item OD and satisfies RCI. See any one of the 16 examples in Table~\ref{tab:rciODTabIntro}.
\end{enumerate}

\subsection{Noncontextuality built up from RCI+OD model instances}
\label{sec:cvhRCIODtoNC}

In this section, it is shown that the convex hull of the RCI+OD model instances is precisely the set of NC model instances. In Appendix~\ref{sec:cvhGeneric}, a more general result is proved, namely that \emph{any} generic model  instance, including \emph{all} QM ones, can be written as a convex combination of 256 (or fewer) OD model instances. The process of constructing interesting families of cpvs from \emph{finite} ensembles of more basic cpvs via convex combinations adds some intuition to the traditional hidden variables representation. In Sec.~\ref{sec:hvCVHEquivalent}, the (almost) equivalence of the two approaches is demonstrated. 

\begin{Theorem}
\label{thm:cvhPDIntro}
Any NC model instance can be written as a convex combination of the 16 RCI+OD model instances (cpvs) shown as \underline{columns} in Table~\ref{tab:rciODTabIntro}.
\end{Theorem}

\begin{proof}
Define the matrix $R$ as follows and consider Eq.~\ref{eq:RtimesRhoTIntro}.
\begin{Definition}
\label{def:rMatrix}
The matrix $R$ is the $16 \times 16$ matrix consisting of the last 16 columns of Table~\ref{tab:rciODTabIntro}. $R$ turns a pmf $\pmb{\rho}=(\rho_1,\rho_2,...,\rho_{16})$ into a NC conditional probability vector (see Eq.~\ref{eq:RtimesRhoTIntro}). $R$ is an example of a \textbf{Bell-type scenario incidence matrix} in~\cite{AbramskyBrandenburger2011}.
\end{Definition}

\begin{equation}
\label{eq:RtimesRhoTIntro}
R \pmb{\rho}^T = 
\left(
\begin{array}{l}
\rho_{1}+\rho_{2}+\rho_{5}+\rho_{6}\\
\rho_{9}+\rho_{10}+\rho_{13}+\rho_{14}\\
\rho_{3}+\rho_{4}+\rho_{7}+\rho_{8}\\
\rho_{11}+\rho_{12}+\rho_{15}+\rho_{16}\\

\rho_{1}+\rho_{2}+\rho_{9}+\rho_{10}\\
\rho_{5}+\rho_{6}+\rho_{13}+\rho_{14}\\
\rho_{3}+\rho_{4}+\rho_{11}+\rho_{12}\\
\rho_{7}+\rho_{8}+\rho_{15}+\rho_{16}\\

\rho_{1}+\rho_{3}+\rho_{5}+\rho_{7}\\
\rho_{9}+\rho_{11}+\rho_{13}+\rho_{15}\\
\rho_{2}+\rho_{4}+\rho_{6}+\rho_{8}\\
\rho_{10}+\rho_{12}+\rho_{14}+\rho_{16}\\

\rho_{1}+\rho_{3}+\rho_{9}+\rho_{11}\\
\rho_{5}+\rho_{7}+\rho_{13}+\rho_{15}\\
\rho_{2}+\rho_{4}+\rho_{10}+\rho_{12}\\
\rho_{6}+\rho_{8}+\rho_{14}+\rho_{16}\\
\end{array}
\right)
\end{equation}

Since $\pmb{\rho}=(\rho_1,\rho_2,...,\rho_{16})$ is a pmf, the expression on the left is a convex combination of the columns of $R$. The column vector on the right constitutes the conditional probabilities of the NC model type -- see Table~\ref{tab:predeterminedTypeIntro} for comparison.
\end{proof}

\begin{figure}[H]
\centering
\includegraphics[width=0.4\linewidth]{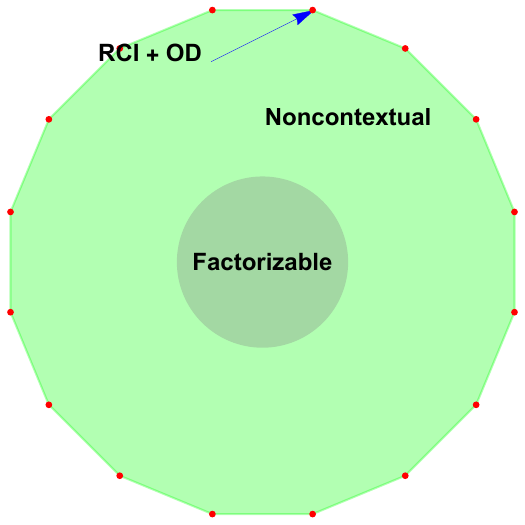}
\caption{The convex hull of the 16 RCI+OD cpvs (red dots) equals the full set of NC cpvs, including the factorizable (Bell local) ones.}
\label{fig:preDetConvexHullIntro}
\end{figure}

Fig.~\ref{fig:preDetConvexHullIntro} illustrates Thm.~\ref{thm:cvhPDIntro}. This is a schematic representation of a polytope in a 9-dimensional subspace of $\mathbb{R}^{16}$. The rank of $R$ was computed using \textit{Mathematica}'s \texttt{MatrixRank} function. 

\begin{Corollary}
\label{cor:rciODNC}
All of the 16 RCI+OD model instances of Table~\ref{tab:rciODTabIntro} are noncontextual.
\end{Corollary}
\begin{proof}
In Eq.~\ref{eq:RtimesRhoTIntro}, for any $k=1,2,...,16$, set $\rho_k=1$ and $\rho_j=0$ for all $j \neq k$. This produces the $k$th RCI+OD cpv, that is, the $k$th column of Table~\ref{tab:rciODTabIntro} (also the $k$th column of $R$, by definition).
\end{proof} 

\begin{Corollary}
All of the 16 RCI+OD model instances of Table~\ref{tab:rciODTabIntro} are factorizable.
\end{Corollary}
\begin{proof}
As an example, it is  proved that the first cpv (column 1) in Table~\ref{tab:rciODTabIntro} is factorizable. The proofs for the other 15 columns are similar. The proof of Corollary~\ref{cor:rciODNC} showed that the first RCI+OD cpv is noncontextual -- simply set $\rho_1=1$ and $\rho_j=0$ for all $j \neq 1$. Then Eq.~\ref{eq:fcImpliesPDIntro} makes it obvious what to do. Set $\alpha_1=\alpha_2=1$, $\beta_1=\beta_2=1$ and $\alpha_j=\beta_j=0$ for all $j \neq 1 \text{ or } 2$. Hence the first RCI+OD model instance is a legitimate instance of the factorizable model type shown in Table~\ref{tab:bellLocalTypeIntro}.
\end{proof} 
The correlations and $s$-functions for the 16 RCI+OD model instances are shown in Table~\ref{tab:odCandSTableIntro}. Clearly all of the CHSH inequalities are satisfied; in fact the $s$-functions attain these extreme values exactly as functions of the correlations of the RCI+OD cpvs.
\begin{table}[H]
\setlength\arraycolsep{2pt}
\caption{The correlations and $s$-functions for the 16 RCI+OD model instances. See Table~\ref{tab:rciODTabIntro} for the corresponding cpv's.}
\label{tab:odCandSTableIntro}
\[
\begin{array}{|c||r|r|r|r|r|r|r|r|r|r|r|r|r|r|r|r|}
\hline
\text{No.} &1 & 2 & 3 & 4 & 5 & 6 & 7 & 8 & 9 & 10 & 11 & 12 & 13 & 14 & 15 & 16 \\
\hline
\hline
w & 1 & 1 & -1 & -1 & 1 & 1 & -1 & -1 & -1 & -1 & 1 & 1 & -1 & -1 & 1 & 1 \\
\hline
x & 1 & 1 & -1 & -1 & -1 & -1 & 1 & 1 & 1 & 1 & -1 & -1 & -1 & -1 & 1 & 1 \\
\hline
y & 1 & -1 & 1 & -1 & 1 & -1 & 1 & -1 & -1 & 1 & -1 & 1 & -1 & 1 & -1 & 1 \\
\hline
z & 1 & -1 & 1 & -1 & -1 & 1 & -1 & 1 & 1 & -1 & 1 & -1 & -1 & 1 & -1 & 1 \\
\hline
\hline
s_1 & 2 & -2 &  2 & -2 &  -2 & -2 &  2 &  2 &    2 & 2 & -2 & -2 &   -2 &  2 &  -2 & 2 \\
\hline
s_2 & 2 & -2 &  2 & -2 &   2 &   2 & -2 & -2 &  -2 & -2 & 2 & 2 &    -2 &  2 &  -2 & 2 \\
\hline
s_3 & 2 &  2 & -2 & -2 &  -2 &  2 & -2 &  2 &    2 & -2 & 2 & -2 &   -2 & -2 &   2 & 2 \\
\hline
s_4 & 2 &  2 & -2 & -2 &   2 &  -2 &  2 & -2 &  -2 & 2 & -2 & 2 &    -2 & -2 &   2 & 2 \\
\hline
\hline
\text{Min and Max} & \multicolumn{16}{c|}{-2 \text{ and } 2} \\
\hline
\end{array}
\]
\end{table}

It is hard to overemphasize the pivotal role played by the NC model type. 

\begin{itemize}
\item The correlations of any NC model instance satisfy all 8 CHSH inequalities.
\item Any factorizable model instance and any outcome deterministic model instance that also satisfies remote context independence are special cases of the NC model type.
\item Noncontextuality is the most general classical notion (among RCI+OD, factorizability, and noncontextuality) that blocks replication of at least some QM statistical predictions. Both factorization and RCI+OD are \emph{sufficient} to derive the CHSH inequalities, but neither is \emph{necessary} (see sec.~\ref{sec:NCDoesNotImplyFC}).
\item In other words, if one assumes RCI+OD or one assumes factorizability, one has implicitly assumed noncontextuality (in the mathematical sense). 
\end{itemize}

\subsection{Noncontextuality does not imply factorizability}
\label{sec:NCDoesNotImplyFC}

To show that the set of factorizable (Bell local) model instances is a \emph{proper} subset of the set of noncontextual model instances, it is sufficient to provide an example of a NC model instance that cannot be factorizable. First make the simple observation that if two model instances have the same conditional probabilities, then they have the same correlations (see Eq.~\ref{eq:defineCorrIntro}). Hence, to produce a counterexample, it is sufficient to find a NC model instance whose correlations cannot be consistent with \emph{any} factorizable model instance. 

\begin{Example}
\label{ex:pdNotFCIntro}
The NC model instance with parameters 
\[\pmb{\rho}=(0,\frac{1}{4},0,0,0,0,0,\frac{1}{4},\frac{1}{4},0,0,0,0,0,\frac{1}{4},0)\]
leads to the generic representation
\[\pmb{\gamma}=(\frac{1}{4},\frac{1}{4},\frac{1}{4},\frac{1}{4},\frac{1}{2},0,0,\frac{1}{2},0,\frac{1}{2},\frac{1}{2},0,\frac{1}{4},\frac{1}{4},\frac{1}{4},\frac{1}{4}),\]
which has correlations $(w,x,y,z)=(0,1,-1,0)$. However, \((0,1,-1,0)\) \emph{cannot} be achieved by any assignment of values to the parameters of a factorizable model. 
\end{Example}

Consider the equations that must be solved.
\begin{equation}
\label{eq:blBRCounterexample}
\left(
\begin{array}{l}
(\alpha_{1}-\alpha_{3})(\beta_{1}-\beta_{3})\\
(\alpha_{1}-\alpha_{3})(\beta_{2}-\beta_{4})\\
(\alpha_{2}-\alpha_{4})(\beta_{1}-\beta_{3})\\
(\alpha_{2}-\alpha_{4})(\beta_{2}-\beta_{4})\\
\end{array}
\right)=
\left(
\begin{array}{c}
0\\
1\\
-1\\
0\\
\end{array}
\right)
\end{equation}

The expressions on the left represent the four correlations $(w,x,y,z)$ for the factorizable model type (see Table~\ref{tab:blCandSTableIntro} for the factorizable correlations). To make the first term zero, either $\alpha_{1}=\alpha_{3}$ or $\beta_{1}=\beta_{3}$. But then either the second or third term (or both) is (are) zero, making one (or both) of the second and third equations false. So it is impossible for a factorizable model instance to achieve this particular set of correlations.\qedhere

As an illustration of what has just been proved, see Fig.~\ref{fig:bellNotLocal} (see Appendix~\ref{sec:corrPlots} for details on correlation plots). It shows a 3D slice of all correlations $(0,x,y,z)$ (corresponding to $w=0$) for the NC region (blue solid) and the factorizable surfaces (green planes). The point \((x,y,z)=(1,-1,0)\) (purple dot) is a vertex of the NC region (blue), but it is way off the factorizable contour (green), i.e., the intersecting planes $x=0$ and $y=0$. Note that, since this is the 3D-slice corresponding to \(w=0\), the 4-tuple of correlations in question is the one specified above, namely \((w,x,y,z)=(0,1,-1,0)\).
 
\begin{figure}[H]
\centering
\includegraphics[width=0.3\linewidth]{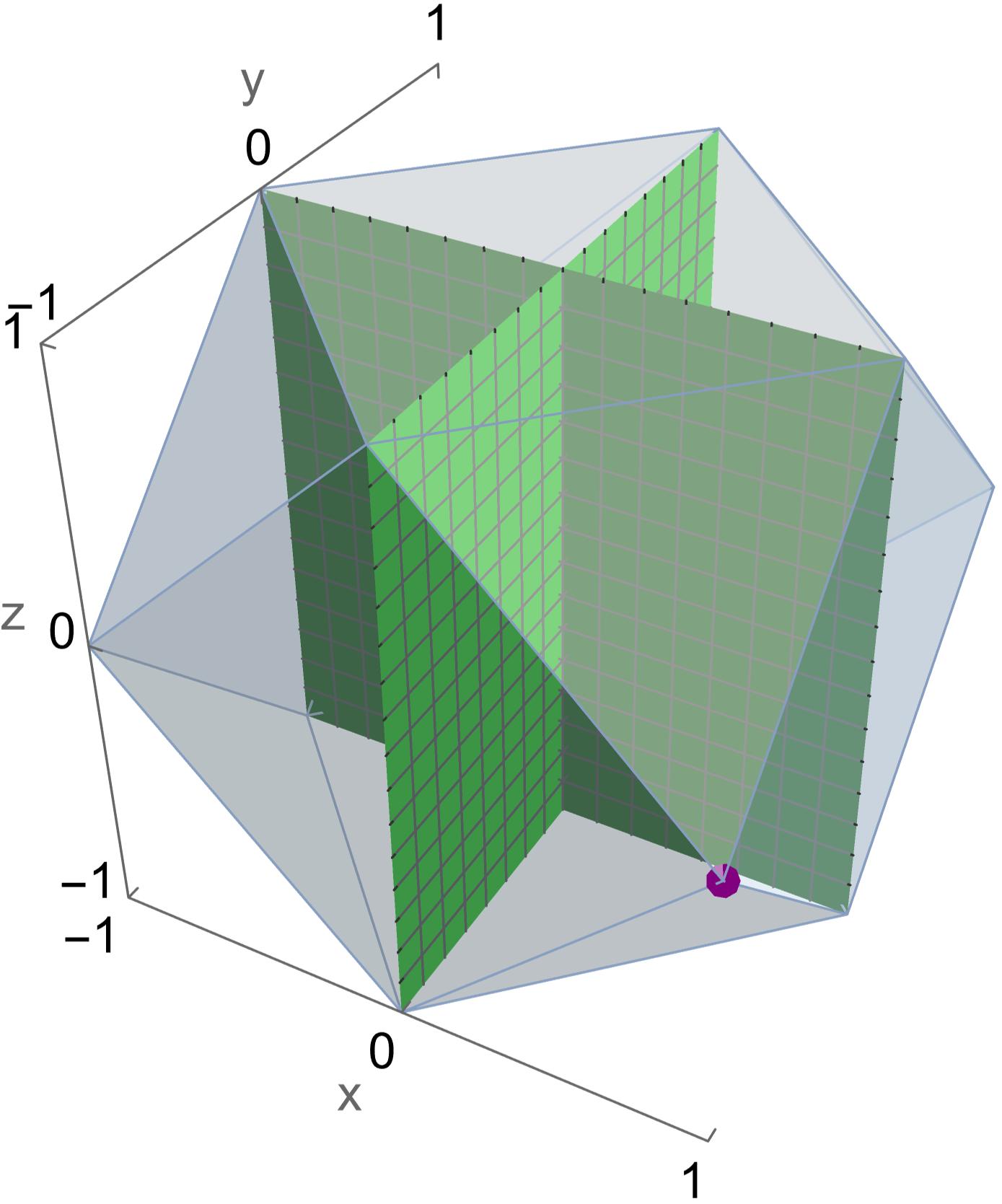}
\caption{$(w,x,y,z)=(0,1,-1,0)$ is an example of a correlation vector consistent with a NC model instance but not consistent with any factorizable model instance. The triple $(x,y,z)=(1,-1,0)$ is shown (purple dot) where $w=0$ is assumed to be fixed.}
\label{fig:bellNotLocal}
\end{figure}

At first glance, this counterexample seems to be in conflict with Proposition 2 in Fine~\cite{Fine1982-1}, since he proves that "\emph{There exists a factorizable stochastic hidden-variables model for a correlation experiment if and only if there exists a deterministic hidden variables model for the experiment}". First let's translate this statement into the language of convex combinations of model instances, namely, "\emph{A model instance is expressible as a convex combination of factorizable model instances iff it is expressible as a convex combination of RCI+OD model instances}". Here is a proof:

\begin{proof}
($\Leftarrow$) Suppose a model instance is a convex combination of the 16 RCI+OD model instances. Since all of these are factorizable (Corollary~\ref{cor:rciODNC}), this is also a convex combination of factorizable model instances.

($\Rightarrow$) Suppose a model instance is a convex combination of factorizable model instances. But each of these factorizable model instances is also NC by Thm.~\ref{thm:fcImpliesPDIntro}, and thus expressible as a convex combination of RCI+OD model instances by Thm.~\ref{thm:cvhPDIntro}. Hence the given model instance is expressible as a convex combination of RCI+OD model instances since a convex combination of convex combinations is again a convex combination.
\end{proof}

So how does Example~\ref{ex:pdNotFCIntro} escape being factorizable? After all, it can be expressed as a convex combination of RCI+OD model instances, since it is NC (Thm.~\ref{thm:cvhPDIntro}). Hence it is expressible as a convex combination of factorizable model instances. However, there is nothing to guarantee that such a convex combination is also factorizable. That is, the factorization property does not generally "survive" the process of constructing convex combinations, or in the traditional language of hidden variables, does not "survive" the process of integration with respect to a measure over the hidden variables. This is a simple consequence of the fact that the integral of a product of functions is usually not the product of the integrals of the functions.

\subsection{CHSH almost implies noncontextuality}
\label{sec:fineProof}

As an example of a (possibly) underappreciated result, consider one of the propositions from Fine's pioneering paper from 1982~\cite{Fine1982-1}, where among other things, he linked the existence of a joint distribution on all observables (commuting or not) to satisfaction of the Bell/CH inequalities. Although he was careful to say \emph{there exists} a factorizable model, or \emph{there exists} a deterministic hidden variables model throughout, it may be not be well-appreciated by everyone that there exist model instances that satisfy all 8 CHSH inequalities, but which are \emph{not} noncontextual (hence also not factorizable). Here is an example of a model instance whose correlations satisfy all of the CHSH inequalities but it is not NC. Note this is one of the 112 OD model instances with this property (see Fig.~\ref{fig:vennDeterminism}). 
\begin{Example}
Consider the OD model instance with parameters 
\[\pmb{\gamma}=(1,0,0,0,1,0,0,0,1,0,0,0,0,0,0,1).\] 
The corresponding correlations are $(w,x,y,z)=(1,1,1,1)$ and the $s$-functions are $(s_{1},s_{2},s_{3},s_{4})=(2,2,2,2)$. However, it is fairly obvious that there is no pmf $\pmb{\rho}=(\rho_1,\rho_2,...,\rho_{16}))$  such that 
\[
\left(
\begin{array}{l}
\rho_{1}+\rho_{2}+\rho_{5}+\rho_{6}\\
\rho_{3}+\rho_{4}+\rho_{7}+\rho_{8}\\
\rho_{9}+\rho_{10}+\rho_{13}+\rho_{14}\\
\rho_{11}+\rho_{12}+\rho_{15}+\rho_{16}\\

\rho_{1}+\rho_{2}+\rho_{9}+\rho_{10}\\
\rho_{3}+\rho_{4}+\rho_{11}+\rho_{12}\\
\rho_{5}+\rho_{6}+\rho_{13}+\rho_{14}\\
\rho_{7}+\rho_{8}+\rho_{15}+\rho_{16}\\

\rho_{1}+\rho_{3}+\rho_{5}+\rho_{7}\\
\rho_{2}+\rho_{4}+\rho_{6}+\rho_{8}\\
\rho_{9}+\rho_{11}+\rho_{13}+\rho_{15}\\
\rho_{10}+\rho_{12}+\rho_{14}+\rho_{16}\\

\rho_{1}+\rho_{3}+\rho_{9}+\rho_{11}\\
\rho_{2}+\rho_{4}+\rho_{10}+\rho_{12}\\
\rho_{5}+\rho_{7}+\rho_{13}+\rho_{15}\\
\rho_{6}+\rho_{8}+\rho_{14}+\rho_{16}\\
\end{array}
\right)= \left(
\begin{array}{l}
1 \\
0 \\
0 \\
0 \\
1 \\
0 \\
0 \\
0 \\
1 \\
0 \\
0 \\
0 \\
0 \\
0 \\
0 \\
1 
\end{array}
\right)
.\]
Just start knocking off all of the $\rho_{k}$'s on the left in rows where 0's appear in the column vector on the right, and pretty soon there is nothing left to make 1's in rows 1,5,9, or 16.
\label{ex:chshNotNC}
\end{Example} 
Thus it is not true that satisfaction of the CHSH inequalities implies noncontextuality. What is true is a slightly weaker statement, namely, if a model instance (noncontextual or not) has correlations that satisfy all 8 CHSH inequalities, then \emph{there exists} a noncontextual model instance with those same correlations. A constructive proof of this result is given, based on convex geometry, which may lend some intuition and emphasis to Fine's earlier result. It will be called "Fine's Theorem" here, even though is is just part of what he did. See Halliwell~\cite{Halliwell2014} for some interesting alternative proofs of this result. The underlying takeaway, although it should be obvious from Eq.~\ref{eq:defineCorrIntro}, is that the mapping from cpvs to correlations is many-to-one.

\begin{figure}
\centering
\includegraphics[width=1.0\linewidth]{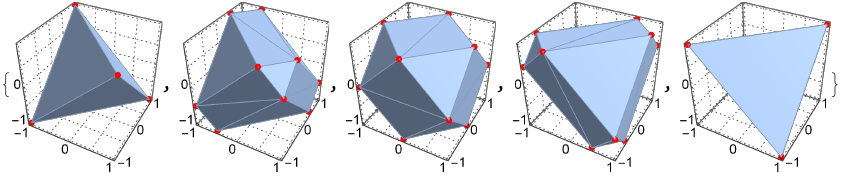}
\caption{Five regions where the CHSH inequalities are satisfied. $w=-1,-\frac{1}{2},0,\frac{1}{2},1$ from left to right.}
\label{fig:bellRealPlots}
\end{figure}

The proof given here is grounded in the geometry of Fig.~\ref{fig:bellRealPlots}, which may add some intuition compared to the original proof in~\cite{Fine1982-1}. These five regions represent 3D slices $(x,y,z)$ through 4D sets of correlations $(w,x,y,z)$ that satisfy the CHSH inequalities, for a given fixed value for $w$ in each case, where $w=-1,-\frac{1}{2},0,\frac{1}{2},1$, from left to right.  Since these are closed and bounded convex regions, any such correlation can be written as a convex combination of the vertices (red dots). To get the parameters $\pmb{\rho}$ of a NC model instance which has the given correlations, one merely computes the same convex combination of a special set of probability vectors.

First note that in each case, the coordinates for the vertices (red dots) can be written as the columns of the last three rows of the following matrices.

\begin{Definition} The \textbf{vertex matrix \(V\)} is defined in three different ways, depending on the value of \(w\). 
\[V=
\left( \begin{array}{rrrrrrrrrrrr}
w & w & w & w & w & w & w & w & w & w & w & w \\
w & -1 & 1 & -w & w & 1 & -1 & -w & 1 & 1 & -1 & -1 \\
1 & -w & w & -1 & -1 & -w & w & 1 & -1 & 1 & 1 & -1 \\
1 & 1 & 1 & 1 & -1 & -1 & -1 & -1 & -w & w & -w & w
\end{array} \right)
\]
\mbox{if $-1<w<1$},

\[V=\left( \begin{array}{rrrr}
-1 & -1 & -1 & -1 \\
-1 &  1 & -1 & 1 \\
1 & -1 & -1 & 1  \\
1 & 1 & -1 & -1
\end{array} \right) \mbox{if $w=-1$},
\] 

\[V=\left( \begin{array}{rrrr}
1 & 1 & 1 & 1 \\
1 & -1 & 1 & -1 \\
1 & -1 & -1 & 1  \\
1 & 1 & -1 & -1
\end{array} \right)  \mbox{if $w=+1$}.
\]
\label{def:V4}
\end{Definition}

Then the following lemma basically says that any correlation vector $\pmb{p}=(w,x,y,z)$ which satisfies all eight CHSH inequalities can be written as a convex combination of the columns of $V$.

\begin{Lemma} Pick an arbitrary set of correlations $\pmb{p}=(w,x,y,z)$ that satisfy CHSH. Then there exists a pmf \(\pmb{\pi }\) such that \(\pmb{p}^T=(w,x,y,z)^T=V \pmb{\pi}^T\). If \(V\) has 12 rows (i.e., \(-1<w<1\)), \(\pmb{\pi}=\left(\pi _1,\pi _2,\pi _3,\pi _4,...,\pi _{12}\right)\), and if it has 4 rows (i.e., \(w=\pm 1\)), \(\pmb{\pi}=\left(\pi
_1,\pi _2,\pi _3,\pi _4\right)\).\label{lemma:lambda}
\end{Lemma}

\begin{proof}
Start by ignoring the first row of \(V\) and consider the last three rows only. The \emph{columns} of the resulting sub-matrix represent the \emph{coordinates} of the vertices of certain 3D volumes defined by the CHSH inequalities. Some examples are shown in Fig.~\ref{fig:bellRealPlots}. Since all of these volumes, for any \(-1\leq w\leq 1\), are formed by the intersection of a finite number of closed half-spaces, the result is a closed convex region. The particular (hyper)planes represented by the CHSH inequalities also make the region bounded. 
Invoking a theorem of Minkowski, an arbitrary point \(\pmb{p}_w=(x,y,z)\) inside any of these bounded, closed, convex regions can be written as a convex combination of the \emph{extreme} points of the region, in this case, the vertices. That is, \(\pmb{p}_w^T= (x,y,z)^T= V(2,3,4)\pmb{\pi}^T\), for some appropriate vertex matrix \(V\) (where $V(2,3,4)$ denotes the last three rows of $V$) and pmf $\pmb{\pi}$. Since $\pmb{\pi}$ is a pmf, \(w=(w,w,...,w)\pmb{\pi}^T\), so by inserting all \(w\){'}s as the first row of \(V(2,3,4)\) to make \(V\), it is easily seen that \(\pmb{p}^T=(w,x,y,z)^T=V \pmb{\pi}^T\) as well.
\end{proof}  \qedhere

The next matrix was discovered by using \emph{Mathematica} to find instances of solutions $M$ to the matrix equation shown in Lemma~\ref{lem:CRMV}. That it worked so well earned it the name "magic".

\begin{Definition} The \textbf{magic matrix} $M$ is defined in three different ways, depending on the value of $w$. $M$ is a (right) stochastic matrix, since the rows consist of non-negative numbers that sum to 1.
\label{def:magicMatrix}\\
\begin{equation*}
\setlength\arraycolsep{4pt}
M= \\
\left( \begin{array}{cccccccccccccccc}
\setlength\arraycolsep{2pt}
\frac{1+w}{2} & 0 & 0 & 0 & 0 & 0 & 0 & 0 & 0 & 0 & 0 & 0 & 0 & \frac{1-w}{2} & 0 & 0 \\
0 & 0 & 0 & 0 & 0 & 0 & 0 & 0 & 0 & 0 & \frac{1+w}{2} & 0 & 0 & \frac{1-w}{2} & 0 & 0 \\
\frac{1+w}{2} & 0 & 0 & 0 & 0 & 0 & 0 & 0 & \frac{1-w}{2} & 0 & 0 & 0 & 0 & 0 & 0 & 0 \\
0 & 0 & 0 & 0 & 0 & \frac{1+w}{2} & 0 & 0 & \frac{1-w}{2} & 0 & 0 & 0 & 0 & 0 & 0 & 0 \\

0 & \frac{1+w}{2} & 0 & 0 & 0 & 0 & 0 & 0 & 0 & 0 & 0 & 0 & \frac{1-w}{2} & 0 & 0 & 0 \\
0 & \frac{1+w}{2} & 0 & 0 & 0 & 0 & 0 & 0 & 0 & \frac{1-w}{2} & 0 & 0 & 0 & 0 & 0 & 0 \\
0 & 0 & 0 & 0 & 0 & 0 & 0 & 0 & 0 & 0 & 0 & \frac{1+w}{2} & \frac{1-w}{2} & 0 & 0 & 0 \\
0 & 0 & 0 & 0 & \frac{1+w}{2} & 0 & 0 & 0 & 0 & \frac{1-w}{2} & 0 & 0 & 0 & 0 & 0 & 0 \\

0 & 0 & 0 & 0 & 0 & 0 & 0 & 0 & 0 & 0 & 0 & \frac{1+w}{2} & 0 & \frac{1-w}{2} & 0 & 0 \\
\frac{1+w}{2} & 0 & 0 & 0 & 0 & 0 & 0 & 0 & 0 & \frac{1-w}{2} & 0 & 0 & 0 & 0 & 0 & 0 \\
0 & \frac{1+w}{2} & 0 & 0 & 0 & 0 & 0 & 0 & \frac{1-w}{2} & 0 & 0 & 0 & 0 & 0 & 0 & 0 \\
0 & 0 & 0 & 0 & 0 & 0 & 0 & 0 & 0 & 0 & \frac{1+w}{2} & 0 & \frac{1-w}{2} & 0 & 0 & 0 
\end{array} \right)  
\end{equation*}
\mbox{if $-1<w<1,$}
\[M=\left( \begin{array}{cccccccccccccccc}
0&0&0&0&0&0&0&0&0&0&0&0&0&1&0&0 \\
0&0&0&0&0&0&0&0&1&0&0&0&0&0&0&0 \\
0&0&0&0&0&0&0&0&0&0&0&0&1&0&0&0 \\
0&0&0&0&0&0&0&0&0&1&0&0&0&0&0&0
\end{array} \right) \mbox{if $w=-1$},
\] 
\[M=\left( \begin{array}{cccccccccccccccc}
0&0&0&0&0&0&0&0&0&0&0&0&0&0&0&1 \\
0&0&0&0&0&0&0&0&0&0&1&0&0&0&0&0 \\
0&0&0&0&0&0&0&0&0&0&0&0&0&0&1&0 \\
0&0&0&0&0&0&0&0&0&0&0&1&0&0&0&0
\end{array} \right)  \mbox{if $w=+1$.}
\]
\end{Definition}

\begin{Lemma}
\label{lem:CRMV}
$CRM^{T}=V$. 
\end{Lemma}

\begin{proof}
This is a straightforward but tedious matrix multiplication, so it is left to the reader. (The definitions of the $C$ and $R$ matrices can be found in Def.'s~\ref{def:cMatrix} and~\ref{def:rMatrix}, respectively.) There are three cases to consider depending on the value of $w$. \emph{Mathematica} or other computational software platform is helpful here! 
\end{proof}    \qedhere

Finally Thm.~\ref{thm:Fine} can be stated and proved. The proof may be of interest due to its compactness and its basis in the geometry exemplified by the polyhedra shown in Fig.~\ref{fig:bellRealPlots}.
\begin{Theorem}
(\textbf{Fine's Theorem}) Given an arbitrary correlation vector \(\pmb{p}=(w,x,y,z)\) that satisfies the CHSH inequalities. Let $\pmb{\pi}$ be a pmf such that \(\pmb{p}^{T}=V \pmb{\pi}^T\) as guaranteed by Lemma~\ref{lemma:lambda}. Define
\begin{itemize}
\item $\pmb{\rho}=(\rho_{1},\rho_{2},...,\rho_{16})= \pmb{\pi} M $, 
\item $\pmb{\gamma}^{T}=(\gamma_{1},\gamma_{2},...,\gamma_{16})^{T}= R \pmb{\rho}^{T}=R M^{T} \pmb{\pi}^{T}$. 
\end{itemize}
Then 
\begin{itemize}
\item $\pmb{\rho}$ is a pmf and therefore $\pmb{\gamma}$ is a set of NC conditional probabilities.
\item The NC conditional probabilities $\pmb{\gamma}$ are consistent with the correlations $\pmb{p}$.
\end{itemize} \label{thm:Fine}
\end{Theorem}

\begin{proof}
For the first conclusion, note that $\pmb{\rho}$ is a pmf because the rows of $M$ are pmf's, $\pmb{\pi}$ is a pmf, therefore $\pmb{\rho}=\pmb{\pi} M$, being a convex combination of pmf's, is also a pmf. Thus $\pmb{\gamma}^{T}=R \pmb{\rho}^{T}$ is a set of NC conditional probabilities by the very definition of $R$ (Def.~\ref{def:rMatrix}).\\
For the second conclusion, invoke Lemma~\ref{lem:CRMV} to write
\begin{equation}
\pmb{p}^{T}=V\pmb{\pi}^{T}= CRM^{T}\pmb{\pi}^{T}=C\pmb{\gamma}^{T},
\end{equation} 
that is, the initial arbitrary correlation vector $\pmb{p}$, that was assumed to satisfy the CHSH inequalities, is equal to the computed correlations $C\pmb{\gamma}^{T}$ based on the NC conditional probabilities $\pmb{\gamma}$. 
\end{proof}

An important consequence of Thm.~\ref{thm:Fine} is that the polyhedra shown in the "Noncontextual" (fourth) row of Fig.~\ref{fig:models5SelectedIntro} correspond to satisfaction of the CHSH inequalities. Here is a formal proof. Assume $-1 \leq w \leq 1$ is fixed and define two polyhedrons as follows:
\begin{equation}
\mathcal{P}_1=\{(x,y,z): (w,x,y,z) \text{ satisfy all of the CHSH inequalities} \},
\end{equation}
and (see Table~\ref{tab:preDetCandSTable2Intro} for the NC correlations)
\begin{align}
\mathcal{P}_2= &\{(x,y,z): \text{there exists a pmf } \pmb{\rho}=(\rho_1,\rho_2,...,\rho_{16}) \text{  such that} \nonumber \\
&\rho_1+\rho_2-\rho_3-\rho_4+\rho_5+\rho_6-\rho_7-\rho_8- \nonumber \\
& \rho_9-\rho_{10}+\rho_{11}+\rho_{12}-\rho_{13}-\rho_{14}+\rho_{15}+\rho_{16}=w, \nonumber \\
&\text{and where } x,y,z \text{ are defined by } \\
&x= \rho_1+\rho_2-\rho_3-\rho_4-\rho_5-\rho_6+\rho_7+\rho_8+ \nonumber \\
& \rho_9+\rho_{10}-\rho_{11}-\rho_{12}-\rho_{13}-\rho_{14}+\rho_{15}+\rho_{16}, \nonumber \\
&y= \rho_1-\rho_2+\rho_3-\rho_4+\rho_5-\rho_6+\rho_7-\rho_8- \nonumber \\
& \rho_9+\rho_{10}-\rho_{11}+\rho_{12}-\rho_{13}+\rho_{14}-\rho_{15}+\rho_{16}, \text{ and }\nonumber \\
&z= \rho_1-\rho_2+\rho_3-\rho_4-\rho_5+\rho_6-\rho_7+\rho_8+ \nonumber \\
& \rho_9-\rho_{10}+\rho_{11}-\rho_{12}-\rho_{13}+\rho_{14}-\rho_{15}+\rho_{16} \}. \nonumber
\end{align}

\begin{Corollary}
$\mathcal{P}_1=\mathcal{P}_2.$
\label{cor:ncPolytopes}
\end{Corollary}
\begin{proof}
$\mathcal{P}_2\subseteq\mathcal{P}_1$ since the correlations of any NC model instance satisfy all of the CHSH inequalities. Conversely, $\mathcal{P}_1\subseteq\mathcal{P}_2$ by  Thm.~\ref{thm:Fine}, since if $(x,y,z) \in \mathcal{P}_1$, then \emph{there exists} a NC cpv (i.e., based on the parameters of some pmf $\pmb{\rho}=(\rho_1,\rho_2,...,\rho_{16})$) that has the correlations $(w,x,y,z)$, hence $(x,y,z) \in \mathcal{P}_2.$
\end{proof}
However, recall that there are cpvs whose correlations satisfy all CHSH inequalities but are \emph{not} NC, for example the 112 OD+CHSH cpvs (see Sec.~\ref{sec:rciOD}). The set of cpvs whose correlations satisfy all of the CHSH inequalities is \emph{strictly larger} than the set of NC cpvs. Corollary~\ref{cor:ncPolytopes} only tells us this: \emph{For any given polyhedron defined by the CHSH inequalities (such as in row 4 of Fig.~\ref{fig:models5SelectedIntro}), it is possible to associate to any point $(x,y,z)$ in the polyhedron} \textbf{some} \textit{NC model instance that has correlations $(w,x,y,z)$ for the appropriate $w$}.

\subsection{Fine's Theorem in action}

\begin{figure}[H]
\centering
\includegraphics[width=0.5\linewidth]{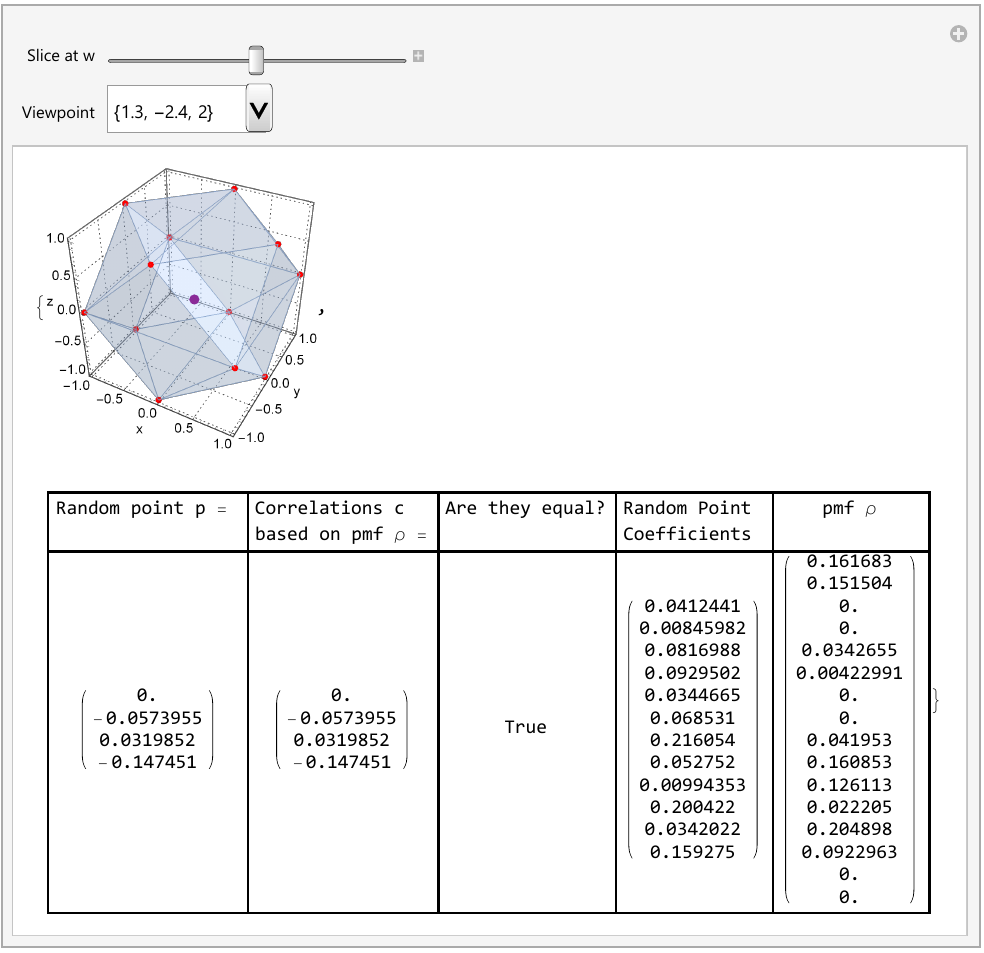}
\caption{\emph{Mathematica} interactive demo for finding a set of NC model instance parameters $\pmb{\rho}$ consistent with a randomly chosen set of correlations $\pmb{p}$ (purple dot) that satisfy CHSH. $w=0$ case is shown. Other cases are obtained by moving the slider in the control area at the top.}
\label{fig:randomPt}
\end{figure}
Fig.~\ref{fig:randomPt} shows example output from a \emph{Mathematica} interactive program that finds the NC parameters $\pmb{\rho}$ that are consistent with a random set of correlations $\pmb{p}$. Moving the slider selects different values for the first correlation $w$, and a triple of the remaining correlations $(x,y,z)$ is chosen randomly (constrained so that $(w,x,y,z)$ satisfies the CHSH inequalities, of course). Both the appropriate CHSH inequality region shape (as a 3D slice) and the random point $(x,y,z)$ are then displayed, along with tables of values showing, respectively from left to right, the coordinates of the random point $\pmb{p}$, the computed correlations based on the NC parameters $\pmb{\rho}$, whether or not they are equal, the coefficients of the point $\pmb{p}$ as a convex combination of the vertices of the polyhedron, and finally the NC model instance parameters $\pmb{\rho}$ that were generated from $\pmb{p}$'s coefficients.

\subsection{No faster-than-light communication}
\label{sec:noFTL}

Despite the fact that QM model instances are not factorizable\footnote{Except for the trivial one. See Table~\ref{tab:trivialType}.} (i.e., are "nonlocal"), it is important to point out that the EPRB experiment cannot be used for faster-than-light transmission of information between Alice and Bob.\footnote{Assuming QM rules, that is.} Of course this is well known, but the parameterized operational modeling approach provides a different viewpoint. Instead of focusing on proving the "no-communication" properties of QM itself, generic operational models are constructed that \emph{do} allow Bob to send messages to Alice, and then show that no QM model instance could possibly have this property. 

To demonstrate this perspective, consider what it takes for Bob to communicate with Alice. There must be something that Bob \emph{controls} that is simultaneously something Alice can \emph{deduce}. Table~\ref{tab:mpExampleIntro} shows an example of how this can be done. This will be called the \textbf{measurement predictable} model type. (See Sec.~\ref{sec:measPredictability} for more details.) The following argument is predicated on the assumption that Alice knows in advance that Table~\ref{tab:mpExampleIntro} specifies the conditional probabilities for the experiment. That is, she knows the experimental arrangement, which is not unreasonable to assume. Furthermore, it is assumed that these probabilities hold for every iteration of the experiment. 

Referring then to Table~\ref{tab:mpExampleIntro}, suppose, for example, that Bob chooses measurement $v=1$. Then whether Alice chooses her measurement $u=1$ or $u=2$, she gets the outcome $s=-1$ corresponding to the only (possibly) nonzero probabilities $\gamma_1,\gamma_2,\gamma_9,\gamma_{10}$. Hence she \emph{deduces} that Bob's measurement choice must be $v=1$.

On the other hand, suppose that Bob chooses measurement $v=2$. Then whether Alice chooses her measurement $u=1$ or $u=2$, she gets the outcome $s=1$ corresponding to the only (possibly) nonzero probabilities $\gamma_7,\gamma_8,\gamma_{15},\gamma_{16}$. Hence she \emph{deduces} that Bob's measurement choice must be $v=2$.

\begin{table}[H]
\caption{Example of a \textbf{measurement predictable (MP)} model type, in which Alice can reliably predict (with probability 1) Bob's measurement. Some of the $\gamma_k$'s have been set to 0 \emph{a priori}, but the measurement context constraints still hold. There are exactly 4 such templates for measurement predictability.}
\label{tab:mpExampleIntro}
\[
\begin{array}{|c|c|c|c|c|}
\hline
\text{Context} & \multicolumn{4}{c|}{\text{Outcome }(s,t)} \\
\hline
(u,v) & (-1,-1)  & (-1,1)   & (1,-1)   & (1,1) \\
\hline
(1,1) & \gamma_1 & \gamma_2 & 0 & 0 \\
\hline
(1,2) & 0 & 0 & \gamma_7 & \gamma_8 \\
\hline
(2,1) & \gamma_9 & \gamma_{10} & 0 & 0 \\
\hline
(2,2) & 0 & 0 & \gamma_{15} & \gamma_{16} \\
\hline
\end{array}
\]
\end{table}

Hence Bob can purposely send a message encoded as a sequence of measurement choices, which Alice can then decode. Of course, there is no QM model instance that can be measurement predictable. Just compare this template to the QM model type of Table~\ref{tab:qmTypeIntro}, in which each row has the pattern "abba". This is inconsistent with each row of Table~\ref{tab:mpExampleIntro}. The same is true of the other three measurement predictable model templates (see Sec.~\ref{sec:measPredictability}). The proof of Thm.~\ref{thm:pdNotMP} shows that no noncontextual model instance can be measurement predictable, either (hence eliminating any factorizable model instances too, of course).

The advantage of this viewpoint is that the set of model instances that allow faster-than-light communication are specifically identified. They are inside the complement of the so-called "no-signaling" model instances (i.e., those that satisfy RCI), of course, but more than that, they constitute a \emph{proper} subset of this complement, thus identifying more precisely the boundary between FTL "signaling" and "no-signaling".

This same technique can be used to produce an "outcome predictable" model type, in which Alice can reliably predict Bob's outcome (see Sec.~\ref{sec:outcomePredictability}). This analysis confirms the fact that only certain QM model instances can be outcome predictable (these are the cases that Bell compared to "Bertlmann's socks"~\cite{Bell1981}). Note that outcome predictability is \emph{not sufficient} for Bob to send messages to Alice, since he does does not \emph{control} his outcomes.

\section{Hidden variables and convex combinations of OD cpvs are (almost) equivalent}
\label{sec:hvCVHEquivalent}

In this paper, finite ensembles of OD cpvs (model instances) are used to build up more complicated cpvs (model instances) -- see Thm.'s~\ref{thm:cvhPDIntro} and~\ref{thm:cvhGeneric}. This is an alternative perspective to the traditional hidden variables approach involving an unspecified measure on a (possibly) infinite probability space of hidden variables. In this section, the (mathematical) equivalence of the two representations is demonstrated, under the assumption of integrability of certain random variables with respect to the hidden variables measure. But first it is helpful to summarize the convex combination approach and to review standard hidden variables, rephrasing it slightly to conform to the notation in this paper.

\subsection{Convex combinations of OD model instances}
\label{sec:cvhODtoGeneric}

\begin{figure}[H]
\centering
\includegraphics[width=0.4\linewidth]{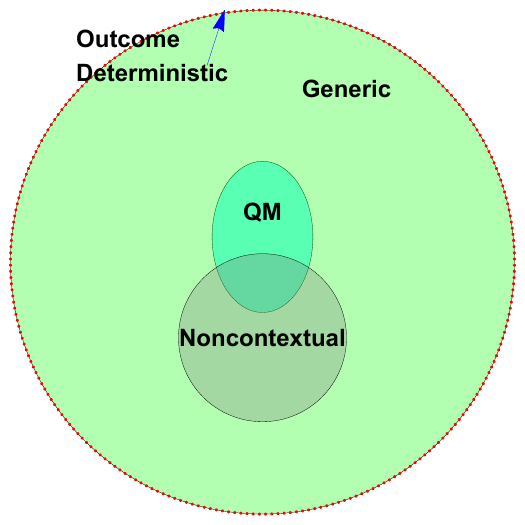}
\caption{The convex hull of all 256 OD conditional probability vectors (red dots on the boundary) equals the full set of generic conditional probability vectors, including the QM and NC ones. See Thm.~\ref{thm:cvhGeneric} in Appendix~\ref{sec:cvhGeneric} for proof.}
\label{fig:odGenericCVH}
\end{figure}

In Sec.~\ref{sec:cvhRCIODtoNC} (Thm.~\ref{thm:cvhPDIntro}) it was shown that the convex of hull the 16 RCI+OD model instances equals the set of noncontextual model instances. This naturally leads to: What is the convex hull of all 256 OD model instances? The proof of Thm.~\ref{thm:cvhGeneric} in Appendix~\ref{sec:cvhGeneric} shows that it equals the set of \emph{all} generic model instances, including QM, factorizable, and noncontextual ones. See Fig.~\ref{fig:odGenericCVH} and compare to Fig.~\ref{fig:preDetConvexHullIntro} in Sec.~\ref{sec:cvhRCIODtoNC}.
In Fig.~\ref{fig:odGenericCVH}, the 256 OD conditional probability vectors are the extreme points on the boundary, shown as red dots (must look closely). The figure is a schematic representation of a polytope in a 13-dimensional subspace of $\mathbb{R}^{16}$. This is because the $16 \times 256$ matrix with the OD conditional probability vectors as columns has rank 13.\footnote{\textit{Mathematica}'s \texttt{MatrixRank} function was used to compute this.}

\subsection{Review of hidden variables}
\label{sec:hvReview}

The "hidden variables" idea is to construct a statistical model of the experiment by combining more basic models, each of which is associated with a "hidden variable" which explains its behavior. These elementary models are then combined via integration over the space of hidden variables to produce the overall final model for the statistics of the experiment. Formally,

\begin{enumerate}
\item There is a probability space $(\Lambda,\mathcal{F},\mu)$ where $\Lambda$ is a set of "hidden variables" (or "complete states"), $\mathcal{F}$ is a suitable $\sigma$-algebra of subsets of $\Lambda$, and $\mu$ is a probability measure on $\mathcal{F}$.
\item For each $\lambda \in \Lambda$, there exists a cpv 
\[\pmb{\gamma}_{\lambda}=(\gamma_1(\lambda),\gamma_2(\lambda),...,\gamma_{16}(\lambda))\]
where it is assumed that each $\gamma_k$ is $\mu$-integrable.
\item The final overall cpv describing the experimental statistics is constructed by integrating over $\Lambda$ (integration is performed componentwise):
\[\pmb{\gamma}=(\gamma_1,\gamma_2,...,\gamma_{16})=\int_{\Lambda}\pmb{\gamma}_{\lambda}\mu(d \lambda).\]
Note that $\pmb{\gamma}$ is a cpv since the measurement context constraints are satisfied. For example 
\begin{align*}
&\sum_{k=1}^{4} \gamma_k =\\
&\sum_{k=1}^{4} \int_{\Lambda} \gamma_k(\lambda) \mu(d\lambda) = \\
&\int_{\Lambda} \sum_{k=1}^{4} \gamma_k(\lambda) \mu(d\lambda)= \\
&\int_{\Lambda} 1 \:\mu(d\lambda)=1,
\end{align*}
and similarly for the other three measurement context constraints.
\end{enumerate}

\subsection{Proof of (almost) equivalence of convex combinations and hidden variables}
\label{sec:cvhEqualsHVProof}

In the proof of Thm.~\ref{thm:cvhEqualsHVProof2}, it is shown that any generic model instance written as a convex combination of OD model instances can be turned into a \emph{finite} hidden variables \emph{model}. No attempt is made to determine what sort of hidden variables \emph{theory} might give rise to this model. 

In the proof of Thm.~\ref{thm:cvhEqualsHVProof} it is shown, that with an extra "reasonable" assumption, the converse is true, namely a hidden variables model can always be turned into a representation as a convex combination of OD cpvs. 
\begin{Theorem}
\label{thm:cvhEqualsHVProof2}
Assume a generic cpv $\pmb{\gamma}$ written as
\begin{equation}
\pmb{\gamma}=\sum_{k=1}^{256} \pi_k \pmb{\gamma}_k
\label{eq:cpvCVX}
\end{equation}
for some pmf $\pmb{\pi}=(\pi_1,\pi_2,...,\pi_{256}))$ and where
$(\pmb{\gamma}_1,\pmb{\gamma}_2,...,\pmb{\gamma}_{256})$
denotes the set of 256 OD cpvs in some order.
Then Eq.~\ref{eq:cpvCVX} can be written in the form of a hidden variables model as
\[\pmb{\gamma}=\int_{\Lambda}\pmb{\gamma}_{\lambda} \mu(d\lambda)\]
over a space $\Lambda$ of hidden variables, where each $\pmb{\gamma}_{\lambda}$ is a cpv.
\end{Theorem}

\begin{proof}
Assume $\pmb{\gamma}=\sum_{k=1}^{256} \pi_k \pmb{\gamma}_k$.
Define $\Lambda=\{\lambda_1,\lambda_2,...,\lambda_{256}\}$, $\mathcal{F}=2^{\Lambda}$, and $\mu(\{\lambda_k\})=\pi_k$ for all $k$. Extend $\mu$ to all of $\mathcal{F}$ in the natural way. Define 
\[\pmb{\gamma}_{\lambda}=\pmb{\gamma}_k\]
whenever $\lambda=\lambda_k$ for all $k=1,2,...,256$. Then summation and integration become the same, and
\begin{equation}
\pmb{\gamma}=\sum_{k=1}^{256} \pi_k \pmb{\gamma}_k=\int_{\Lambda}\pmb{\gamma}_{\lambda} \mu(d\lambda).
\label{eq:equivalenceCVHHV1}
\end{equation}
\end{proof}

\begin{Theorem}
\label{thm:cvhEqualsHVProof}
Assume a generic cpv can be written in the form
\begin{equation}
\pmb{\gamma}=\int_{\Lambda}\pmb{\gamma}_{\lambda} \mu(d\lambda)
\label{eq:cpvHV}
\end{equation}
over a space $\Lambda$ of hidden variables, where $\pmb{\gamma}_{\lambda}$ is a cpv for each $\lambda$. By Thm.~\ref{thm:cvhGeneric}, for each $\lambda \in \Lambda$, there exists a pmf $(\eta_1(\lambda),\eta_2(\lambda),...,\eta_{256}(\lambda))$ such that
\[\pmb{\gamma}_{\lambda}=\sum_{k=1}^{256} \eta_k(\lambda) \pmb{\gamma}_k,\]
where $(\pmb{\gamma}_1,\pmb{\gamma}_2,...,\pmb{\gamma}_{256})$
denotes the set of 256 OD cpvs in some order. It is assumed that each $\eta_k$ is $\mu$-measurable.\footnote{This is the "reasonable" assumption. It is certainly true, for example, if $\Lambda$ is finite (and $\mathcal{F}$ is the power set of $\Lambda$).}
Then
\[\pi_k \doteq \int_{\Lambda} \eta_k(\lambda) \mu(d\lambda)\]
is well-defined for each $k=1,2,...,256.$
Finally Eq.~\ref{eq:cpvHV} can be written as a convex combination of cpvs as follows:
\[\pmb{\gamma}=\sum_{k=1}^{256} \pi_k \pmb{\gamma}_k.\]
\end{Theorem}
\begin{proof}
Assume $\pmb{\gamma}=\int_{\Lambda}\pmb{\gamma}_{\lambda} \mu(d\lambda)$.
It was assumed that each random variable $\eta_k(.):\Lambda\longrightarrow [0,1]$ is $\mu$-measurable.  Since a bounded, Lebesgue-measurable function defined on a domain of finite measure is Lebesgue-integrable, this yields $\mu$-integrability for each $\eta_k$. Hence the $\pi_k$'s are well-defined. Also, since $(\eta_1(\lambda),\eta_2(\lambda),...,\eta_{256}(\lambda))$ is a pmf for each $\lambda$,
\begin{align}
&\sum_{k=1}^{256} \pi_k=\sum_{k=1}^{256} \int_{\Lambda} \eta_k(\lambda) \mu(d\lambda) = \nonumber \\
&\int_{\Lambda} \sum_{k=1}^{256} \eta_k(\lambda) \mu(d\lambda) = \\
& \int_{\Lambda} 1\: \mu(d\lambda) =1, \nonumber
\end{align}
hence $(\pi_1,\pi_2,...,\pi_{256})$ is a pmf as well. Finally,
\begin{align}
&\pmb{\gamma}=\int_{\Lambda}\sum_{k=1}^{256} \eta_k(\lambda) \pmb{\gamma}_k \mu(d\lambda)= \nonumber \\
&\sum_{k=1}^{256} \pmb{\gamma}_k \int_{\Lambda} \eta_k(\lambda) \mu(d\lambda) = \nonumber \\
&\sum_{k=1}^{256} \pi_k \pmb{\gamma}_k.
\end{align}
\end{proof}

\section{QM model instances can be both contextual and noncontextual}
\label{sec:hvConvexCombo}

In this section, examples are given showing how to write a noncontextual QM model instance and a contextual QM model instance explicitly as a convex combination of OD model instances. Following these examples, the essential difference between noncontextuality and contextuality is discussed using a fable.

\subsection{QM model instance that is also noncontextual}
\label{sec:hvPD}

Table~\ref{tab:pdQMExample} shows a model instance that is both QM and NC. Table~\ref{tab:pdQMExampleWeights} shows how to write it as a convex combination of the 4 RCI+OD model instances in columns 2,8,9, and 15 from Table~\ref{tab:rciODTabIntro}.

\begin{table}[H]
\caption{Example of QM model instance that is also NC. The QM parameters are $\theta_1=135^{\circ},\theta_2=0^{\circ},\theta_3=90^{\circ},\theta_4=45^{\circ}$, and the NC parameters are $\rho_k=\frac{1}{4}$ for $k=2,8,9,15,$ and $\rho_k=0$ otherwise.}
\label{tab:pdQMExample}
\[
\begin{array}{|c|c|c|c|c|}
\hline
\text{Context} & \multicolumn{4}{c|}{\text{Outcome }(s,t)} \\
\hline
(u,v) & (-1,-1)  & (-1,1)   & (1,-1)   & (1,1) \\
\hline
(1,1) & \frac{1}{4} & \frac{1}{4} & \frac{1}{4} & \frac{1}{4} \\
\hline
(1,2) & \frac{1}{2} & 0 & 0 & \frac{1}{2} \\
\hline
(2,1) & 0 & \frac{1}{2} & \frac{1}{2} & 0 \\
\hline
(2,2) & \frac{1}{4} & \frac{1}{4} & \frac{1}{4} & \frac{1}{4} \\
\hline
\end{array}
\]
\end{table}

\begin{table}[H]
\setlength\arraycolsep{2pt}
\caption{The model instance in Table~\ref{tab:pdQMExample} can be written as the convex combination of the RCI+OD model instances in columns 1-4 below, with weights $\pi_k=\frac{1}{4}$ for $k=1,2,3,4$. These conditional probability vectors are columns 2,8,9,15 from Table~\ref{tab:rciODTabIntro}.}

\label{tab:pdQMExampleWeights}
\[
\begin{array}{|r|r|r|r||c|c|c|c|}
\hline
s & t & u & v & 1 & 2 & 3 & 4   \\
\hline
\hline
-1 & -1 & 1 & 1 & 1 & 0 & 0 & 0   \\
\hline
-1 & 1 & 1 & 1 & 0 & 0 & 1 & 0   \\
\hline
1 & -1 & 1 & 1 & 0 & 1 & 0 & 0  \\
\hline
1 & 1 & 1 & 1 & 0 & 0 & 0 & 1   \\
\hline

-1 & -1 & 1 & 2 & 1 & 0 & 1 & 0   \\
\hline
-1 & 1 & 1 & 2 & 0 & 0 & 0 & 0   \\
\hline
1 & -1 & 1 & 2 & 0 & 0 & 0 & 0  \\
\hline
1 & 1 & 1 & 2 & 0 & 1 & 0 & 1   \\
\hline

-1 & -1 & 2 & 1 & 0 & 0 & 0 & 0   \\
\hline
-1 & 1 & 2 & 1 & 0 & 0 & 1 & 1   \\
\hline
1 & -1 & 2 & 1 & 1 & 1 & 0 & 0   \\
\hline
1 & 1 & 2 & 1 & 0 & 0 & 0 & 0   \\
\hline

-1 & -1 & 2 & 2 & 0 & 0 & 1 & 0  \\
\hline
-1 & 1 & 2 & 2 & 0 & 0 & 0 & 1  \\
\hline
1 & -1 & 2 & 2 & 1 & 0 & 0 & 0   \\
\hline
1 & 1 & 2 & 2 & 0 & 1 & 0 & 0   \\
\hline
\multicolumn{4}{|c|}{\text{ Weights }} & \frac{1}{4} & \frac{1}{4} & \frac{1}{4} & \frac{1}{4}  \\
\hline
\end{array}
\]
\end{table} 

\subsection{QM model instance that is also contextual}
\label{sec:hvNotPD}

Consider the infamous QM model instance with an $s_1$-function value of $2\sqrt{2}$. This is obviously not NC because one of the CHSH inequalities is violated. It has  QM parameters $\theta_1=67.5^{\circ}$ and $\theta_k=22.5^{\circ}$ for $k=2,3,4$. Table~\ref{tab:odQMExample} shows how to write this as a convex combination of 13 OD model instances.\footnote{\emph{Mathematica} was used to find this solution.} None of the OD model instances satisfy RCI, 8 come from the subset of 112 OD model instances that satisfy CHSH but do not satisfy RCI, and 5 come from the subset of 128 model instances that have an $s$-function that exceeds a Tsirelson bound (see Fig.~\ref{fig:vennDeterminism}). The corresponding weights $\pi_k$ in the bottom row are defined in Eq.~\ref{eq:qmODWeights}.

\begin{table}[H]
\setlength\arraycolsep{2pt}
\caption{The QM model instance with $\theta_1=67.5^{\circ}$ and $\theta_k=22.5^{\circ}$ for $k=2,3,4$ can be written as a convex combination of these 13 OD model instances. The weights $\pi_k$ are defined in Eq.~\ref{eq:qmODWeights}.}
\label{tab:odQMExample}
\[
\begin{array}{|r|r|r|r||c|c|c|c|c|c|c|c|c|c|c|c|c|}
\hline
s & t & u & v & 1 & 2 & 3 & 4 & 5 & 6 & 7 & 8 & 9 & 10 & 11 & 12 & 13  \\
\hline
\hline
-1 & -1 & 1 & 1 & 1 & 1 & 1 & 0 & 0 & 0 & 0 & 0 & 0 & 0 & 0 & 0 & 0  \\
\hline
-1 & 1 & 1 & 1 & 0 & 0 & 0 & 1 & 1 & 1 & 1 & 0 & 0 & 0 & 0 & 0 & 0  \\
\hline
1 & -1 & 1 & 1 & 0 & 0 & 0 & 0 & 0 & 0 & 0 & 1 & 1 & 1 & 1 & 1 & 0  \\
\hline
1 & 1 & 1 & 1 & 0 & 0 & 0 & 0 & 0 & 0 & 0 & 0 & 0 & 0 & 0 & 0 & 1  \\
\hline

-1 & -1 & 1 & 2 & 1 & 1 & 0 & 0 & 0 & 0 & 0 & 1 & 0 & 0 & 0 & 0 & 1  \\
\hline
-1 & 1 & 1 & 2 & 0 & 0 & 0 & 0 & 0 & 0 & 0 & 0 & 1 & 1 & 0 & 0 & 0  \\
\hline
1 & -1 & 1 & 2 & 0 & 0 & 0 & 1 & 1 & 0 & 0 & 0 & 0 & 0 & 1 & 1 & 0  \\
\hline
1 & 1 & 1 & 2 & 0 & 0 & 1 & 0 & 0 & 1 & 1 & 0 & 0 & 0 & 0 & 0 & 0  \\
\hline

-1 & -1 & 2 & 1 & 1 & 0 & 1 & 0 & 0 & 0 & 0 & 1 & 0 & 0 & 0 & 0 & 1  \\
\hline
-1 & 1 & 2 & 1 & 0 & 0 & 0 & 0 & 0 & 0 & 0 & 0 & 1 & 0 & 1 & 0 & 0  \\
\hline
1 & -1 & 2 & 1 & 0 & 0 & 0 & 1 & 0 & 1 & 0 & 0 & 0 & 1 & 0 & 1 & 0  \\
\hline
1 & 1 & 2 & 1 & 0 & 1 & 0 & 0 & 1 & 0 & 1 & 0 & 0 & 0 & 0 & 0 & 0  \\
\hline

-1 & -1 & 2 & 2 & 0 & 1 & 1 & 0 & 0 & 0 & 0 & 1 & 0 & 0 & 0 & 0 & 1 \\
\hline
-1 & 1 & 2 & 2 & 0 & 0 & 0 & 0 & 0 & 0 & 0 & 0 & 1 & 0 & 0 & 1 & 0  \\
\hline
1 & -1 & 2 & 2 & 0 & 0 & 0 & 0 & 1 & 1 & 0 & 0 & 0 & 1 & 1 & 0 & 0  \\
\hline
1 & 1 & 2 & 2 & 1 & 0 & 0 & 1 & 0 & 0 & 1 & 0 & 0 & 0 & 0 & 0 & 0  \\
\hline
\multicolumn{4}{|c|}{\text{ Weights }} & \pi_1 & \pi_2 & \pi_3 & \pi_4 & \pi_5 & \pi_6 & \pi_7 & \pi_8  & \pi_9 & \pi_{10} & \pi_{11} & \pi_{12} & \pi_{13} \\
\hline
\end{array}
\]
\end{table} 

The weights $\pmb{\pi}=(\pi_1,\pi_2,...,\pi_{13})$ form a pmf, where
\begin{align}
&\pi_k=\frac{1}{6}\cos^2 67.5^{\circ} \text{ for } k=1,2,3,10,11,12, \nonumber  \\
&\pi_k=\frac{1}{12}\cos^2 67.5^{\circ} \text{ for } k=4,5,6, \nonumber \\
&\pi_7=\frac{1}{8}(1-3\cos(2 \times 67.5^{\circ})), \label{eq:qmODWeights} \\ 
&\pi_8=-\frac{1}{6}(1+4\cos(2 \times 67.5^{\circ})), \nonumber \\
&\pi_9=\frac{1}{3}\cos^2 67.5^{\circ}, \nonumber \\
&\pi_{13}=\frac{1}{2}\cos^2 67.5^{\circ}. \nonumber
\end{align}

It is straightforward to check that the correlations and $s$-functions for this model instance are given by, respectively,
\begin{align*}
&(w,x,y,z)=\frac{1}{2}\sqrt{2}(-1,1,1,1), \text{ and } \\
&(s_1,s_2,s_3,s_4)=(2\sqrt{2},0,0,0).
\end{align*}

As was shown in general in Sec.~\ref{sec:cvhEqualsHVProof}, this construction can be turned into a hidden variables \emph{model} $(\Lambda,\mathcal{F},\mu)$ (no claims of a \emph{theory}) by setting
\begin{align}
&\Lambda=\{\lambda_1,\lambda_2,...,\lambda_{13}\}, \: \mathcal{F}=2^{\Lambda}, \nonumber \\
&\mu(\{\lambda_k\})=\pi_k, \text{ (extend to } \mathcal{F} \text{ in the natural way), and}  \\
&\pmb{\gamma}_{\lambda}=\pmb{\gamma}_k \text{ whenever } \lambda=\lambda_k, \nonumber
\label{eq:infamousHV}
\end{align}
where $\pmb{\gamma}_k$ is the $k$th column of Table~\ref{tab:odQMExample} for $k=1,2,...,13$.
This is a perfectly legitimate \emph{outcome deterministic} hidden variables model, but it is not \emph{local}, either in the sense of RCI or factorizability. It is interesting to note here that
\begin{itemize}
\item Each of the 13 individual components of the construction are OD but do not satisfy RCI.
\item The QM model instance that is the culmination of the convex combination (equivalent to a hidden variables' construction) is \emph{not} OD but \emph{does} satisfy RCI.
\end{itemize}

\subsection{A fable illustrating the difference between contextuality and noncontextuality}
\label{sec:soFable}

Intuitively, what is the difference between the \emph{noncontextual} QM model instance in Sec.~\ref{sec:hvPD} and the corresponding \emph{contextual} QM model instance in Sec.~\ref{sec:hvNotPD}? After all, in both cases, the QM model instance can be written as a convex combination of OD model instances. Therefore, it is not unreasonable to hope that, in both cases, one could reproduce the statistics of the target QM model instance in a "classical" way. For example, imagine a preparation process that, on each iteration of the experiment, produces one of the basic OD model instances with a probability equal to its weight in the convex combination. The long-term statistics should then approach those of the target QM model instance. 

\begin{figure}[H]
\centering
\includegraphics[width=0.7\linewidth]{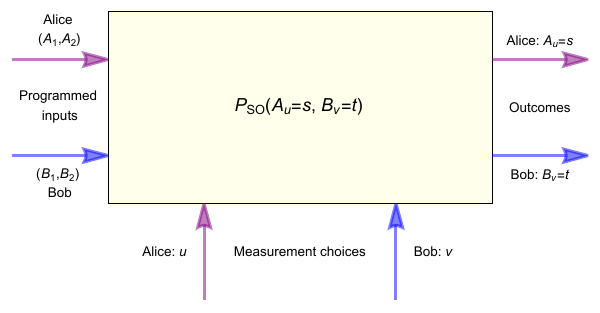}
\caption{The imaginary scenario in which Eve attempts to fool Alice and Bob into thinking they are observing data from an EPRB experiment. Eve is in control of the 4-tuple of inputs $(A_1,A_2,B_1,B_2)$.}
\label{fig:preArrBlackBox}
\end{figure}

Putting this into the form of a fable, imagine that Eve (Alice and Bob's friend) has been asked to try and fool Alice and Bob into thinking they are getting data from a real EPRB experiment. She sends out four values $(A_1,A_2,B_1,B_2)$ on each iteration of the experiment. $(A_1,A_2)$ go to Alice and $(B_1,B_2)$ go to Bob. Also suppose that if Alice chooses her first measurement option $M_A=1$, she gets the result $A=A_1$ and if she chooses her second $M_A=2$, she gets $A=A_2$. Similarly for Bob's result $B$ with respect to his own measurement options. Fig.~\ref{fig:preArrBlackBox} illustrates this setup. Borrowing suggestive terminology from Fine's excellent (and more general) paper~\cite{Fine1982-2} on joint distributions, the random variables $(A_1,A_2,B_1,B_2)$ will be called "statistical observables". 

Recall that in the actual EPRB experiment, Alice and Bob can only (jointly) record statistics on \emph{compatible pairs} $(A_u,B_v)$ for $u,v=1 \text{ or } 2$. It is not possible to perform simultaneous measurements of non-commuting observables. Therefore, Eve's task is to send out the statistical observables $(A_1,A_2,B_1,B_2)$ in such a way that the resulting \emph{occurrence frequencies of compatible pairs} $(A_1,B_1),(A_1,B_2),(A_2,B_1),(A_2,B_2)$ approach the corresponding expected frequencies for the target QM model instance. Here are two strategies she could try.

\begin{enumerate}
\item \underline{Strategy 1}. Eve adopts a full joint probability distribution 
\[P_{SO}(A_1=s,A_2=s',B_1=t,B_2=t')\] 
on $(A_1,A_2,B_1,B_2).$ See Table~\ref{tab:pmfPA}. The long-term frequency with which she sends out a given 4-tuple $(s,s',t,t')$ is determined entirely by the associated $\rho_k$. For example,
\[P_{SO}(A_1=1,A_2=-1,B_1=-1,B_2=1)=\rho_7,\]
so in the long-term, Alice and Bob will be presented with the 4-tuple $(1,-1,-1,1)$ approximately $100\rho_7\%$ of the time, from which they choose a compatible pair, based on their measurement choices.
\begin{table}[H]
\caption{A full joint pmf $\pmb{\rho}=(\rho_1,\rho_2,...,\rho_{16})$ for the imaginary scenario. Each $\rho_k$ corresponds to a full joint probability $P_{SO}(A_1=s,A_2=s',B_1=t,B_2=t')$. These determine the approximate long-term frequency with which Eve sends out any given 4-tuple $(s,s',t,t')$.}
\label{tab:pmfPA}
\[
\begin{array}{|c|c|c|c|c|}
\hline
\text{} & \multicolumn{4}{c|} {(s,s')} \\
\hline
(t,t') & (-1,-1)  & (-1,1)   & (1,-1)   & (1,1) \\
\hline
(-1,-1) & \rho_1 & \rho_2 & \rho_3 & \rho_4 \\
\hline
(-1,1) & \rho_5 & \rho_6 & \rho_7 & \rho_8 \\
\hline
(1,-1) & \rho_9 & \rho_{10} & \rho_{11} & \rho_{12} \\
\hline
(1,1) & \rho_{13} & \rho_{14} & \rho_{15} & \rho_{16} \\
\hline
\end{array}
\]
\end{table}

Then a straightforward marginal probability calculation yields the compatible double marginals in Table~\ref{tab:preArrMarginals}. Obviously these double marginals are exactly the same form as the noncontextual double conditional probabilities of Table~\ref{tab:predeterminedTypeIntro}.

\begin{table}[H]
\setlength\arraycolsep{2pt}
\caption{The statistical observables' double marginals $P_{SO}(A_u=s,B_v=t)$. These are identical to the NC double conditional probabilities of Table~\ref{tab:predeterminedTypeIntro}.}
\label{tab:preArrMarginals}
\[
\begin{array}{|c|c|c|c|c|}
\hline
\text{ Indices } & \multicolumn{4}{c|}{\text{ Outcomes }(s,t)} \\
\hline
(u,v) & (-1,-1)  & (-1,1)   & (1,-1)   & (1,1) \\
\hline
(1,1) & \rho_1+\rho_2+\rho_5+\rho_6 & \rho_9+\rho_{10}+\rho_{13}+\rho_{14} & \rho_3+\rho_4+\rho_7+\rho_8 & \rho_{11}+\rho_{12}+\rho_{15}+\rho_{16} \\
\hline
(1,2) & \rho_1+\rho_2+\rho_9+\rho_{10} & \rho_5+\rho_{6}+\rho_{13}+\rho_{14} & \rho_3+\rho_4+\rho_{11}+\rho_{12} & \rho_{7}+\rho_{8}+\rho_{15}+\rho_{16} \\
\hline
(2,1) & \rho_1+\rho_3+\rho_5+\rho_7 & \rho_9+\rho_{11}+\rho_{13}+\rho_{15} & \rho_2+\rho_4+\rho_6+\rho_8 & \rho_{10}+\rho_{12}+\rho_{14}+\rho_{16} \\
\hline
(2,2) & \rho_1+\rho_3+\rho_9+\rho_{11} & \rho_5+\rho_{7}+\rho_{13}+\rho_{15} & \rho_2+\rho_4+\rho_{10}+\rho_{12} & \rho_{6}+\rho_{8}+\rho_{14}+\rho_{16} \\
\hline
\end{array}
\]
\end{table}

Therefore, Eve can mimic the statistics of the QM model instance in Sec.~\ref{sec:hvPD} but \emph{not} the one in Sec.~\ref{sec:hvNotPD}. 

\item \underline{Strategy 2}. On each iteration of the experiment, Eve \emph{guesses} what Alice and Bob's joint measurement choice will be, then sends out a 4-tuple $(s,s',t,t')$ with a probability based on that assumption. This strategy is illustrated for an arbitrary QM model instance in the case where she guesses that both Alice and Bob will make their \emph{first} measurement choice. In this case, Eve sends out (where $"*"$ denotes "don't care")
\begin{align}
(-1,*,-1,*) \text{ with probability } &\frac{1}{2}\cos^2 \theta_1, \nonumber \\
(-1,*,1,*)  \text{ with probability } &\frac{1}{2}\sin^2 \theta_1, \nonumber \\
(1,*,-1,*)  \text{ with probability } &\frac{1}{2}\sin^2 \theta_1, \\
(1,*,1,*)   \text{ with probability } &\frac{1}{2}\cos^2 \theta_1. \nonumber
\end{align}
Then \emph{if} Alice and Bob actually make their first measurement choices, thus choosing the values in the first (Alice's) and third (Bob's) places in each of these 4-tuples, in the long-term they will get the values shown with a frequency close to the those predicted by the QM model type as shown in Table~\ref{tab:qmTypeIntro}. Similarly for the other three measurement choice combinations for Alice and Bob. The problem with this strategy of course is that Eve will often be wrong in her guess about Alice and Bob's upcoming measurement choices, so the values she sends them will occur with the "wrong" frequencies. In fact, she can only reliably make this strategy work if the QM model instance is also the trivial model instance. See Table~\ref{tab:trivialType}. Since this is the completely uniform, all zero-correlation situation, it is not of much value or interest.
\end{enumerate}

\begin{table}[H]
\caption{The trivial model instance is the only one that is simultaneously QM ($\theta_1=135^{\circ},\theta_k=45^{\circ}$ for $k=2,3,4$), factorizable ($\alpha_k=\beta_k=\frac{1}{2}$ for all $k=1,2,3,4$), and noncontextual ($\rho_k=\frac{1}{16}$ for all $k=1,2,...,16$). All of its correlations and $s$-functions are 0.}
\label{tab:trivialType}
\[
\begin{array}{|c|c|c|c|c|}
\hline
\text{Context} & \multicolumn{4}{c|}{\text{Outcome }(s,t)} \\
\hline
(u,v) & (-1,-1)  & (-1,1)   & (1,-1)   & (1,1) \\
\hline
(1,1) & \frac{1}{4} & \frac{1}{4} & \frac{1}{4} & \frac{1}{4} \\
\hline
(1,2) & \frac{1}{4} & \frac{1}{4} & \frac{1}{4} & \frac{1}{4} \\
\hline
(2,1) & \frac{1}{4} & \frac{1}{4} & \frac{1}{4} & \frac{1}{4} \\
\hline
(2,2) & \frac{1}{4} & \frac{1}{4} & \frac{1}{4} & \frac{1}{4} \\
\hline
\end{array}
\]
\end{table}  

As a last hope, what about sending out the 13 OD model instances of Table~\ref{tab:odQMExample} with frequencies given by the weights in Eq.~\ref{eq:qmODWeights}? The problem with this idea is that at least one (in fact all) of the OD model instances in Table~\ref{tab:odQMExample} fail to satisfy RCI, hence cannot be noncontextual. In other words, for every one of these OD model instance, there is no full joint probability distribution on $(A_{1},A_{2},B_{1},B_{2})$ whose marginals match the double conditionals of that OD model instance.

It is interesting to pause and contemplate what it means that neither of these strategies always works. In strategy 1, the overall joint probability distribution (for both compatible and incompatible statistical observables) basically \emph{encodes} the desired correlations into the experimental results, \emph{regardless of future measurement choices}. Strategy 2 \emph{requires} foreknowledge of the measurement choices. In other words, in a QM world, in which the behavior can be contextual, it is not always possible to fake the results in a way that ignores the future measurements. 

\section{Predictability}
\label{sec:predictability}

Correlations are one thing, actual transfer of information quite another. The parameterized operational modeling framework is very useful in exploring this concept. "Predictability" is all about Alice's ability "predict" or "know" something about Bob's wing of the experiment. It comes in two different flavors. The first, "outcome predictability" (OP), indicates that Alice can reliably predict (i.e., with probability 1) what Bob's outcome will be, based solely on her  measurement choice and outcome.  "Measurement predictability" (MP) indicates that Alice can reliably deduce what Bob's measurement choice is, again based solely on her measurement choice and outcome. No \emph{direct} knowledge of Bob's measurement choice is required in either case. In the "measurement predictability" case,  Bob can send information to Alice through his choice of measurement.

In Sec.~\ref{sec:qmOP}, it will be shown that \emph{some} QM model instances are outcome predictable. But in Sec.~\ref{sec:qmNoComm}, it will be shown that \emph{no} QM model instance is measurement predictable -- alas no faster-than-light communication in a Bell/Aspect experiment under QM rules. On the other hand, it will be shown that there are infinitely many generic model instances which \emph{are} outcome predictable and/or measurement predictable.

The ability of Alice to deduce either Bob's outcome or measurement choice is predicated on the assumptions 
\begin{itemize}
\item The generic parameters $\pmb{\gamma}=(\gamma_1,\gamma_2,...,\gamma_{16})$ are fixed throughout the experiment,
\item The generic parameters $\pmb{\gamma}$ conform to one of several special "patterns" of zero and non-zero entries,
\item Alice knows this pattern.
\end{itemize}
These assumptions merely state that Alice knows the experimental setup, and that it is fixed for all iterations of the experiment.

\subsection{Outcome predictability}
\label{sec:outcomePredictability}

OP implies that every Alice outcome-measurement choice pair $(A,M_{A})=(s,u)$ is mapped to only one of Bob's outcomes $B=t$. Table~\ref{tab:predOutcomeTable} shows the 16 possible "patterns" for generic parameters that fulfill this condition.

\begin{table}[H]
\setlength\arraycolsep{2pt}
\caption{The 16 \textbf{outcome predictable (OP)} model type patterns. Alice can always deduce what Bob's outcome is, given her own measurement choice and outcome. This is predicated on the assumption that Alice knows in advance which conditional probabilities are zero.}
\label{tab:predOutcomeTable}
\[
\begin{array}{|r|r|r|r||c|c|c|c|c|c|c|c|c|c|c|c|c|c|c|c|}
\hline
s & t & u & v & 1 & 2 & 3 & 4 & 5 & 6 & 7 & 8 & 9 & 10 & 11 & 12 & 13 & 14 & 15 & 16 \\
\hline
\hline
-1 & -1 & 1 & 1 & \gamma_1 & \gamma_1 & \gamma_1 & \gamma_1 & \gamma_1 & \gamma_1 & \gamma_1 & \gamma_1 & 0 & 0 & 0 & 0 & 0 & 0 & 0 & 0 \\
\hline
-1 & 1 & 1 & 1 & 0 & 0 & 0 & 0 & 0 & 0 & 0 & 0 & \gamma_2 & \gamma_2 & \gamma_2 & \gamma_2 & \gamma_2 & \gamma_2 & \gamma_2 & \gamma_2 \\
\hline
1 & -1 & 1 & 1 & \gamma_3 & \gamma_3 & \gamma_3 & \gamma_3 & 0 & 0 & 0 & 0 & \gamma_3 & \gamma_3 & \gamma_3 & \gamma_3 & 0 & 0 & 0 & 0 \\
\hline
1 & 1 & 1 & 1 & 0 & 0 & 0 & 0 & \gamma_4 & \gamma_4 & \gamma_4 & \gamma_4 & 0 & 0 & 0 & 0 & \gamma_4 & \gamma_4 & \gamma_4 & \gamma_4 \\
\hline

-1 & -1 & 1 & 2 & \gamma_5 & \gamma_5 & \gamma_5 & \gamma_5 & \gamma_5 & \gamma_5 & \gamma_5 & \gamma_5 & 0 & 0 & 0 & 0 & 0 & 0 & 0 & 0 \\
\hline
-1 & 1 & 1 & 2 & 0 & 0 & 0 & 0 & 0 & 0 & 0 & 0 & \gamma_6 & \gamma_6 & \gamma_6 & \gamma_6 & \gamma_6 & \gamma_6 & \gamma_6 & \gamma_6 \\
\hline
1 & -1 & 1 & 2 & \gamma_7 & \gamma_7 & \gamma_7 & \gamma_7 & 0 & 0 & 0 & 0 & \gamma_7 & \gamma_7 & \gamma_7 & \gamma_7 & 0 & 0 & 0 & 0 \\
\hline
1 & 1 & 1 & 2 & 0 & 0 & 0 & 0 & \gamma_8 & \gamma_8 & \gamma_8 & \gamma_8 & 0 & 0 & 0 & 0 & \gamma_8 & \gamma_8 & \gamma_8 & \gamma_8 \\
\hline

-1 & -1 & 2 & 1 & \gamma_9 & \gamma_9 & 0 & 0 & \gamma_9 & \gamma_9 & 0 & 0 & \gamma_9 & \gamma_9 & 0 & 0 & \gamma_9 & \gamma_9 & 0 & 0 \\
\hline
-1 & 1 & 2 & 1 & 0 & 0 & \gamma_{10} & \gamma_{10} & 0 & 0 & \gamma_{10} & \gamma_{10} & 0 & 0 & \gamma_{10} & \gamma_{10} & 0 & 0 & \gamma_{10} & \gamma_{10} \\
\hline
1 & -1 & 2 & 1 & \gamma_{11} & 0 & \gamma_{11} & 0 & \gamma_{11} & 0 & \gamma_{11} & 0 & \gamma_{11} & 0 & \gamma_{11} & 0 & \gamma_{11} & 0 & \gamma_{11} & 0 \\
\hline
1 & 1 & 2 & 1 & 0 & \gamma_{12} & 0 & \gamma_{12} & 0 & \gamma_{12} & 0 & \gamma_{12} & 0  & \gamma_{12} & 0 & \gamma_{12} & 0 & \gamma_{12} & 0 & \gamma_{12}\\
\hline

-1 & -1 & 2 & 2 & \gamma_{13} & \gamma_{13} & 0 & 0 & \gamma_{13} & \gamma_{13} & 0 & 0 & \gamma_{13} & \gamma_{13} & 0 & 0 & \gamma_{13} & \gamma_{13} & 0 & 0 \\
\hline
-1 & 1 & 2 & 2 & 0 & 0 & \gamma_{14} & \gamma_{14} & 0 & 0 & \gamma_{14} & \gamma_{14} & 0 & 0 & \gamma_{14} & \gamma_{14} & 0 & 0 & \gamma_{14} & \gamma_{14} \\
\hline
1 & -1 & 2 & 2 & \gamma_{15} & 0 & \gamma_{15} & 0 & \gamma_{15} & 0 & \gamma_{15} & 0 & \gamma_{15} & 0 & \gamma_{15} & 0 & \gamma_{15} & 0 & \gamma_{15} & 0 \\
\hline
1 & 1 & 2 & 2 & 0 & \gamma_{16} & 0 & \gamma_{16} & 0 & \gamma_{16} & 0 & \gamma_{16} &  0 & \gamma_{16} & 0 & \gamma_{16} & 0 & \gamma_{16} & 0 & \gamma_{16} \\
\hline
\end{array}
\]
\end{table} 
To see this, note that every one of the 16 patterns in Table~\ref{tab:predOutcomeTable} is like a standard set of generic parameters, except that certain $\gamma_{k}$'s are set to zero \emph{a priori}. It is assumed that the measurement context constraints (Eq.~\ref{eq:contextualIntro}) still hold. 

Consider the first model instance in the column marked "1". Suppose Alice chooses her measurement $M_A=u=1$ and gets outcome $A=s=-1$. This can only happen in rows 1 and/or 5, since the other possibilities are associated with  $\gamma_2=0$ and $\gamma_6=0$. And in these cases (where at least one of $\gamma_1$ and $\gamma_5$ must be nonzero), Alice can see that Bob's outcome is always $B=t=-1$. Similarly suppose Alice chooses her measurement $M_A=u=1$ and gets outcome $A=s=+1$. This can only happen in rows 3 and/or 7, since the other possibilities are associated with  $\gamma_4=0$ and $\gamma_8=0$. And in these cases (where at least one of $\gamma_3$ and $\gamma_7$ must be nonzero), Alice can see that Bob's outcome is again always $B=t=-1$. A similar argument can be made for the cases where $(A,M_{A})=(s,u)=(-1,2)$  or $(1,2)$. Extend the same arguments to all of the other columns.

In other words, Alice can always deduce Bob's outcome because any ambiguities were preemptively removed by strategically zeroing out certain $\gamma_k$'s while retaining the measurement context constraints. 

\subsection{Can a QM model instance be outcome predictable?}
\label{sec:qmOP}

\begin{Theorem}
There are only four OP model instances that are also QM model instances.
\label{thm:qmOutPred}
\end{Theorem}
\begin{proof}
The (generic) parameters for any QM model instance must satisfy the two conditions (see for example Table~\ref{tab:qmTypeIntro}):

\begin{itemize}
\item Each block of four parameters must have the pattern "abba", and 
\item All parameters must be between $0$ and $1/2$. 
\end{itemize}

The only OP model instances in Table~\ref{tab:predOutcomeTable} that could possibly satisfy the first condition are numbers 6,7,10, and 11. Combining the second condition with the usual measurement context constraints (see Eq.~\ref{eq:contextualIntro}) implies that all nonzero $\gamma_{k}$'s must be equal to $1/2$. Therefore the only such cases are as shown in Table~\ref{tab:opQMTable}.
\end{proof}

Examples of QM parameter assignments that yield instances 1,2,3, and 4 of Table~\ref{tab:opQMTable}, respectively, are 
\begin{align}
\label{eq:qmOutcomePred}
&1.\: \theta_{1}=\theta_{2}=\theta_{3}=\theta_{4}=0^{\circ}, \nonumber \\
&2.\: \theta_{1}=0^{\circ},\theta_{2}=0^{\circ},\theta_{3}=90^{\circ},\theta_{4}=-90^{\circ}, \\
&3.\: \theta_{1}=90^{\circ},\theta_{2}=90^{\circ},\theta_{3}=0^{\circ},\theta_{4}=0^{\circ}, \nonumber \\
&4.\: \theta_{1}=270^{\circ},\theta_{2}=90^{\circ},\theta_{3}=90^{\circ},\theta_{4}=90^{\circ}. \nonumber
\end{align}

These QM model instances are all perfectly correlated, as their sets of correlations are, respectively
\begin{align}
\label{eq:qmOutcomePredCorr}
&1. \:(1,1,1,1), \nonumber \\
&2. \:(1,1,-1,-1),\\
&3. \:(-1,-1,1,1), \nonumber\\
&4. \:(-1,-1,-1,-1). \nonumber
\end{align}

One suspects there are RCI+OD model instances that can achieve the same correlations.  Table~\ref{tab:odQMOPInstance} shows a comparison of an RCI+OD model instance ("OD" column) and a QM model instance ("QM" column), both of which achieve the perfect correlations 
\[
(w,x,y,z)=(-1,-1,-1,-1) 
\]
with $s$-functions 
\[
(s_1,s_2,s_3,s_4)=(-2,-2,-2,-2).
\]
Parameter assignments which produce the two model instances in~Table~\ref{tab:odQMOPInstance} are:

\begin{align} 
&\text{ OD: } \gamma_2=\gamma_6=\gamma_{10}=\gamma_{14}=1 \text{ and all other } \gamma_k=0, \nonumber \\
&\text{ QM: } \theta_1=270^{\circ}, \theta_2=\theta_3=\theta_4=90^{\circ}.
\end{align} 

Once again, this illustrates the fact that the mapping from conditional probabilities to correlations is "many-to-one".

\begin{table}[H]
\caption{The four QM model instances that are also outcome predictable.}
\label{tab:opQMTable}
\[
\begin{array}{|r|r|r|r||c|c|c|c|}
\hline
s & t & u & v & 1 & 2 & 3 & 4  \\
\hline
\hline
-1 & -1 & 1 & 1 & \frac{1}{2} & \frac{1}{2} & 0 & 0  \\
\hline
-1 & 1 & 1 & 1 & 0 & 0 & \frac{1}{2} & \frac{1}{2}  \\
\hline
1 & -1 & 1 & 1 & 0 & 0 & \frac{1}{2} & \frac{1}{2}  \\
\hline
1 & 1 & 1 & 1 & \frac{1}{2} & \frac{1}{2} & 0 & 0  \\
\hline

-1 & -1 & 1 & 2 & \frac{1}{2} & \frac{1}{2} & 0 & 0  \\
\hline
-1 & 1 & 1 & 2 & 0 & 0 & \frac{1}{2} & \frac{1}{2}  \\
\hline
1 & -1 & 1 & 2 & 0 & 0 & \frac{1}{2} & \frac{1}{2}  \\
\hline
1 & 1 & 1 & 2 & \frac{1}{2} & \frac{1}{2} & 0 & 0  \\
\hline

-1 & -1 & 2 & 1 & \frac{1}{2} & 0 & \frac{1}{2} & 0  \\
\hline
-1 & 1 & 2 & 1 & 0 & \frac{1}{2} & 0 & \frac{1}{2}  \\
\hline
1 & -1 & 2 & 1 & 0 & \frac{1}{2} & 0 & \frac{1}{2}  \\
\hline
1 & 1 & 2 & 1 &  \frac{1}{2} & 0 & \frac{1}{2} & 0   \\
\hline

-1 & -1 & 2 & 2 & \frac{1}{2} & 0 & \frac{1}{2} & 0  \\
\hline
-1 & 1 & 2 & 2 & 0 & \frac{1}{2} & 0 & \frac{1}{2}  \\
\hline
1 & -1 & 2 & 2 & 0 & \frac{1}{2} & 0 & \frac{1}{2} \\
\hline
1 & 1 & 2 & 2 & \frac{1}{2} & 0 & \frac{1}{2} & 0  \\
\hline
\end{array}
\]
\end{table} 

\begin{table}[H]
\caption{RCI+OD and QM model instances which both achieve the correlations $(w,x,y,z)=(-1,-1,-1,-1)$.}
\label{tab:odQMOPInstance}
\[
\begin{array}{|r|r|r|r||c|c|}
\hline
s & t & u & v & OD & QM  \\
\hline
\hline
-1 & -1 & 1 & 1 & 0 & 0   \\
\hline
-1 & 1 & 1 & 1 & 1 &  \frac{1}{2}  \\
\hline
1 & -1 & 1 & 1 & 0 & \frac{1}{2}  \\
\hline
1 & 1 & 1 & 1 &  0 & 0  \\
\hline

-1 & -1 & 1 & 2 &  0 & 0  \\
\hline
-1 & 1 & 1 & 2 & 1 & \frac{1}{2}  \\
\hline
1 & -1 & 1 & 2 & 0 & \frac{1}{2}   \\
\hline
1 & 1 & 1 & 2 & 0 & 0  \\
\hline

-1 & -1 & 2 & 1 & 0 & 0  \\
\hline
-1 & 1 & 2 & 1 & 1 & \frac{1}{2}  \\
\hline
1 & -1 & 2 & 1 & 0 & \frac{1}{2}   \\
\hline
1 & 1 & 2 & 1 &  0 & 0   \\
\hline

-1 & -1 & 2 & 2 & 0 &  0  \\
\hline
-1 & 1 & 2 & 2 & 1 & \frac{1}{2}   \\
\hline
1 & -1 & 2 & 2 & 0 & \frac{1}{2}  \\
\hline
1 & 1 & 2 & 2 &  0 &  0  \\
\hline
\end{array}
\]
\end{table} 

\subsection{Outcome predictable instances that are not QM}

Inspection of Table~\ref{tab:predOutcomeTable} yields the following three theorems.

\begin{Theorem}
There are an infinite number of outcome predictable model instances that are also NC.
\label{thm:opPredInfinite}
\end{Theorem}
\begin{proof}
Many examples can be generated from the 16 patterns in Table~\ref{tab:predOutcomeTable}. Consider model instance pattern number 6, for example. Then for the corresponding NC model parameters, set
\[\rho_{k}=0 \text{ for all } k \text{ except } 1 > \rho_{1}>0 \text{ and } \rho_{16}=1-\rho_{1}.\]

Now plug these into the NC  parameter list in Table~\ref{tab:predeterminedTypeIntro}. This produces a set of generic parameters that match the pattern of the 6th column of Table~\ref{tab:predOutcomeTable}, namely
\begin{align*}
&(\gamma_{1},0,0,\gamma_{4},\gamma_{5},0,0,\gamma_{8},\gamma_{9},0,0,\gamma_{12},\gamma_{13},0,0,\gamma_{16}) = \\
&(\rho_{1},0,0,\rho_{16},\rho_{1},0,0,\rho_{16},\rho_{1},0,0,\rho_{16},\rho_{1},0,0,\rho_{16}).
\end{align*}
\end{proof}       \qedhere

\begin{Theorem}
There are an infinite number of outcome predictable model instances that satisfy CHSH.
\end{Theorem}
\begin{proof}
The example from Thm.~\ref{thm:opPredInfinite} is NC, hence satisfies all CHSH inequalities. Obviously there are an infinite number of choices for the two defining parameters $(\rho_1,\rho_{16})$. As another example, consider model instance number 7 in Table~\ref{tab:predOutcomeTable}. The correlations are $(w,x,y,z)=(\gamma_{1}+\gamma_{4},\gamma_{5}+\gamma_{8},-\gamma_{10}-\gamma_{11},-\gamma_{14}-\gamma_{15})=(1,1,-1,-1)$. Hence the $s$-functions are $(-2,-2,2,2)$, all of which satisfy CHSH. Obviously infinitely many values can be assigned to the non-zero $\gamma_{k}$'s which yield these same $s$-function values.
\end{proof}      \qedhere

\begin{Theorem}
There are an infinite number of outcome predictable model instances that exceed at least one Tsirelson bound.
\end{Theorem}
\begin{proof}
Consider model instance pattern number 1 in Table~\ref{tab:predOutcomeTable}. The correlations are $(w,x,y,z)=(\gamma_{1}-\gamma_{3},\gamma_{5}-\gamma_{7},\gamma_{9}-\gamma_{11},\gamma_{13}-\gamma_{15})$. One possible assignment is $\gamma_{1}=0,\gamma_{3}=1,\gamma_{5}=1,\gamma_{7}=0,\gamma_{9}=1,\gamma_{11}=0,\gamma_{13}=1,\gamma_{15}=0$, which yields the correlations $(-1,1,1,1)$. This results in the $s$-functions $(4,0,0,0)$, so one $s$-function exceeds a Tsirelson bound. To obtain infinitely many assignments of values to the $\gamma_{k}$'s so that at least one $s$-functions exceeds a Tsirelson bound, introduce an $0<\epsilon<1$ to write the $\gamma_{k}$'s as follows: $\gamma_{1}=\epsilon,\gamma_{3}=1-\epsilon,\gamma_{5}=1,\gamma_{7}=0,\gamma_{9}=1,\gamma_{11}=0,\gamma_{13}=1,\gamma_{15}=0$. Then the correlations become 
\begin{align}
&(w,x,y,z)=(\gamma_{1}-\gamma_{3},\gamma_{5}-\gamma_{7},\gamma_{9}-\gamma_{11},\gamma_{13}-\gamma_{15})=\\ \nonumber
&(2\epsilon-1,1,1,1)
\end{align}
which yields $s$-functions 

\begin{equation}
(4-2\epsilon,2\epsilon,2\epsilon,2\epsilon).
\end{equation}

The first $s$-function exceeds a Tsirelson bound if $\epsilon<2-\sqrt{2} \approx 0.59$. For example, if $\epsilon=0.5$, then the $s$-functions are $(3,1,1,1)$, and the first one exceeds the upper Tsirelson bound, i.e. $3>2\sqrt{2}$. 
\end{proof}      \qedhere

\subsection{Measurement predictability}
\label{sec:measPredictability}

MP implies that every Alice outcome-measurement choice pair $(A,M_{A})=(s,u)$ is mapped to only one of Bob's measurements $M_B=v$. Table~\ref{tab:predMeasTable} shows the 4 possible "patterns" for generic parameters that fulfill this condition.

\begin{table}[H]
\caption{The 4 \textbf{measurement predictable (MP)} model type patterns. Alice can always deduce what Bob's measurement is, given her own measurement choice and outcome. This is predicated on the assumption that Alice knows in advance which conditional probabilities are zero.}
\label{tab:predMeasTable}
\[
\begin{array}{|r|r|r|r||c|c|c|c|}
\hline
s & t & u & v & 1 & 2 & 3 & 4  \\
\hline
\hline
-1 & -1 & 1 & 1 & \gamma_1 & \gamma_1 & 0 & 0  \\
\hline
-1 & 1 & 1 & 1 & \gamma_2 & \gamma_2 & 0 & 0  \\
\hline
1 & -1 & 1 & 1 & 0 & 0 & \gamma_3 & \gamma_3  \\
\hline
1 & 1 & 1 & 1 &  0 & 0 & \gamma_4 & \gamma_4 \\
\hline

-1 & -1 & 1 & 2 & 0 & 0 & \gamma_5 & \gamma_5  \\
\hline
-1 & 1 & 1 & 2 & 0 & 0 & \gamma_6 & \gamma_6  \\
\hline
1 & -1 & 1 & 2 & \gamma_7 & \gamma_7 & 0 & 0  \\
\hline
1 & 1 & 1 & 2 & \gamma_8 & \gamma_8 & 0 & 0  \\
\hline

-1 & -1 & 2 & 1 & \gamma_9 & 0 & \gamma_9 & 0  \\
\hline
-1 & 1 & 2 & 1 & \gamma_{10} & 0 & \gamma_{10} & 0  \\
\hline
1 & -1 & 2 & 1 & 0 & \gamma_{11} & 0 & \gamma_{11}  \\
\hline
1 & 1 & 2 & 1 &  0 & \gamma_{12} & 0 & \gamma_{12}   \\
\hline

-1 & -1 & 2 & 2 & 0 & \gamma_{13} & 0 & \gamma_{13}  \\
\hline
-1 & 1 & 2 & 2 & 0  & \gamma_{14} & 0 & \gamma_{14}  \\
\hline
1 & -1 & 2 & 2 &  \gamma_{15} & 0 & \gamma_{15} & 0 \\
\hline
1 & 1 & 2 & 2 & \gamma_{16} & 0 & \gamma_{16} & 0  \\
\hline
\end{array}
\]
\end{table}

To see this, note that every one of the 4 patterns in Table~\ref{tab:predMeasTable} is like a standard set of generic parameters, except that certain $\gamma_{k}$'s are set to zero \emph{a priori}. It is assumed that the measurement context constraints (Eq.~\ref{eq:contextualIntro}) still hold. Consider the first model instance in the column marked "1". Suppose Alice chooses her measurement $M_A=u=1$ and gets outcome $A=s=-1$. This can only happen in rows 1 and/or 2, since the other possibilities are associated with  $\gamma_5=0$ and $\gamma_6=0$. And in the cases where either or both $\gamma_1$ and $\gamma_2$ are nonzero, Alice can see that Bob's measurement is always $M_B=v=1$. Now suppose Alice chooses her measurement $M_A=u=1$ and gets outcome $A=s=+1$. This can only happen in rows 7 and/or 8, since the other possibilities are associated with  $\gamma_3=0$ and $\gamma_4=0$. And in the cases where either or both $\gamma_7$ and $\gamma_8$ are nonzero, Alice can see that Bob's measurement is always $M_B=v=2$. A similar argument can be made for the cases where $(A,M_{A})=(s,u)=(-1,2)$ or $(1,2)$. Extend the same arguments to all of the other columns.

Alice can always deduce Bob's measurement because any ambiguities were preemptively removed by strategically zeroing out certain $\gamma_k$'s. There are 16 cpvs that have this property, but after the zeroing process, 12 fail to satisfy the measurement context constraints, so only 4 are left. 

\subsection{The no-communication theorem for QM}
\label{sec:qmNoComm}

It would be a surprise if any of these four model instance patterns could also be QM. The next theorem says they cannot, and therefore confirms the no-communication theorem for QM (in the context of a Bell/Aspect-like experiment, that is).

\begin{Theorem}
\label{thm:qmNotMP}
No QM model instance can also be measurement predictable.
\end{Theorem}

\begin{proof}
Consider the first four entries $(\gamma_1,\gamma_2,0,0)$ in the first measurement predictable pattern of Table~\ref{tab:predMeasTable}. Recall these 4 numbers must sum to 1, so necessarily $\gamma_1+\gamma_2=1$. In order to be a QM model instance, the pattern of these first four entries must be "abba" (see Table~\ref{tab:qmTypeIntro}). This means $\gamma_1=0$. Then $\gamma_2=1$. But then the second and third entries $1 \text{ and } 0$ do not match, so this cannot be a QM model instance. A similar argument can be used to show the other three patterns could not possibly conform to the QM pattern either.
\end{proof} 
In the literature, the model instances that satisfy RCI are often called "no-signaling", presumably implying that any "signaling" (i.e. faster-than-light communication capable) model instances must lie in the set  that do not satisfy RCI. The set of measurement predictable model instances are inside this RCI complement, of course, but more than that, they form a \emph{proper} subset. That is, measurement predictability defines more precisely the boundary between "no-signaling" and "signaling". Consider the following OD model instance:\footnote{This is also Ex.~\ref{ex:chshNotNC} in Sec.~\ref{sec:NCDoesNotImplyFC}, repeated here for convenience.}

\begin{Example}
$\pmb{\gamma}=(\gamma_1,\gamma_2,...,\gamma_{16})$ where $\gamma_k=1$ for $k=1,5,9,16$ and $\gamma_k=0$ otherwise.
\label{ex:notRCInotMP}
\end{Example} 
This OD model instance does not satisfy RCI (it is not one of the 16 listed in Table~\ref{tab:rciODTabIntro}). It also cannot be measurement predictable. Consider Table~\ref{tab:predMeasTable}, and simply check that there is no way to assign 1's and 0's to the non-zero $\gamma_k$'s in any of the 4 columns to produce Ex.~\ref{ex:notRCInotMP}.\footnote{This is true for Bob sending messages to Alice. The corresponding table for Alice sending messages to Bob is different, and must be checked as well (exercise for reader).} 

\subsection{Measurement predictability and noncontextuality}
\label{sec:mpAndPD}

Thm.~\ref{thm:qmNotMP} shows that no QM model instance can be MP. The next theorem shows a similar result for NC model instances.

\begin{Theorem}
\label{thm:pdNotMP}
There are no measurement predictable model instances that are NC, hence none that are factorizable either. 
\end{Theorem}
\begin{proof}
Consider the table of MP instances in Table~\ref{tab:predMeasTable}. The proof consists of a straightforward check that no instance in that table can be NC.  That is,  in Eq.~\ref{eqn:measPredBR}, can the column vector on the left be equal to any of the four column vectors on the right?
\begin{equation}
\label{eqn:measPredBR}
\left(
\begin{array}{l}
\rho_{1}+\rho_{2}+\rho_{5}+\rho_{6}\\
\rho_{9}+\rho_{10}+\rho_{13}+\rho_{14}\\
\rho_{3}+\rho_{4}+\rho_{7}+\rho_{8}\\
\rho_{11}+\rho_{12}+\rho_{15}+\rho_{16}\\

\rho_{1}+\rho_{2}+\rho_{9}+\rho_{10}\\
\rho_{5}+\rho_{6}+\rho_{13}+\rho_{14}\\
\rho_{3}+\rho_{4}+\rho_{11}+\rho_{12}\\
\rho_{7}+\rho_{8}+\rho_{15}+\rho_{16}\\

\rho_{1}+\rho_{3}+\rho_{5}+\rho_{7}\\
\rho_{9}+\rho_{11}+\rho_{13}+\rho_{15}\\
\rho_{2}+\rho_{4}+\rho_{6}+\rho_{8}\\
\rho_{10}+\rho_{12}+\rho_{14}+\rho_{16}\\

\rho_{1}+\rho_{3}+\rho_{9}+\rho_{11}\\
\rho_{5}+\rho_{7}+\rho_{13}+\rho_{15}\\
\rho_{2}+\rho_{4}+\rho_{10}+\rho_{12}\\
\rho_{6}+\rho_{8}+\rho_{14}+\rho_{16}\\
\end{array}
\right)= \left(
\begin{array}{llll}
\gamma_1 & \gamma_1 & 0 & 0  \\

\gamma_2 & \gamma_2 & 0 & 0  \\

0 & 0 & \gamma_3 & \gamma_3  \\

0 & 0 & \gamma_4 & \gamma_4 \\

0 & 0 & \gamma_5 & \gamma_5  \\

0 & 0 & \gamma_6 & \gamma_6  \\

\gamma_7 & \gamma_7 & 0 & 0  \\

\gamma_8 & \gamma_8 & 0 & 0  \\

\gamma_9 & 0 & \gamma_9 & 0  \\

\gamma_{10} & 0 & \gamma_{10} & 0  \\

0 & \gamma_{11} & 0 & \gamma_{11}  \\

0 & \gamma_{12} & 0 & \gamma_{12}   \\

0 & \gamma_{13} & 0 & \gamma_{13}  \\

0  & \gamma_{14} & 0 & \gamma_{14}  \\

\gamma_{15} & 0 & \gamma_{15} & 0 \\

\gamma_{16} & 0 & \gamma_{16} & 0  \\
\end{array}
\right)
\end{equation}

Pick the first column on the right. Just start knocking off all of the $\rho_{k}$'s on the left in rows where 0's appear in the column vector on the right, and pretty soon there is nothing left to make \emph{any} non-zero $\gamma_{k}$'s in rows 1,2,7,8,9,10,15, or 16. There can be no such factorizable instances either, since every factorizable model instance is also NC (see Thm.~\ref{thm:fcImpliesPDIntro}). A similar argument works for the other three column vectors on the right.
\end{proof}

\subsection{Measurement predictability, CHSH and the Tsirelson bounds}
\label{sec:mpCHSH}

In spite of Thm.~\ref{thm:pdNotMP}, there are infinitely many MP model instances that satisfy CHSH. These instances show that even NC is not \emph{necessary} to derive the CHSH inequalities.

\begin{Theorem}
There are an infinite number of measurement predictable model instances that satisfy all CHSH inequalities and an infinite number that violate at least one CHSH inequality. In fact there are instances that exceed a Tsirelson bound.
\label{thm:qmTsirelson}
\end{Theorem}
\begin{proof}
Consider the first MP model pattern from Table~\ref{tab:predMeasTable}. An easy calculation shows the correlations to be 

\begin{align}
&(w,x,y,z)=(\gamma_1-\gamma_2,\gamma_8-\gamma_7,\gamma_9-\gamma_{10},\gamma_{16}-\gamma_{15})= \nonumber \\
&(2\gamma_1-1,2\gamma_8-1,2\gamma_9-1,2\gamma_{16}-1),
\end{align}

with $s$-functions given by

\begin{align}
&(s_1,s_2,s_3,s_4)=
(2(-1-\gamma_1+\gamma_8+\gamma_9+\gamma_{16}),
2(-1+\gamma_1-\gamma_8+\gamma_9+\gamma_{16}), \nonumber \\
&2(-1+\gamma_1+\gamma_8-\gamma_9+\gamma_{16}),
2(-1+\gamma_1+\gamma_8+\gamma_9-\gamma_{16})),
\end{align}
where $\gamma_1,\gamma_8,\gamma_9,\gamma_{16}$ are independently adjustable parameters between $0$ and $1$. Here are some examples for obtaining specific $(s_1,s_2 ,s_3,s_4)$ values:
\begin{itemize}
\item $(-4,0,0,0)$ by setting $\gamma_1=1,\gamma_8=\gamma_9=\gamma_{16}=0$.
\item $(-2,2,-2,2)$ by setting $\gamma_1=1,\gamma_8=0,\gamma_9=1,\gamma_{16}=0$.
\item $(-1,-3,1,1)$ by setting $\gamma_1=\frac{1}{2},\gamma_8=1,\gamma_9=0,\gamma_{16}=0$.
\item $(0,0,0,0)$ by setting $\gamma_1=\gamma_8=\gamma_9=\gamma_{16}=\frac{1}{2}$.
\end{itemize}
\end{proof}

\section{Summary and discussion}
\label{sec:summary} 

\begin{figure}[H]
\includegraphics[width=0.9\linewidth]{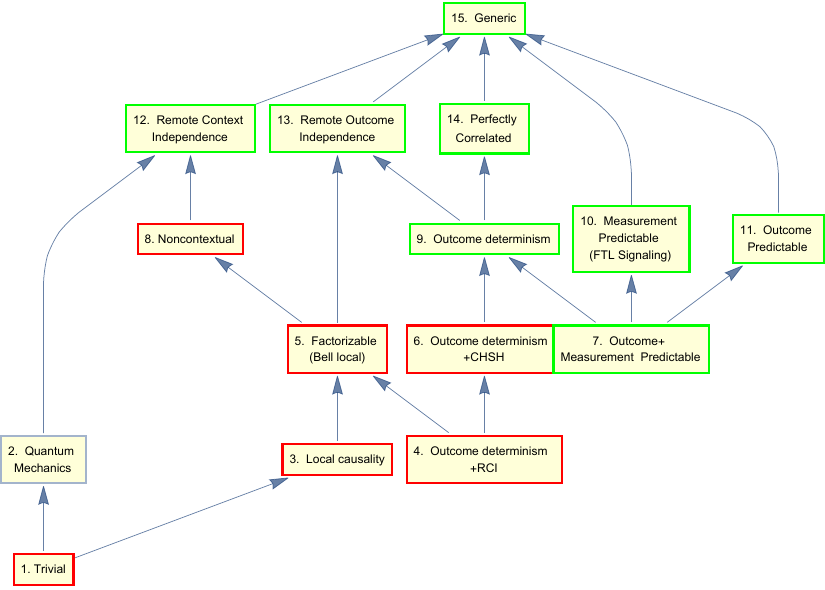}
\caption{Relationships among sets of model instances for selected types. Arrows indicate set inclusion. Any box outlined in red indicates that all such model instances satisfy all of the CHSH inequalities. A box outlined in green indicates that at least some model instances may not only violate a CHSH inequality, but even exceed a Tsirelson bound.}
\label{fig:modelsGraphPlotNew}
\end{figure}

The original motivation for this paper was to understand why there continue to be so many papers concerning the meaning and significance of Bell's theorems even after several decades. The thinking was that ambiguous everyday words, such as "locality", "determinism", and "realism" must be represented by assumptions inside of precise mathematical models, and the originally intended meanings may get lost in translation, either going into or coming out of the model. The operational modeling viewpoint, as represented by Tables~\ref{tab:genericTypeIntro},~\ref{tab:qmTypeIntro},~\ref{tab:predeterminedTypeIntro},~\ref{tab:bellLocalTypeIntro}, and~\ref{tab:trivialType}, represents properties such as "noncontextual", "factorizable", "QM", "outcome determinism", etc. in a uniform way that enables direct comparisons. This led ultimately to the taxonomy of notions and concepts (not necessarily complete) shown in Fig.~\ref{fig:modelsGraphPlotNew}.

Each node of the graph consists of a set of conditional probability vectors (aka model instances) which satisfy a precise mathematical definition that is intended to capture the notion embodied in the name of the node. For example, box 5, labeled "Factorizable", represents the set of all factorizable model instances of the form shown in Table~\ref{tab:bellLocalTypeIntro}. The arrows represent set inclusion. 

Every model instance in any box outlined in red satisfies all of the CHSH inequalities in Eq.~\ref{eq:chshIntro}. Some model instances in any box outlined in green not only violate at least one CHSH inequality, but also exceed a Tsirelson bound ($\pm 2\sqrt{2}$). In other words, none of the concepts represented by any of these red or green boxes are completely consistent with standard quantum mechanics. Put another way, hypothetical alternatives to quantum mechanics have been considered that not only "fall short" (i.e. satisfy CHSH) but are "excessive", i.e. have correlations that exceed even what quantum mechanics predicts.

\begin{figure}[H]
\centering
\includegraphics[width=0.5\linewidth]{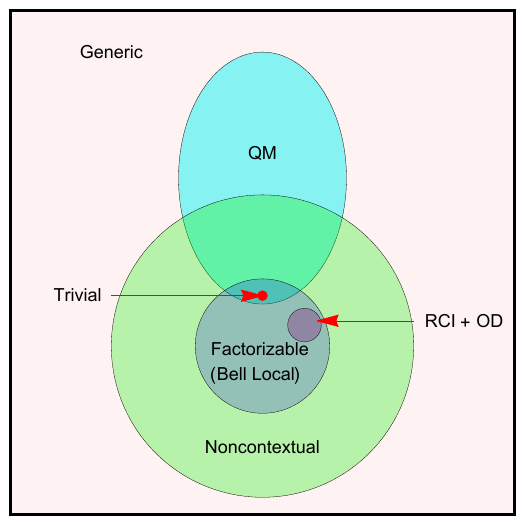}
\caption{Relationships among sets of instances of the main operational model types. The sets of QM and NC model instances have a nonempty intersection, but neither is a subset of the other. The only QM model instance that is also factorizable is the trivial one (Table~\ref{tab:trivialType}).}
\label{fig:vennGeneric}
\end{figure}

Fig.~\ref{fig:vennGeneric} illustrates some of the relationships among the major model types that may not be completely evident from Fig.~\ref{fig:modelsGraphPlotNew}, such as the fact that the QM and noncontextual model instances have a nonempty intersection. The "boundary" between the noncontextual and contextual QM model instances  is defined by those that satisfy all CHSH inequalities and those that do not.

The top-down, operational modeling perspective developed in this paper unifies many known facts in an intuitive way, suggests more compact or revealing proofs of some of these facts, highlights some facts which are perhaps underappreciated, and lends itself to the production of unique figures and plots that illustrate them. Some new insights resulting from this approach include

\begin{itemize}
\item The 256 outcome deterministic model instances can be classified according to whether or not they satisfy RCI, are factorizable, satisfy all CHSH inequalities, violate at least one CHSH inequality, or even exceed a Tsirelson bound.

\item Convex combinations of subsets of the 256 OD model instances can be translated into a traditional hidden variables model (not necessarily a hidden variables theory) for any generic model instance, including all QM ones.

\item The convex hull of the 16 RCI+OD model instances is equal to the set of noncontextual model instances, including the factorizable ones. This implies the existence of a local hidden variables model for any noncontextual model instance (including some QM ones).

\item The convex hull of all 256 OD model instances equals the set of all generic model instances, including all QM ones. From this construction, a hidden variables model can be derived for any QM model instance, whether contextual or not. Of course for a contextual QM model instance, at least some of the OD model instances in the convex combination cannot satisfy RCI or factorizability.

\item It seems that at least some of the plots have not appeared in the literature before. See Fig.'s~\ref{fig:models5SelectedIntro},~\ref{fig:models5}, and~\ref{fig:models5Combined}. One usually sees the polyhedral plots corresponding to the CHSH inequalities (the same as the noncontextual plots). But the QM or factorizable (Bell local) plots (and their various intersections) appear to be new. The QM plots are interesting in that their contours roughly "mimic" the shapes of the solid noncontextual plots, especially on the ends, where $w=\pm1$. The factorizable plots are also not solids, but contours such as hyperbolic paraboloids (on the ends where $w=\pm1$), which happen to fit neatly inside the noncontextual solids, visually affirming that factorizability is a special case of noncontextuality.
\end{itemize}

As for interesting proofs of known facts or particular emphasis on perhaps underappreciated facts, here are some examples.
\begin{itemize}
\item The fact that every factorizable model instance and every RCI+OD model instance is noncontextual is easily proved by "matching" the parameters in the symbolic forms of their cpvs, as in Thm.~\ref{thm:fcImpliesPDIntro} and Corollary~\ref{cor:rciODNC}.

\item A compact, geometrically-motivated proof of one of Fine's theorems was given that emphasizes that fact that if a model instance has correlations that satisfy all 8 CHSH inequalities, then \textit{there exists} a NC model instance with those same correlations. The original model instance may or may not be NC. In other words, just because a model instance satisfies all CHSH inequalities does not necessarily mean it is noncontextual (hence not necessarily factorizable or RCI+OD, either). Many examples of model instances that satisfy all CHSH inequalities but are not noncontextual have been given (for example, there are 112 such OD model instances).

\item Operational model types that allow Alice to deduce Bob's outcomes or that allow Alice to deduce Bob's measurements have been constructed. It then becomes trivial to show that some QM model instances can be outcome predictable, but none can be measurement predictable. The latter result means that the EPRB experimental setup cannot be used for faster-than-light communication (assuming the world is governed by QM rules, that is). These are well-known results, but not only does the operational modeling perspective make these proofs very intuitive, the method naturally extends to all model types. In fact it was shown that there are no noncontextual measurement predictable model instances either, but there is an infinite number of contextual (but not QM!) model instances that are not only measurement predictable, but also satisfy all of the CHSH inequalities.

\end{itemize}

\subsection{Final thoughts on (non)contextuality}
\label{sec:thoughts}

It is important to point out that the use of the name "noncontextual" for the operational model type in Table~\ref{tab:predeterminedTypeIntro} was intended to be suggestive but is in fact quite arbitrary. It could have been called "Model Type 3", and none of the subsequent proofs, examples, or correlation plots would change (except for the names). It would still be the case that all of the model instances belonging to the "Model Type 3" class would have correlations that satisfy all of the CHSH inequalities. So why give it the name "noncontextual" in the first place?

As discussed in Sec.~\ref{sec:soFable}, one way to arrive at the NC model type is to assume definite, preexisting values $A_1,A_2,B_1,B_2$, just waiting to be revealed by Alice and Bob's measurements, and equipped with a full joint pmf on all 4 r.v.'s. This leads directly to Table~\ref{tab:predeterminedTypeIntro} through marginalization. Then it is clear, for example, that if Alice chooses her first measurement $M_A=1$ she gets $A=A_1$ \emph{regardless} of Bob's measurement choice. Similarly for Alice's other measurement choice and both of Bob's. In other words, the measurement of each of their observables is independent of context, which is the standard definition of noncontextuality.

Bell himself argued in~\cite{Bell1966} against such an assumption. His line of reasoning is also described in~\cite{StanfordBell2021}, where it is noted that assumptions like this would later be dubbed "noncontextual". Of course, Bell did show that the assumption of "locality" as embedded in the mathematical form RCI+OD leads to a Bell inequality~\cite{Bell1964} and later showed that "locality" in the mathematical form of factorization also leads to a Bell inequality~\cite{Bell1976}. And as has been shown in this paper, these are both special cases of noncontextuality, in the \emph{mathematical} sense. So what gives? Did Bell \emph{implicitly} assume noncontextuality (which he himself rejected as a legitimate starting point), thus effectively assuming what he was trying to prove? 

He had legitimate physical reasons to make these assumptions. The \emph{mathematical} assumption of RCI, i.e. the probability of Alice's outcome should depend only on her own (local) measurement (and similarly for Bob), followed from the \emph{physical} assumption of separability. Likewise, the \emph{mathematical} assumption of factorizability was derived from the \emph{physical} assumption of local causality, namely that, given space-like separation,  the probability of Alice's outcome should be independent of \emph{both} Bob's measurement choice and outcome, and similarly for Bob with respect to Alice's wing of the experiment. In other words, Bell's physical assumptions were not explicitly "noncontextual". Nevertheless, once these are translated into a \emph{mathematical} model with conditional probabilities as in Tables~\ref{tab:bellLocalTypeIntro} and~\ref{tab:rciODTabIntro}, it is clear that these are special cases of the noncontextual model type of Table~\ref{tab:predeterminedTypeIntro} (see Thm.~\ref{thm:fcImpliesPDIntro} and Corollary~\ref{cor:rciODNC} of Thm.~\ref{thm:cvhPDIntro}). Whatever one's opinion about what Bell explicitly or implicitly assumed, the fact remains that noncontextuality is at the \emph{mathematical} core of these notions of locality.\\

\textbf{Funding} This research received no external funding.

\textbf{Acknowledgments} Thanks to the math/physics discussion group at Bellevue Community College for suffering through my first talks on this topic, and in particular to Victor Polinger, who believed that I had something to say. I would like to thank Doug Stoll for reading early drafts and providing helpful comments, Benjamin Schumacher for comments, suggestions, and encouragement, and Howard Wiseman for being a sponsor to get early versions on the arXiv. Long discussions with Frank Lad and Karl Hess were very valuable in refining my thoughts on noncontextuality vs. locality in Bell's theorems. It should be noted that neither of these two agrees entirely with the views expressed in this paper, but I am grateful for their critiques, because they forced careful rethinking at several junctures, that wouldn't have happened otherwise. 

\textbf{Conflicts of interest} The author declares no conflict of interest.

\appendix
\appendixpage
\addappheadtotoc

\section{Correlations, s-functions,  and plots}
\label{sec:correlations}

In this section correlations and $s$-functions tables are displayed for several different model types, including QM, together with plots of correlations (Appendices~\ref{sec:corrPlots} and~\ref{sec:corrPlotsCompared}). These plots provide visual affirmation (not necessarily proof in all cases) that

\begin{itemize}
\item The intersection of the sets of QM and NC model instances is non-empty, but neither is a subset of the other.
\item Every RCI+OD model instances and every Bell local (factorizable) model instance is also NC.
\item There are NC model instances that are not factorizable.
\end{itemize}

The advantage of showing sets of correlations (dimension 4) rather than sets of the original conditional probabilities (dimension 16) is that only one dimension needs to be fixed in order to produce human-understandable 3D plots. The disadvantage is that the "mapping" from generic parameters to correlations is "many-to-one" (see Eq.~\ref{eq:defineCorrIntro}), and one must be careful when interpreting the correlation visualizations. In Appendix~\ref{sec:corrPlots} individual examples are shown, and in the following Appendix~\ref{sec:corrPlotsCompared}, selected comparisons.

\subsection{Correlations and s-functions}
\label{sec:modelCorrandSFun}

\subsubsection{Generic}

See Table~\ref{tab:genericCandSTable}, which shows the correlations and $s$-functions of the generic model type of Table~\ref{tab:genericTypeIntro}. Since the $\gamma_k$'s are nonnegative and sum to 4, the $s$-functions are obviously between -4 and +4. There exist generic model instances that not only violate at least one of the CHSH inequalities, but exceed one of the Tsirelson bounds as well.
\begin{table}[H]
\caption{The correlations and $s$-functions for the generic type.}
\label{tab:genericCandSTable}
\[
\begin{array}{|c||l|}
\hline
w & \gamma_1-\gamma_2-\gamma_3+\gamma_4 \\
\hline
x & \gamma_5-\gamma_6-\gamma_7+\gamma_8 \\
\hline
y & \gamma_9-\gamma_{10}-\gamma_{11}+\gamma_{12} \\
\hline
z & \gamma_{13}-\gamma_{14}-\gamma_{15}+\gamma_{16} \\
\hline
\hline
s_1 & -\gamma_1+\gamma_2+\gamma_3-\gamma_4+\gamma_5-\gamma_6-\gamma_7+\gamma_8+ \\
& \gamma_9-\gamma_{10}-\gamma_{11}+\gamma_{12}+\gamma_{13}-\gamma_{14}-\gamma_{15}+\gamma_{16} \\
\hline
s_2 & \gamma_1-\gamma_2-\gamma_3+\gamma_4-\gamma_5+\gamma_6+\gamma_7-\gamma_8+ \\
& \gamma_9-\gamma_{10}-\gamma_{11}+\gamma_{12}+\gamma_{13}-\gamma_{14}-\gamma_{15}+\gamma_{16} \\
\hline
s_3 & \gamma_1-\gamma_2-\gamma_3+\gamma_4+\gamma_5-\gamma_6-\gamma_7+\gamma_8- \\
& \gamma_9+\gamma_{10}+\gamma_{11}-\gamma_{12}+\gamma_{13}-\gamma_{14}-\gamma_{15}+\gamma_{16} \\
\hline
s_4 & \gamma_1-\gamma_2-\gamma_3+\gamma_4+\gamma_5-\gamma_6-\gamma_7+\gamma_8+ \\
& \gamma_9-\gamma_{10}-\gamma_{11}+\gamma_{12}-\gamma_{13}+\gamma_{14}+\gamma_{15}-\gamma_{16} \\
\hline
\hline
\text{Min and Max} & \multicolumn{1}{c|}{-4 \text{ and } 4} \\
\hline
\end{array}
\]
\end{table}

\subsubsection{QM}
See Table~\ref{tab:qmCandSTable}, which shows the correlations and $s$-functions of the QM type of Table~\ref{tab:qmTypeIntro}. Not all QM model instances satisfy all the CHSH inequalities. The parameterization of a QM model that maximally violates one CHSH inequality is given by

\begin{equation}
\theta_{1}=67.5^{\circ} \text{ and } \theta_{k}=22.5^{\circ} \text{ for } k=2,3,4.
\end{equation}

This QM model instance has correlations

\begin{equation}
(w,x,y,z)=\frac{1}{2}\sqrt{2}(-1,1,1,1)
\end{equation}

and $s$-functions

\begin{equation}
(s_1,s_2,s_3,s_4)=(2\sqrt{2},0,0,0).
\end{equation}

Obviously the first one ($s_{1}$) does not satisfy one of the CHSH inequalities.

\begin{table}[H]
\caption{The correlations and $s$-functions for the QM type.}
\label{tab:qmCandSTable}
\[
\begin{array}{|c||l|}
\hline
w & \cos 2\theta_1 \\
\hline
x & \cos 2\theta_2 \\
\hline
y & \cos 2\theta_3 \\
\hline
z & \cos 2\theta_4 \\
\hline
\hline
s_1 & -\cos 2\theta_1+\cos 2\theta_2+\cos 2\theta_3+\cos 2\theta_4 \\
\hline
s_2 & \cos 2\theta_1-\cos 2\theta_2+\cos 2\theta_3+\cos 2\theta_4 \\
\hline
s_3 & \cos 2\theta_1+\cos 2\theta_2-\cos 2\theta_3+\cos 2\theta_4 \\
\hline
s_4 & \cos 2\theta_1+\cos 2\theta_2+\cos 2\theta_3-\cos 2\theta_4 \\
\hline
\hline
\text{Min and Max} & \multicolumn{1}{c|}{-2\sqrt{2} \text{ and } 2\sqrt{2}}\\
\hline
\end{array}
\]
\end{table}

\subsubsection{Noncontextual}
\label{sec:pdCorrelations}
See Table~\ref{tab:preDetCandSTable2}, which shows the correlations and $s$-functions of the NC model type of Table~\ref{tab:predeterminedTypeIntro}. Since the $\rho_k$'s form a pmf, it is obvious that all of the $s$-functions must be between -2 and 2. Note how the parameterized modeling framework makes derivation of the CHSH inequalities trivial.

\begin{table}[H]
\caption{The correlations and $s$-functions for the NC type.}
\label{tab:preDetCandSTable2}
\[
\begin{array}{|c||l|}
\hline
w & \rho_1+\rho_2-\rho_3-\rho_4+\rho_5+\rho_6-\rho_7-\rho_8- \\
& \rho_9-\rho_{10}+\rho_{11}+\rho_{12}-\rho_{13}-\rho_{14}+\rho_{15}+\rho_{16} \\
\hline
x & \rho_1+\rho_2-\rho_3-\rho_4-\rho_5-\rho_6+\rho_7+\rho_8+ \\
& \rho_9+\rho_{10}-\rho_{11}-\rho_{12}-\rho_{13}-\rho_{14}+\rho_{15}+\rho_{16} \\
\hline
y & \rho_1-\rho_2+\rho_3-\rho_4+\rho_5-\rho_6+\rho_7-\rho_8- \\
& \rho_9+\rho_{10}-\rho_{11}+\rho_{12}-\rho_{13}+\rho_{14}-\rho_{15}+\rho_{16} \\
\hline
z & \rho_1-\rho_2+\rho_3-\rho_4-\rho_5+\rho_6-\rho_7+\rho_8+ \\
& \rho_9-\rho_{10}+\rho_{11}-\rho_{12}-\rho_{13}+\rho_{14}-\rho_{15}+\rho_{16} \\
\hline
\hline
s_1 & 2(\rho_1-\rho_2+\rho_3-\rho_4-\rho_5-\rho_6+\rho_7+\rho_8+ \\ 
& \rho_9+\rho_{10}-\rho_{11}-\rho_{12}-\rho_{13}+\rho_{14}-\rho_{15}+\rho_{16}) \\
\hline
s_2 & 2(\rho_1-\rho_2+\rho_3-\rho_4+\rho_5+\rho_6-\rho_7-\rho_8- \\
& \rho_9-\rho_{10}+\rho_{11}+\rho_{12}-\rho_{13}+\rho_{14}-\rho_{15}+\rho_{16}) \\
\hline
s_3 & 2(\rho_1+\rho_2-\rho_3-\rho_4-\rho_5+\rho_6-\rho_7+\rho_8+ \\
& \rho_9-\rho_{10}+\rho_{11}-\rho_{12}-\rho_{13}-\rho_{14}+\rho_{15}+\rho_{16}) \\
\hline
s_4 & 2(\rho_1+\rho_2-\rho_3-\rho_4+\rho_5-\rho_6+\rho_7-\rho_8- \\
& \rho_9+\rho_{10}-\rho_{11}+\rho_{12}-\rho_{13}-\rho_{14}+\rho_{15}+\rho_{16}) \\
\hline
\hline
\text{Min and Max} & \multicolumn{1}{c|}{-2 \text{ and } 2}\\
\hline
\end{array}
\]
\end{table}

\subsubsection{Factorizable (Bell local)}
\label{sec:blCorrelations}
See Table~\ref{tab:blCandSTable}, which shows the correlations and $s$-functions of the factorizable model type of Table~\ref{tab:bellLocalTypeIntro}. As a special case of the NC model type (Thm.~\ref{thm:fcImpliesPDIntro} in Sec.~\ref{sec:definitionsAndOperModel}), all factorizable (Bell local) model instances also satisfy all CHSH inequalities.

\begin{table}[H]
\caption{The correlations and $s$-functions for the factorizable (Bell local) type.}
\label{tab:blCandSTable}
\[
\begin{array}{|c||l|}
\hline
w & (\alpha_3-\alpha_1)(\beta_3-\beta_1) \\
\hline
x & (\alpha_3-\alpha_1)(\beta_4-\beta_2) \\
\hline
y & (\alpha_4-\alpha_2)(\beta_3-\beta_1) \\
\hline
z & (\alpha_4-\alpha_2)(\beta_4-\beta_2) \\
\hline
\hline
s_1 & -(\alpha_3-\alpha_1)(\beta_3-\beta_1)+(\alpha_3-\alpha_1)(\beta_4-\beta_2)+ \\& (\alpha_4-\alpha_2)(\beta_3-\beta_1)+(\alpha_4-\alpha_2)(\beta_4-\beta_2) \\
\hline
s_2 & (\alpha_3-\alpha_1)(\beta_3-\beta_1)-(\alpha_3-\alpha_1)(\beta_4-\beta_2)+ \\
& \alpha_4-\alpha_2)(\beta_3-\beta_1)+(\alpha_4-\alpha_2)(\beta_4-\beta_2) \\
\hline
s_3 & (\alpha_3-\alpha_1)(\beta_3-\beta_1)+(\alpha_3-\alpha_1)(\beta_4-\beta_2)- \\
& (\alpha_4-\alpha_2)(\beta_3-\beta_1)+(\alpha_4-\alpha_2)(\beta_4-\beta_2) \\
\hline
s_4 & (\alpha_3-\alpha_1)(\beta_3-\beta_1)+(\alpha_3-\alpha_1)(\beta_4-\beta_2)+ \\
& (\alpha_4-\alpha_2)(\beta_3-\beta_1)-(\alpha_4-\alpha_2)(\beta_4-\beta_2) \\
\hline
\hline
\text{Min and Max} & \multicolumn{1}{c|}{-2 \text{ and } 2}\\
\hline
\end{array}
\]
\end{table}

Note that since the single conditional expectations are given by
\begin{align}
&E[A|M_{A}=1]=\alpha_3-\alpha_1 \; & E[B|M_{B}=1]=\beta_3-\beta_1,  \nonumber \\
&E[A|M_{A}=2]=\alpha_4-\alpha_2 \; & E[B|M_{B}=2]=\beta_4-\beta_2,
\end{align}
each correlation (conditional expectation of the product) is the product of the corresponding single conditional expectations.
\begin{align}
\label{eq:fcEquation}
&w=E[AB|M_A=1,M_B=1]=E[A|M_A=1] \times E[B|M_B=1], \nonumber \\
&x=E[AB|M_A=1,M_B=2]=E[A|M_A=1] \times E[B|M_B=2],  \\ 
&y=E[AB|M_A=2,M_B=1]=E[A|M_A=2] \times E[B|M_B=1], \nonumber \\
&z=E[AB|M_A=2,M_B=2]=E[A|M_A=2] \times E[B|M_B=2]. \nonumber
\end{align}

From either Eq.~\ref{eq:fcEquation} or Table~\ref{tab:blCandSTable} it is easy to see that $wz=xy$ and hence, for any fixed $w$, the following equation determines the shape of the sets of factorizable correlations in 3-space. See Fig.~\ref{fig:models5SelectedIntro} or~\ref{fig:models5}.
\begin{equation}
\left\{ \begin{array}{ll}
               z=w^{-1}{xy}  &  \mbox{if $w \neq 0$} \\
               xy=0            & \mbox{if $w=0$.}
        \end{array}
\right.
\label{eq:wz=xy}
\end{equation} 

However, the \emph{unconditional} correlation is equal to the product of the individual unconditional expectations, that is,
\begin{equation}
E[AB] =E[A] \times E[B]
\end{equation}
\emph{only if} all joint parameter setting probabilities are equal, that is,
\begin{align}
&P(M_A=1,M_B=1)=P(M_A=1,M_B=2)= \nonumber \\
&P(M_A=2,M_B=1)=P(M_A=2,M_B=2)=\frac{1}{4}.
\end{align}

\subsubsection{Remote context independence plus outcome determinism}
\label{sec:odCorrelations}
See Table~\ref{tab:odCandSTable}, which shows the correlations and $s$-functions of the 16 OD model instances from Table~\ref{tab:rciODTabIntro}. It is clear that they satisfy all 8 CHSH inequalities. It was shown in Sec.~\ref{sec:introduction} that these model instances also satisfy RCI. Note the CHSH inequality bounds are tight in this case.

\begin{table}[H]
\setlength\arraycolsep{2pt}
\caption{The correlations and $s$-functions for the 16 RCI+OD model instances. See Table~\ref{tab:rciODTabIntro} for the underlying cpvs.}
\label{tab:odCandSTable}
\[
\begin{array}{|c||r|r|r|r|r|r|r|r|r|r|r|r|r|r|r|r|}
\hline
\text{No.} &1 & 2 & 3 & 4 & 5 & 6 & 7 & 8 & 9 & 10 & 11 & 12 & 13 & 14 & 15 & 16 \\
\hline
\hline
w & 1 & 1 & -1 & -1 & 1 & 1 & -1 & -1 & -1 & -1 & 1 & 1 & -1 & -1 & 1 & 1 \\
\hline
x & 1 & 1 & -1 & -1 & -1 & -1 & 1 & 1 & 1 & 1 & -1 & -1 & -1 & -1 & 1 & 1 \\
\hline
y & 1 & -1 & 1 & -1 & 1 & -1 & 1 & -1 & -1 & 1 & -1 & 1 & -1 & 1 & -1 & 1 \\
\hline
z & 1 & -1 & 1 & -1 & -1 & 1 & -1 & 1 & 1 & -1 & 1 & -1 & -1 & 1 & -1 & 1 \\
\hline
\hline
s_1 & 2 & -2 &  2 & -2 &  -2 & -2 &  2 &  2 &    2 & 2 & -2 & -2 &   -2 &  2 &  -2 & 2 \\
\hline
s_2 & 2 & -2 &  2 & -2 &   2 &   2 & -2 & -2 &  -2 & -2 & 2 & 2 &    -2 &  2 &  -2 & 2 \\
\hline
s_3 & 2 &  2 & -2 & -2 &  -2 &  2 & -2 &  2 &    2 & -2 & 2 & -2 &   -2 & -2 &   2 & 2 \\
\hline
s_4 & 2 &  2 & -2 & -2 &   2 &  -2 &  2 & -2 &  -2 & 2 & -2 & 2 &    -2 & -2 &   2 & 2 \\
\hline
\hline
\text{Min and Max} & \multicolumn{16}{c|}{-2 \text{ and } 2} \\
\hline
\end{array}
\]
\end{table}

\subsubsection{Trivial}
\label{sec:trivialCorrelations}
See Table~\ref{tab:trivialCandSTable}. It is obvious that all CHSH inequalities are satisfied. This is the completely uniform, no-information case.

\begin{table}[H]
\caption{The correlations and $s$-functions for the trivial type.}
\label{tab:trivialCandSTable}
\[
\begin{array}{|c||c|}
\hline
w & 0 \\
\hline
x & 0 \\
\hline
y & 0 \\
\hline
z & 0 \\
\hline
\hline
s_1 & 0 \\
\hline
s_2 & 0 \\
\hline
s_3 & 0 \\
\hline
s_4 & 0 \\
\hline
\text{Min and Max} & 0 \text{ and } 0\\
\hline
\end{array}
\]
\end{table}

\subsection{Correlation plots}
\label{sec:corrPlots}

\begin{figure}[H]
\centering
\includegraphics[width=0.75\linewidth]{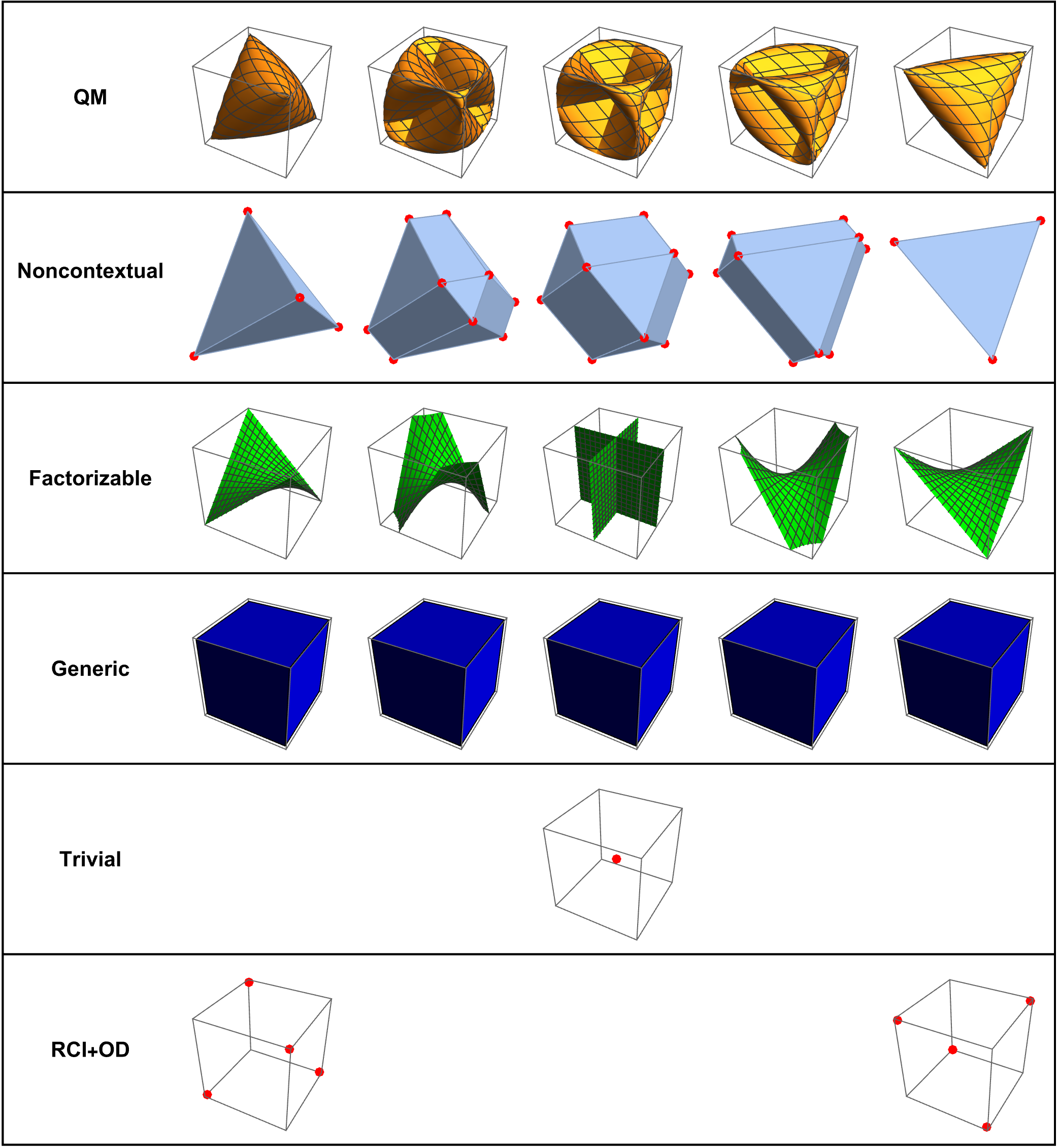}
\caption{Sets of correlations $(x,y,z)$ of some model types. $w$ is fixed at either $-1,-\frac{1}{2},0,\frac{1}{2}, \text{ or } 1$ from left to right in each row.}
\label{fig:models5}
\end{figure}

Fig.~\ref{fig:models5} shows different sets of correlations corresponding to six different model types. Each of the five (or fewer) pictures in any row is a 3D-slice $(x,y,z)$ through the set of possible correlations $(w,x,y,z)$ for the given model type, where $w$ is fixed in each case. In these figures, $w=-1,-\frac{1}{2},0,\frac{1}{2},1$ from left to right across the 5 columns.

\underline{QM (row 1)}: Each contour represents the set of points 

\begin{align}
&(x,y,z)=(\cos2\theta_2,\cos2\theta_3,\cos2\theta_4) \text{ where } \theta_1=\theta_2+\theta_3+\theta_4, \nonumber \\
&\theta_1=\frac{1}{2}\arccos{w}, \text{ and } -90^{\circ} \leq \theta_2, \theta_3, \theta_4 \leq 90^{\circ}.
\end{align}

\underline{Noncontextuality (row 2)}: Each of the five figures is a representation of all possible correlations $(x,y,z)$ for a NC model type with $w$ fixed. The figures at the far left ($w=-1$) and the far right ($w=1$) are tetrahedrons with 4 vertices and 4 equilateral triangular faces. The polyhedra in between have 12 vertices and 6 rectangular faces and 8 triangular faces. The one in the middle ($w=0$) is called a "cuboctahedron", a semi-regular polyhedron where the rectangles are congruent squares and the triangles are congruent equilateral triangles. The points on the boundary and in the interior of each of these polyhedra represent those correlations $(w,x,y,z)$ that satisfy the CHSH inequalities for a given fixed value of $w$. These pictures can be viewed as more detailed representations of the "correlation polytope" in Fig. 3 of Pitowski~\cite{Pitowski2005}. For more detailed connections between the Bell inequalities and the Platonic and Archimedean solids, see~\cite{LasonKosinski2021,GisinTavakoli2020}.

\underline{Factorizable (row 3)}: At the far left ($w=-1$) and far right ($w=1$), the surfaces are hyperbolic paraboloids. All 5 surfaces can be represented by the equation

\begin{equation}
\left\{ \begin{array}{ll}
               z=w^{-1}{xy}  &  \mbox{if $w \neq 0$} \\
               xy=0            & \mbox{if $w=0$.}
        \end{array}
\right.
\end{equation} 

\underline{Generic (row 4)}: These fill the entire "correlation cube" ($[-1,1] \times [-1,1] \times [-1,1]$). This is because each of the four correlations is determined by a set of four parameters completely independent of the others, hence they all can independently take any value between -1 and 1. 

\underline{Trivial (row 5)}: This can only appear in the middle ($w=0$) because all of its correlations are 0. 

\underline{RCI+OD (row 6)}: These can only appear on the far left ($w=-1$) or far right ($w=1$) because any OD model instance has perfect correlations ($\pm1$).

\subsection{Correlation plots compared}
\label{sec:corrPlotsCompared}

\begin{figure}[H]
\centering
\includegraphics[width=0.75\linewidth]{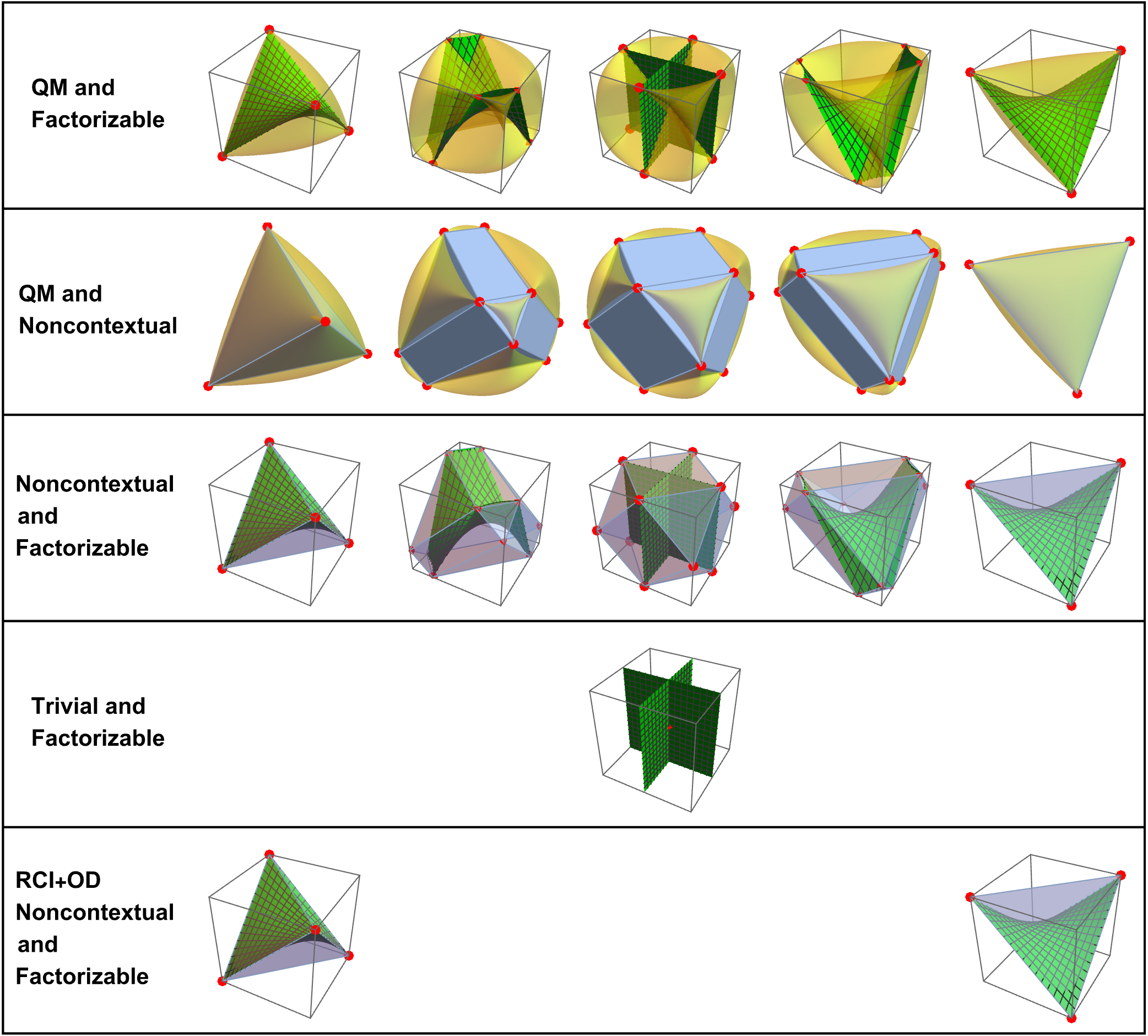}
\caption{Sets of correlations $(x,y,z)$ of some pairs of model instances. Fixed $w=-1,-\frac{1}{2},0,\frac{1}{2}, \text{ or } 1$ from left to right in each row.}
\label{fig:models5Combined}
\end{figure}

Now let's do some comparisons. See Fig.~\ref{fig:models5Combined}. In the first and second rows, QM correlations are compared to factorizable and NC ones. In both cases, there is a non-empty intersection, but neither is a subset of the other. In the end cases ($w=\pm1$), the surfaces (or region in the NC case) stretch out to the RCI+OD vertices shown as red dots. 

In the third row (noncontextuality and factorization) it is clear that in every case, the Bell local (factorizable) correlations are completely contained in the corresponding NC correlations. This is a direct consequence of the fact that every factorizable model instance is a special case of the NC model type (Thm.~\ref{thm:fcImpliesPDIntro} in Sec.~\ref{sec:introduction}). 

The first and last plots ($w=\pm1$) in rows 3 and 5 show that the perfect correlations (red dots) of the OD model types can be achieved by some NC and Bell local (factorizable) model instances. This reflects the fact that every RCI+OD model instance is also factorizable and NC. 

\begin{figure}
\centering
\includegraphics[width=0.3\linewidth]{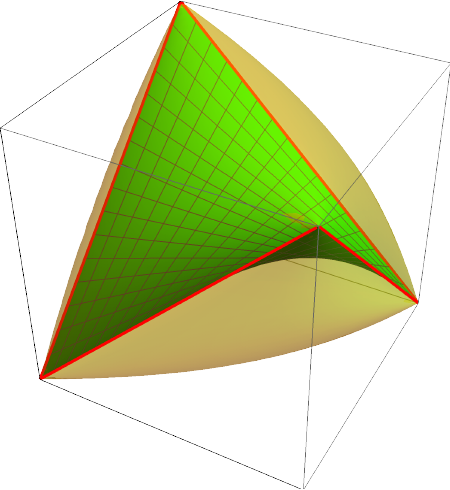}
\includegraphics[width=0.3\linewidth]{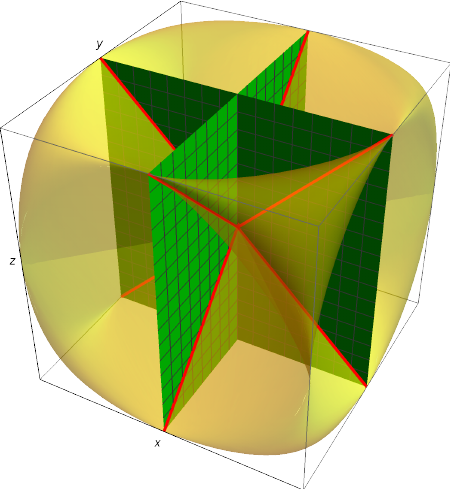}
\includegraphics[width=0.3\linewidth]{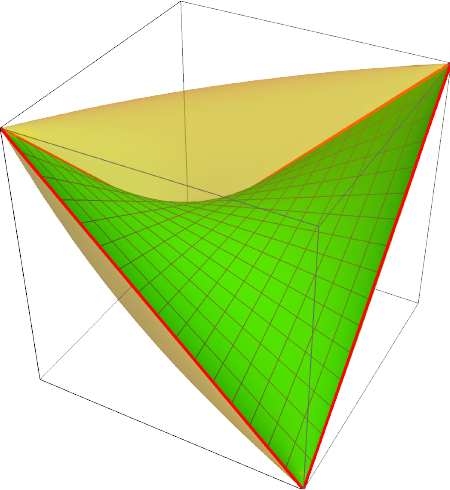}
\caption{Red lines indicate correlations $(x,y,z)$ mutually achievable by \emph{different} QM and factorizable model instances. Fixed $w=-1,0,1$ from left to right.}
\label{fig:bellLocalQMPlots}
\end{figure}

Fig.~\ref{fig:bellLocalQMPlots} shows the first, middle and last plot of row 1 in Fig.~\ref{fig:models5Combined} "blown up". The red lines have been added to highlight the intersections of the QM and factorizable correlation plots. How does this square with the fact that the \emph{only} model instance that is simultaneously QM and factorizable is the trivial one (corresponding to the point at the very center of the middle plot, at the intersection of the diagonals)? This is a dramatic demonstration of the fact that the mapping from model instance parameters to correlations is many-to-one. That is, every point on the each of the red lines (i.e., a set of correlations) is achievable by \emph{some} QM model instance, and it is achievable by \emph{some} factorizable model instance, but in all but one case, the conditional probabilities (generic parameters) of these two model instances are \emph{different}.

\section{QM model s-function max and min}
\label{sec:qmMaxMin}

Define a function 
$
\bar{s}_{1}\colon\left[\frac{-\pi}{2}, \frac{\pi}{2}\right]\times\left[\frac{-\pi}{2}, \frac{\pi}{2}\right] \times \left[\frac{-\pi}{2}, \frac{\pi}{2}\right]\rightarrow R
$ where 
\[
\bar{s}_{1}(\theta_{2},\theta_{3},\theta_{4})=-\cos2(\theta_{2}+\theta_{3}+\theta_{4})+ \cos2 \theta_{2}+ \cos2 \theta_{3}+ \cos2 \theta_{4}.
\] 
$\bar{s}_{1}$ is related to $s_{1}$ by $\bar{s}_{1}(\theta_{2},\theta_{3},\theta_{4})=s_{1}(\cos2\theta_{1},\cos2\theta_{2},\cos2\theta_{3},\cos2\theta_{4})$ where 
$\theta_{1}=\theta_{2}+\theta_{3}+\theta_{4}$.

Tables~\ref{tab:critAngles},~\ref{tab:critCorr}, and~\ref{tab:critValGradHess} show the results of using Mathematica to search for stationary points. Of the 20 stationary points, two are local maxima, two are local minima, and the other sixteen are neither.

\begin{table}[H]
\caption{Critical point angles.}
\label{tab:critAngles}
\[
\begin{array}{|c||c|c|c|c|}
\hline
\text{Pt} & \theta_1 & \theta_2 & \theta_3 & \theta_4 \\
\hline
\hline
1 & \pm 67.5^{\circ} & \pm 22.5^{\circ} & \pm 22.5^{\circ} & \pm 22.5^{\circ}  \\
\hline
2 & \pm 202.5^{\circ} & \pm 67.5^{\circ} & \pm 67.5^{\circ} & \pm 67.5^{\circ}  \\
\hline
3 &  0^{\circ} &  0^{\circ} &  0^{\circ} &  0^{\circ}  \\
\hline
4 & \pm 90^{\circ} & 0^{\circ} & 0^{\circ} & \pm 90^{\circ}  \\
\hline

5 & \pm 90^{\circ} & 0^{\circ} & \pm 90^{\circ} & 0^{\circ}   \\
\hline
6 & \pm 90^{\circ} & \pm 90^{\circ} & 0^{\circ} & 0^{\circ}   \\
\hline
7 & \pm 180^{\circ} & 0^{\circ} & \pm 90^{\circ} & \pm 90^{\circ}   \\
\hline
8 & \pm 180^{\circ} & \pm 90^{\circ} & 0^{\circ} & \pm 90^{\circ}   \\
\hline
9 & \pm 180^{\circ} & \pm 90^{\circ} & \pm 90^{\circ} & \pm 0^{\circ}   \\
\hline
10 & \pm 270^{\circ} & \pm 90^{\circ} & \pm 90^{\circ} & \pm 90^{\circ}   \\
\hline
\end{array}
\]
\end{table}

\begin{table}[H]
\caption{Critical point correlations.}
\label{tab:critCorr}
\[
\begin{array}{|c||r|r|r|r|}
\hline
\text{Pt} & w & x & y & z  \\
\hline
\hline
1  & -\frac{1}{2} \sqrt{2} & \frac{1}{2} \sqrt{2} &\frac{1}{2} \sqrt{2} &\frac{1}{2} \sqrt{2} \\
\hline
2 & \frac{1}{2} \sqrt{2} & -\frac{1}{2} \sqrt{2} & -\frac{1}{2} \sqrt{2} & -\frac{1}{2} \sqrt{2} \\
\hline
3  & 1 & 1 & 1 & 1 \\
\hline
4 & -1 & 1 & 1 & -1 \\
\hline

5 & -1 & 1 & -1 & 1 \\
\hline
6 & -1 & -1 & 1 & 1 \\
\hline
7 & 1 & 1 & -1 & -1 \\
\hline
8 & 1 & -1 & 1 & -1 \\
\hline
9 & 1 & -1 & -1 & 1 \\
\hline
10 & -1 & -1 & -1 & -1 \\
\hline
\end{array}
\]
\end{table}

\begin{table}[H]
\caption{Critical point value, gradient, and Hessian.}
\label{tab:critValGradHess}
\[
\begin{array}{|c||r|c|c|}
\hline
\text{Pt} & \text{Value} & \text{Gradient} & \text{Hessian} \\
\hline
\hline
1 & 2 \sqrt{2} & (0,0,0) & \text{Negative Definite} \\
\hline
2 & -2 \sqrt{2} & (0,0,0) & \text{Positive Definite} \\
\hline
3 & 2  & (0,0,0) & \text{Indefinite} \\
\hline
4 & 2  & (0,0,0) & \text{Indefinite} \\
\hline

5 & 2  & (0,0,0) & \text{Indefinite} \\
\hline
6 & 2  & (0,0,0) & \text{Indefinite} \\
\hline
7 & -2  & (0,0,0) & \text{Indefinite} \\
\hline
8 & -2 & (0,0,0) & \text{Indefinite} \\
\hline
9 & -2  & (0,0,0) & \text{Indefinite} \\
\hline
10 & -2  & (0,0,0) & \text{Indefinite} \\
\hline
\end{array}
\]
\end{table}

\begin{figure}[H]
\centering
\includegraphics[height=10cm]{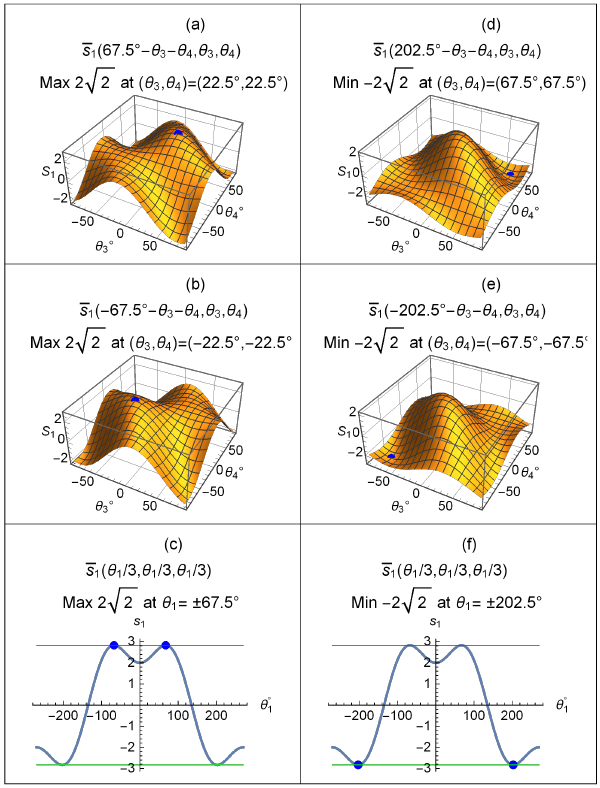}
\caption{Maxima and minima for the QM $\bar{s}_{1}$ function.}
\label{tab:qmMaxMin}
\end{figure}

Fig.~\ref{tab:qmMaxMin} illustrates the local maxima and minima indicated by rows 1 and 2 of Tables~\ref{tab:critAngles},~\ref{tab:critCorr}, and~\ref{tab:critValGradHess}. The first column (panels (a), (b), and (c)) indicate the maxima and the second column (panels (d), (e), and (f)) indicate the minima (both shown as blue dots). 
In panels (a) and (b), \(\theta _1\) is held fixed at $\pm $67.5${}^{\circ}$, respectively, and \(\theta _3\) and \(\theta _4\) are free
parameters. In (c), only \(\theta _1\) is free, and the maxima are seen to coincide at \(\left(\theta _1,\theta _2,\theta _3,\theta _4\right)=(\pm
67.5^{\circ},\pm 22.5^{\circ},\pm 22.5^{\circ},\pm 22.5^{\circ})\).
In panels (d) and (e), \(\theta _1\) is held fixed at $\pm $202.5$^{\circ}$, respectively, and \(\theta _3\) and \(\theta _4\) are free
parameters. In (f), only \(\theta _1\) is free, and the minima are seen to coincide at \(\left(\theta _1,\theta _2,\theta _3,\theta _4\right)=(\pm
202.5^{\circ},\pm 67.5^{\circ},\pm 67.5^{\circ},\pm 67.5^{\circ})\).

\subsection{Bell's theorem in a nutshell}
\label{sec:nutShell}

\begin{figure}[hbtp]
\centering
\includegraphics[width=0.70\linewidth]{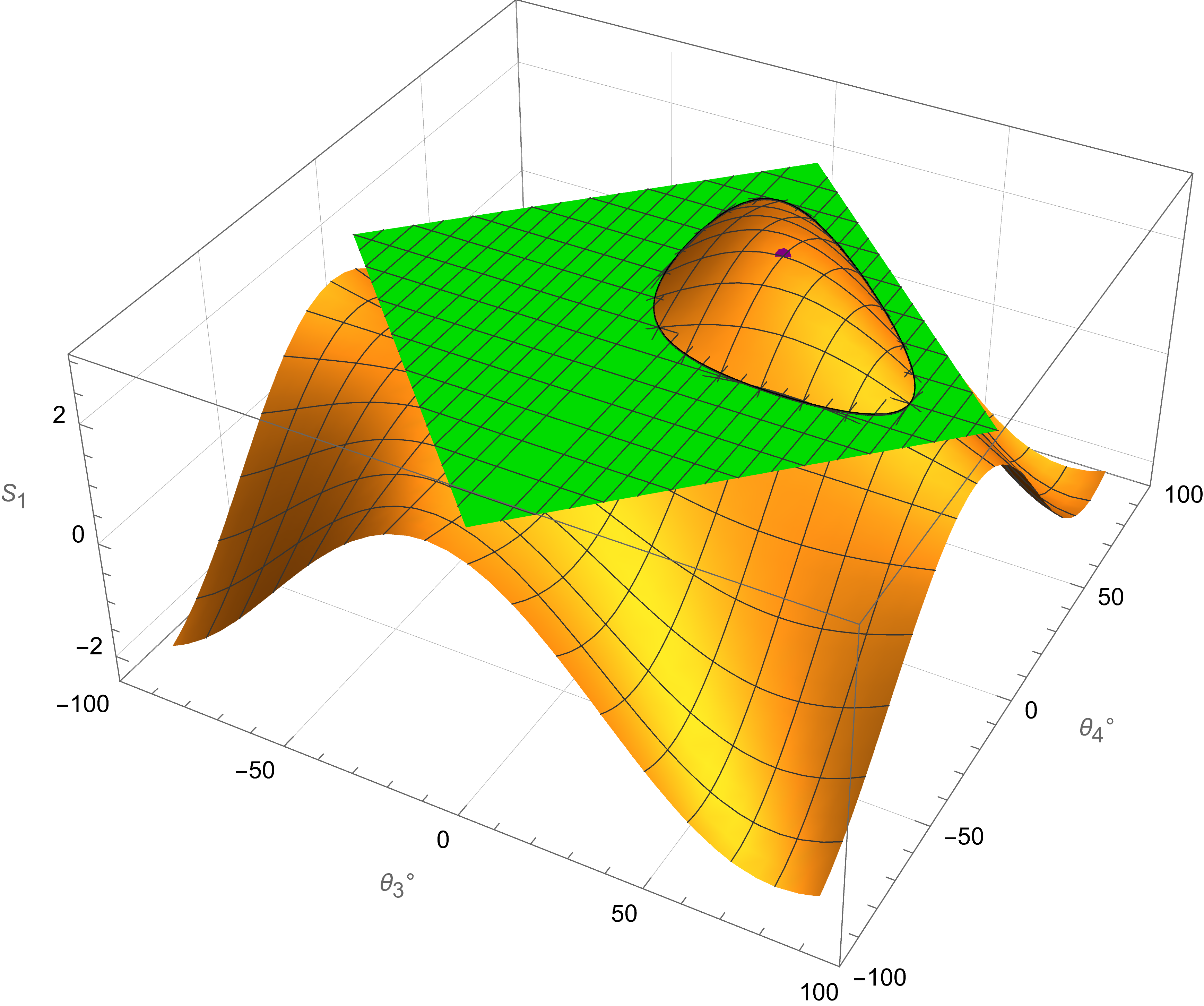}
\caption{Illustration of the violation of one of the CHSH inequalities by the $s_{1}$-function of QM correlations ($\theta_{2}=22.5^{\circ}$ is held fixed).}
\label{fig:qmBellNutshell}
\end{figure}

Fig.~\ref{fig:qmBellNutshell} is an illustration of the relationship between the first QM $s$-function, as a function of measurement difference angles $(\theta_3,\theta_4)$ and one of the CHSH bounds. The green square represents the CHSH inequality bound of 2, and the yellow surface represents the QM $s_{1}$-function, as a function of the two free parameters $\theta_{3}$ and $\theta_{4}$, based on the two constraints $\theta_{1}=\theta_{2}+\theta_{3}+\theta_{4}$ and $\theta_{2}=22.5^{\circ}$. The small hill above the green square indicates where the CHSH inequality is violated. The purple dot at the top of the hill shows the maximum of $2\sqrt{2}$ when $\theta _2=\theta _3=\theta _4=22.5^{\circ}$ and $\theta _1=67.5^{\circ}$.

\subsection{The Tsirelson bounds}
\label{sec:tsirelson}

The QM $s$-function max and min of $\pm{2}\sqrt{2}$ are known as the \textit{Tsirelson bound(s)}. If all four of the parameters $\theta_{1},\theta_{2},\theta_{3},\text{ and }\theta_{4}$ were free, then the max and min of the s-functions would be +4 and -4, just like the generic model -- see Table~\ref{tab:genericTypeIntro}. It is the simple geometric relationship among the four measurement difference angles that imposes a linear constraint among these angles (see Fig.~\ref{fig:thetaConstraintsIntro}). Note that these would not be \emph{completely} generic models, however, since the form of the QM conditional probabilities imposes 16 extra constraints on the generic parameters $\gamma_{1},\gamma_{2},...,\gamma_{16}$, such as $\gamma_{1}+\gamma_{2}=\frac{1}{2}\sin^{2}(\theta_{1})+\frac{1}{2}\cos^{2}(\theta_{1})=\frac{1}{2}$.

\section{Convex hull of OD model instances equals the set of all generic model instances}
\label{sec:cvhGeneric}

The proof of Thm.~\ref{thm:cvhPDIntro} shows that the convex hull of the 16 RCI+OD model instances equals the class of noncontextual model instances. This motivates the question: What is the convex hull of all 256 OD conditional probability vectors? The next theorem answers this question.
\begin{Theorem}
\label{thm:cvhGeneric}
Let $G$ be the set of generic conditional probability vectors and $K$ the set of 256 OD conditional probability vectors. Then any $\pmb{\gamma} \in G$ can be written as a convex combination of the members of $K$. See Fig.~\ref{fig:odGenericCVH}.
\end{Theorem}
\begin{proof}
A theorem of Minkowski states that any convex, closed, and bounded set in $\mathbb{R}^n$ is equal to the convex hull of its extreme points. Therefore the plan is to show that $G$ is convex, closed, and bounded, and that the members of $K$ comprise all of the extreme points of $G$. 
\begin{itemize}
\item $G$ is convex because if $\pmb{\gamma}=\pi \pmb{\gamma}_1 +(1-\pi) \pmb{\gamma}_2$ for any $\pmb{\gamma}_1,\pmb{\gamma}_2 \in G$ and $0 \leq \pi \leq 1$, it is clear that all of the entries of $\pmb{\gamma}$ are nonnegative and satisfy the measurement context constraints, hence $\pmb{\gamma} \in G$.

\item $G$ is closed since for any sequence $\pmb{\gamma}_n=(\gamma_1^{(n)},\gamma_2^{(n)},...,\gamma_{16}^{(n)}) \in G$ converging to $\pmb{\gamma}=(\gamma_1,\gamma_2,...,\gamma_{16})$, 
\[
\gamma_1+\gamma_2+\gamma_3+\gamma_4=\lim_{n \longrightarrow \infty} (\gamma_1^{(n)}+\gamma_2^{(n)}+\gamma_3^{(n)}+\gamma_4^{(n)})=\lim_{n \longrightarrow \infty} 1=1,
\] 
since each $\pmb{\gamma}_n \in G$. Similarly for the other successive blocks of four entries in $\pmb{\gamma}$. Hence $\pmb{\gamma} \in G$ and thus $G$ is closed.
\item $G$ is bounded because if $\pmb{\gamma}=(\gamma_1,\gamma_2,...,\gamma_{16}) \in G$, then $\parallel \pmb{\gamma} \parallel = \sqrt{\sum \gamma_k^2} \leq \sqrt{\sum \gamma_k} = \sqrt{4}=2$, so $G$ is contained in a sphere of radius 2 centered at the origin in $\mathbb{R}^{16}$.
\item Now it is shown that $K$ is the set of all extreme points of $G$. It is obvious that $K \subseteq G$ since every OD model instance has nonnegative entries that satisfy the measurement context constraints. Next it is demonstrated that every member of $K$ is an extreme point of $G$. To simplify notation, the proof is restricted to showing a particular member of $K$ is an extreme point of $G$. It should be clear that the argument is not affected by the exact positions of the four 1's in any of the other 255 OD conditional probability vectors. Thus, consider this representative member of $K$: 
\[
\pmb{o}=(1,0,0,0,1,0,0,0,1,0,0,0,1,0,0,0).
\] 
Suppose $\pmb{o}=\frac{1}{2}(\pmb{\gamma}+\pmb{\gamma}')$ where $\pmb{\gamma},\pmb{\gamma}' \in G$. Since the entries of $\pmb{\gamma},\pmb{\gamma}'$ are nonnegative, it is pretty clear that all entries in each, except for positions $1,5,9,13$, must be 0, and that leaves only the choice of 1 in those positions for both $\pmb{\gamma},\pmb{\gamma}'$. In other words, $\pmb{o}$ cannot be the midpoint of a non-trivial line in $G$, hence it must be an extreme point. 

Now it is proved that any $\pmb{\gamma}=(\gamma_1,\gamma_2,...,\gamma_{16}) \in G \setminus K$ cannot be an extreme point of $G$. In one of the successive blocks of four entries, there must be one with at least two entries not equal to either 0 or 1. Wlog, assume that $0 < \gamma_1 \leq \gamma_2 <1$. Choose a real $x$ such that $0<x< \gamma_1$, and define
\begin{align*}
&\pmb{\gamma}_1=(2 \gamma_1-x,\gamma_2-\gamma_1+x,\gamma_3,...,\gamma_{16}) \text{ and } \\
&\pmb{\gamma}_2=(x,\gamma_1+\gamma_2-x,\gamma_3,...,\gamma_{16}).
\end{align*}
Then it is not too hard to show that $\pmb{\gamma}_1,\pmb{\gamma}_2 \in G \setminus K$
are distinct and therefore, since $G$ is convex, 
\[\pi \pmb{\gamma}_1+(1-\pi)\pmb{\gamma}_2\]
for $0 \leq \pi \leq 1$ is a non-trivial line segment completely contained in $G$. In other words, 
\[\pmb{\gamma}=\frac{1}{2}(\pmb{\gamma}_1+\pmb{\gamma}_2)\]
is the midpoint of a non-trivial line segment entirely inside $G$, so it cannot be an extreme point.
\end{itemize}
Finally the since set of all extreme points of $G$ is exactly $K$, the set $G$ of generic model instances is equal to the convex hull of $K$, the 256 OD model instances. 
\end{proof}

\section{Filling in the gaps}
\label{sec:looseEnds}

This appendix has been reserved for proofs and counterexamples related to some of the details of Fig.~\ref{fig:modelsGraphPlotNew} (which shows the relationships among the different model types and conditions) that haven't been proved or illustrated elsewhere. This was done in order not to interrupt the flow in the main body of the paper.

The first order of business is to address the issue of possibly undefined conditional probabilities, which affects both the underlying definitions of model types in general and the LC model type in particular.

\subsection{Undefined conditional probabilities}
\label{sec:undefinedConditionals}

In the definition of conditional probabilities, for example, 
\[P(E|F)=\frac{P(E	\cap F)}{P(F)}\]
for events $E \text{ and } F$, it is assumed that $P(F)>0$, otherwise the conditional probability is undefined. In this paper, this tradition is respected, which requires mention of two situations where the issue of zero denominators might arise.

\subsubsection{Measurement choice probabilities}
\label{sec:measChoice}
Consider Table~\ref{tab:operationalModelTable}, which shows the \textit{unconditional} probabilities 
\[P(A=s,B=t,M_A=u,M_B=v)=P(A=s,B=t|M_A=u,M_B=v) \delta_{uv}\] 
for the generic model type, where the joint measurement choice probabilities are given by

\begin{equation}
\delta_{uv}=P(M_A=u,M_B=v) \text{ for } u,v=1,2.
\end{equation}

\begin{table}[H]
\caption{PMF for the generic model type. These are the unconditional probabilities based on measurement choice probabilities (assumed frequencies).}
\label{tab:operationalModelTable}
\[
\begin{array}{|c|c|c|c|c|}
\hline
\text{Context} & \multicolumn{4}{c|}{\text{Outcome }(s,t)} \\
\hline
(u,v) & (-1,-1)  & (-1,1)   & (1,-1)   & (1,1) \\
\hline
(1,1) & \gamma_1 \delta_{11} & \gamma_2 \delta_{11} & \gamma_3 \delta_{11} & \gamma_4 \delta_{11} \\
\hline
(1,2) & \gamma_5 \delta_{12} & \gamma_6 \delta_{12} & \gamma_7 \delta_{12} & \gamma_8 \delta_{12} \\
\hline
(2,1) & \gamma_9 \delta_{21} & \gamma_{10} \delta_{21} & \gamma_{11} \delta_{21} & \gamma_{12} \delta_{21} \\
\hline
(2,2) & \gamma_{13} \delta_{22} & \gamma_{14} \delta_{22}  & \gamma_{15} \delta_{22} & \gamma_{16} \delta_{22} \\
\hline
\end{array}
\]
\end{table}

If the joint  measurement choice probabilities satisfy $0<\delta_{11},\delta_{12},\delta_{21}, \delta_{22}<1$ strictly, then the parameters $\pmb{\gamma}=(\gamma_{1},\gamma_{2},...,\gamma_{16})$ can be identified with the conditional probabilities 
$P(A,B|M_A,M_B)$, e.g., 
\[P(A=-1,B=-1|M_A=1,M_B=1)=\frac{\gamma_{1}\delta_{11}}{(\gamma_{1}+\gamma_{2}+\gamma_{3}+\gamma_{4})\delta_{11}}=\gamma_{1},\]
since $\gamma_{1}+\gamma_{2}+\gamma_{3}+\gamma_{4}=1$ by the measurement context constraint. In this example, if $\delta_{11}=0$, then the conditional probability on the LHS is undefined, whereas $\gamma_{1}$ presumably is defined. For this reason, it will be assumed that the  measurement choice probabilities are all strictly between 0 and 1 unless explicitly stated otherwise. The implication for any actual experiment is that every one of the 4 possible joint measurement combinations by Alice and Bob must occur at least once.

\subsubsection{Local causality conditionals}
\label{sec:lcDefineCond}
Recall the definitions of various forms of locality (RCI, factorizability, ROI, LC) in Sec.~\ref{sec:introduction}. Note that the first three involve conditioning on $M_A$ and/or $M_B$ only, whereas LC involves conditioning on either $(A,M_A,M_B)$ or $(B,M_A,M_B)$. For example, in terms of the generic parameters, conditional probabilities such as the following can be computed.

\begin{align}
&P(A=-1|B=-1,M_A=1,M_B=1)=\frac{\gamma_1}{\gamma_1+\gamma_3} \\
&P(B=-1|A=-1,M_A=1,M_B=1)=\frac{\gamma_1}{\gamma_1+\gamma_2}. \nonumber
\end{align}

There are 32 conditional probabilities like this that will be undefined unless the corresponding denominators are nonzero. It turns out that a \emph{necessary} condition for being well-defined is that at least 12 $\gamma_k$'s must be nonzero. This is the reason that "Local causality" (box 3) is shown below "Factorizable (Bell local)" (box 5) in Fig.~\ref{fig:modelsGraphPlotNew}. That is, the LC model instances form a subset of factorizable model instances that just happen to have 12 or more non-zero generic parameters.

\subsection{Examples and counterexamples}
\label{sec:counterExamples}

In this section, examples and counterexamples are constructed that justify some of the relationships among model types and properties shown in Fig.~\ref{fig:modelsGraphPlotNew}.

\subsubsection{QM and NC but not factorizable}
Table~\ref{tab:pdNotMQBL} shows parameterizations for a model instance that is both NC and QM, but it is not Bell local (factorizable). It shows that a model instance does not have to be factorizable to satisfy all of the CHSH inequalities.

\begin{table}[H]
\caption{Example of a model instance that is QM and NC but not factorizable (Bell local).}
\label{tab:pdNotMQBL}
\[
\begin{array}{|c||l|}
\hline
\text{Model type} & \text{Parameter assignment} \\
\hline
\hline
\text{Generic} & \gamma_k=\frac{1}{4} \text{ for } k=1,2,3,4,13,14,15,16, \\
& \gamma_k=\frac{1}{2} \text{ for } k=5,8,10,11,  \\ 
& 0 \text{ otherwise } \\
\hline
\text{QM} & \theta_1=135^{\circ}, \theta_2=0^{\circ}, \theta_3=90^{\circ},\theta_4=45^{\circ} \\
\hline
\text{NC} & \rho_k=\frac{1}{4} \text{ for } k=2,8,9,15 \text{ and } \rho_k=0 \text{ otherwise }.\\
\hline
\end{array}
\]
\end{table}

\subsubsection{NC but neither QM nor factorizable}
Table~\ref{tab:pdNotBellLocal} shows a NC model instance that does not belong to either the set of QM model instances or Bell local (factorizable) model instances.

\begin{table}[H]
\caption{Example of a model instance that is NC but not QM or factorizable (Bell local).}
\label{tab:pdNotBellLocal}
\[
\begin{array}{|c||c|}
\hline
\text{Model type} & \text{Parameter assignment} \\
\hline
\hline
\text{Generic} & \gamma_3=1, \\ 
& \gamma_k=\frac{1}{3} \text{ for } k=8,9,13,15,16, \\ 
& \gamma_k=\frac{2}{3} \text{ for } k=7,11, \\
&  0 \text{ otherwise } \\
\hline
\text{NC} & \rho_k=\frac{1}{3} \text{ for } k=3,4,8 \text{ and } \rho_k=0 \text{ otherwise }\\
\hline
\end{array}
\]
\end{table}

\subsubsection{QM but not NC}
Table~\ref{tab:qmNotPD} shows a QM model instance that is not NC, hence also not factorizable.

\begin{table}[H]
\caption{Example of a model instance that is QM but not NC.}
\label{tab:qmNotPD}
\[
\begin{array}{|c||c|}
\hline
\text{Model type} & \text{Parameter assignment} \\
\hline
\hline
\text{Generic}
& \gamma_k=\frac{1}{2}\cos^2 67.5^{\circ} \text{ for } k=1,4,6,7,10,11,14,15, \text{ and }\\ 
& \gamma_k=\frac{1}{2} \sin^2 67.5^{\circ} \text{ for } k=2,3,5,8,9,12,13,16. \\
\hline
\text{QM} & \theta_1=67.5^{\circ} \text{ and } \theta_k=22.5^{\circ} \text{ for } k=2,3,4.\\
\hline
\end{array}
\]
\end{table}

\subsubsection{The trivial model instance}
Table~\ref{tab:trivialMI} shows that there are parameterizations for the generic, QM, NC, and factorizable model types that produce the trivial model instance.

\begin{table}[H]
\caption{The only model instance achievable by all five main operational model types is the trivial one.}
\label{tab:trivialMI}
\[
\begin{array}{|c||c|}
\hline
\text{Model type} & \text{Parameter assignment} \\
\hline
\hline
\text{Generic} & \gamma_k=\frac{1}{4} \text{ for all } k \\
\hline
\text{QM} & \theta_1=135^{\circ} \text{ and } \theta_k=45^{\circ} \text{ for } k=2,3,4\\
\hline
\text{NC} & \rho_k=\frac{1}{16} \text{ for } k=1,2,...,16 \\
\hline
\text{Factorizable} & \alpha_k=\beta_k=\frac{1}{2} \text{ for } k=1,2,3,4 \\
\hline
\text{Trivial} & \text{ All entries } =\frac{1}{4} \\
\hline
\end{array}
\]
\end{table}

\subsection{All QM model instances satisfy RCI but only one satisfies ROI}
\label{sec:qmRCINotROI}

\begin{Theorem}
Every QM model instance satisfies RCI but only one satisfies ROI (the trivial model instance).
\label{thm:qmImpliesRCI}
\end{Theorem}

\begin{proof}
See Table~\ref{tab:qmRCI} for the proof that QM model instances satisfy RCI. Since all of the conditional probabilities are equal to $\frac{1}{2}$, it is obvious that Alice'a outcome probabilities do not depend on Bob's measurement, and similarly Bob's outcome probabilities do not depend on Alice's measurement.

\begin{table}[H]
\caption{The QM model type satisfies RCI. Both Alice's conditionals (next-to-last column) and Bob's conditionals (last column) are obviously independent of the other's measurement choice.}
\label{tab:qmRCI}
\[
\begin{array}{|c|c|c|c||c|c|}
\hline
s & t & u & v & P(A=s|M_A=u,M_B=v) & P(B=t|M_A=u,M_B=v)\\
\hline
\hline
-1 & -1 & 1 & 1 & \frac{1}{2} & \frac{1}{2} \\
\hline
-1 & 1 &  1 & 1 & \frac{1}{2} & \frac{1}{2} \\
\hline
1 & -1 & 1 & 1 & \frac{1}{2} & \frac{1}{2} \\
\hline
1 &  1 & 1 & 1 & \frac{1}{2} & \frac{1}{2} \\
\hline

-1 & -1 & 1 & 2 & \frac{1}{2} & \frac{1}{2} \\
\hline
-1 & 1 &  1 & 2 & \frac{1}{2} & \frac{1}{2} \\
\hline
1 & -1 & 1 & 2 & \frac{1}{2} & \frac{1}{2} \\
\hline
1 &  1 & 1 & 2 & \frac{1}{2} & \frac{1}{2} \\
\hline

-1 & -1 & 2 & 1 & \frac{1}{2} & \frac{1}{2} \\
\hline
-1 & 1 &  2 & 1 & \frac{1}{2} & \frac{1}{2} \\
\hline
1 & -1 & 2 & 1 & \frac{1}{2} & \frac{1}{2} \\
\hline
1 &  1 & 2 & 1 & \frac{1}{2} & \frac{1}{2} \\
\hline

-1 & -1 & 2 & 2 & \frac{1}{2} & \frac{1}{2} \\
\hline
-1 & 1 &  2 & 2 & \frac{1}{2} & \frac{1}{2} \\
\hline
1 & -1 & 2 & 2 & \frac{1}{2} & \frac{1}{2} \\
\hline
1 &  1 & 2 & 2 & \frac{1}{2} & \frac{1}{2} \\
\hline
\end{array}
\]
\end{table}

As for satisfying ROI, consult the definition of ROI (Def.~\ref{def:ROI}) and then refer to Tables~\ref{tab:qmTypeIntro} and~\ref{tab:qmRCI}. These imply that 
\[
\frac{1}{2}\cos^2 \theta_k=\frac{1}{2}\sin^2 \theta_k=\frac{1}{4}
\]
for all $k=1,2,3,4$. This works only if the measurement difference angles are all of the form 
\[\theta_{k}=(2j_{k}+1)\frac{\pi}{4}\]
for integers $j_{k}$. This is the definition of the trivial model type (Table~\ref{tab:trivialType}). The only QM model instance that satisfies ROI is the trivial one. The QM model type is highly "nonlocal" in this sense.
\end{proof}

\subsection{All NC model instances satisfy RCI but only the factorizable ones satisfy ROI}
\label{sec:ncRCINotROI}

\begin{Theorem}
Every NC model instance satisfies RCI, but not all satisfy ROI.
\label{thm:predetImpliesRCI}
\end{Theorem}
\begin{proof}
See Tables~\ref{tab:rciBRTableAlice} and~\ref{tab:rciBRTableBob}. Note the probabilities in each pair of rows are equal whenever $(s,u)$ is the same (for Alice), or whenever $(t,v)$ is the same (for Bob), no matter the value of the other's measurement. That is, the measurement in either wing of the experiment does not affect the outcome probability in the other.

\begin{table}[H]
\caption{The NC model type satisfies RCI. Alice's are conditionals shown. The probabilities in each pair of rows in which $(s,u)$ is the same are equal.}
\label{tab:rciBRTableAlice}
\[
\begin{array}{|r|r|r|l|}
\hline
s & u & v & P(A=s|M_A=u,M_B=v) \\
\hline
-1 & 1 & 1 & \rho_1+\rho_2+\rho_5+\rho_6+\rho_9+\rho_{10}+\rho_{13}+\rho_{14} \\
\hline
-1 & 1 & 2 & \rho_1+\rho_2+\rho_5+\rho_6+\rho_9+\rho_{10}+\rho_{13}+\rho_{14} \\
\hline
-1 & 2 & 1 & \rho_1+\rho_3+\rho_5+\rho_7+\rho_9+\rho_{11}+\rho_{13}+\rho_{15} \\
\hline
-1 & 2 & 2 & \rho_1+\rho_3+\rho_5+\rho_7+\rho_9+\rho_{11}+\rho_{13}+\rho_{15} \\
\hline

1 & 1 & 1 & \rho_3+\rho_4+\rho_7+\rho_8+\rho_{11}+\rho_{12}+\rho_{15}+\rho_{16} \\
\hline
1 & 1 & 2 & \rho_3+\rho_4+\rho_7+\rho_8+\rho_{11}+\rho_{12}+\rho_{15}+\rho_{16} \\
\hline
1 & 2 & 1 & \rho_2+\rho_4+\rho_6+\rho_8+\rho_{10}+\rho_{12}+\rho_{14}+\rho_{16} \\
\hline
1 & 2 & 2 & \rho_2+\rho_4+\rho_6+\rho_8+\rho_{10}+\rho_{12}+\rho_{14}+\rho_{16} \\
\hline
\end{array}
\]
\end{table}

\begin{table}[H]
\caption{The NC model type satisfies RCI. Bob's conditionals are shown. The probabilities in each pair of rows in which $(t,v)$ is the same are equal.}
\label{tab:rciBRTableBob}
\[
\begin{array}{|r|r|r|l|}
\hline
t & u & v & P(B=t|M_A=u,M_B=v) \\
\hline
-1 & 1 & 1 & \rho_1+\rho_2+\rho_3+\rho_4+\rho_5+\rho_6+\rho_7+\rho_8 \\
\hline
-1 & 2 & 1 &  \rho_1+\rho_2+\rho_3+\rho_4+\rho_5+\rho_6+\rho_7+\rho_8 \\
\hline
-1 & 1 & 2 & \rho_1+\rho_2+\rho_3+\rho_4+\rho_9+\rho_{10}+\rho_{11}+\rho_{12} \\
\hline
-1 & 2 & 2 & \rho_1+\rho_2+\rho_3+\rho_4+\rho_9+\rho_{10}+\rho_{11}+\rho_{12} \\
\hline

1 & 1 & 1 & \rho_9+\rho_{10}+\rho_{11}+\rho_{12}+\rho_{13}+\rho_{14}+\rho_{15}+\rho_{16} \\
\hline
1 & 2 & 1 &  \rho_9+\rho_{10}+\rho_{11}+\rho_{12}+\rho_{13}+\rho_{14}+\rho_{15}+\rho_{16} \\
\hline
1 & 1 & 2 & \rho_5+\rho_6+\rho_7+\rho_8+\rho_{13}+\rho_{14}+\rho_{15}+\rho_{16} \\
\hline
1 & 2 & 2 & \rho_5+\rho_6+\rho_7+\rho_8+\rho_{13}+\rho_{14}+\rho_{15}+\rho_{16} \\
\hline
\end{array}
\]
\end{table}

To see that a NC model instance does not satisfy ROI in general, see Table~\ref{tab:pdNotROI}. These represent the equations that define ROI for the NC model type (see Def.~\ref{def:ROI} and Tables~\ref{tab:predeterminedTypeIntro},~\ref{tab:rciBRTableAlice}, and~\ref{tab:rciBRTableBob}). There are 16 nonlinear equations that must be solved for $\pmb{\rho}=(\rho_1,\rho_2,...,\rho_{16})$, where of course $\pmb{\rho}$ must be a pmf.

It turns out that these are exactly the equations which must be solved in order for a NC model instance to be factorizable. This should not be a surprise. Since every NC model instance is already RCI, finding a solution to these equations would make it ROI as well, hence factorizable (Thm.~\ref{thm:fcRCIROI}).

\begin{table}[H]
\caption{A NC model instance satisfies ROI only if these 16 equations are satisfied.}
\label{tab:pdNotROI}
\[
\begin{array}{|c|c|c|}
\hline
(\rho_1+\rho_2+\rho_5+\rho_6+\rho_9+\rho_{10}+\rho_{13}+\rho_{14}) & \\ \times (\rho_1+\rho_2+\rho_3+\rho_4+\rho_5+\rho_6+\rho_7+\rho_8)= & \rho_1+\rho_2+\rho_5+\rho_6  \\
\hline
(\rho_1+\rho_2+\rho_5+\rho_6+\rho_9+\rho_{10}+\rho_{13}+\rho_{14}) & \\ \times (\rho_9+\rho_{10}+\rho_{11}+\rho_{12}+\rho_{13}+\rho_{14}+\rho_{15}+\rho_{16})= & \rho_9+\rho_{10}+\rho_{13}+\rho_{14} \\
\hline
(\rho_2+\rho_4+\rho_6+\rho_8+\rho_{10}+\rho_{12}+\rho_{14}+\rho_{16}) & \\ \times (\rho_5+\rho_6+\rho_7+\rho_8+\rho_{13}+\rho_{14}+\rho_{15}+\rho_{16})= & \rho_6+\rho_8+\rho_{14}+\rho_{16} \\
\hline
(\rho_3+\rho_4+\rho_7+\rho_8+\rho_{11}+\rho_{12}+\rho_{15}+\rho_{16}) & \\ \times (\rho_1+\rho_2+\rho_3+\rho_4+\rho_5+\rho_6+\rho_7+\rho_8) = & \rho_3+\rho_4+\rho_7+\rho_8 \\
\hline
(\rho_3+\rho_4+\rho_7+\rho_8+\rho_{11}+\rho_{12}+\rho_{15}+\rho_{16}) & \\ \times (\rho_9+\rho_{10}+\rho_{11}+\rho_{12}+\rho_{13}+\rho_{14}+\rho_{15}+\rho_{16}) = & \rho_{11}+\rho_{12}+\rho_{15}+\rho_{16} \\
\hline
(\rho_1+\rho_2+\rho_5+\rho_6+\rho_9+\rho_{10}+\rho_{13}+\rho_{14}) & \\ \times (\rho_1+\rho_2+\rho_3+\rho_4+\rho_9+\rho_{10}+\rho_{11}+\rho_{12}) = & \rho_1+\rho_2+\rho_9+\rho_{10} \\
\hline
(\rho_1+\rho_2+\rho_5+\rho_6+\rho_9+\rho_{10}+\rho_{13}+\rho_{14}) & \\ \times (\rho_5+\rho_6+\rho_7+\rho_8+\rho_{13}+\rho_{14}+\rho_{15}+\rho_{16}) = & \rho_5+\rho_6+\rho_{13}+\rho_{14} \\
\hline
(\rho_3+\rho_4+\rho_7+\rho_8+\rho_{11}+\rho_{12}+\rho_{15}+\rho_{16}) & \\ \times (\rho_1+\rho_2+\rho_3+\rho_4+\rho_9+\rho_{10}+\rho_{11}+\rho_{12}) = & \rho_3+\rho_4+\rho_{11}+\rho_{12} \\
\hline
(\rho_3+\rho_4+\rho_7+\rho_8+\rho_{11}+\rho_{12}+\rho_{15}+\rho_{16}) & \\ \times (\rho_5+\rho_6+\rho_7+\rho_8+\rho_{13}+\rho_{14}+\rho_{15}+\rho_{16}) = & \rho_7+\rho_8+\rho_{15}+\rho_{16} \\
\hline
(\rho_1+\rho_3+\rho_5+\rho_7+\rho_9+\rho_{11}+\rho_{13}+\rho_{15}) & \\ \times (\rho_1+\rho_2+\rho_3+\rho_4+\rho_5+\rho_6+\rho_7+\rho_8) = & \rho_1+\rho_3+\rho_5+\rho_7 \\
\hline
(\rho_1+\rho_3+\rho_5+\rho_7+\rho_9+\rho_{11}+\rho_{13}+\rho_{15}) & \\ \times (\rho_9+\rho_{10}+\rho_{11}+\rho_{12}+\rho_{13}+\rho_{14}+\rho_{15}+\rho_{16}) = & \rho_9+\rho_{11}+\rho_{13}+\rho_{15} \\
\hline
(\rho_2+\rho_4+\rho_6+\rho_8+\rho_{10}+\rho_{12}+\rho_{14}+\rho_{16}) & \\ \times (\rho_1+\rho_2+\rho_3+\rho_4+\rho_5+\rho_6+\rho_7+\rho_8) = & \rho_2+\rho_4+\rho_6+\rho_8 \\
\hline
(\rho_2+\rho_4+\rho_6+\rho_8+\rho_{10}+\rho_{12}+\rho_{14}+\rho_{16}) & \\ \times (\rho_9+\rho_{10}+\rho_{11}+\rho_{12}+\rho_{13}+\rho_{14}+\rho_{15}+\rho_{16}) = & \rho_{10}+\rho_{12}+\rho_{14}+\rho_{16} \\
\hline
(\rho_1+\rho_3+\rho_5+\rho_7+\rho_9+\rho_{11}+\rho_{13}+\rho_{15}) & \\ \times (\rho_1+\rho_2+\rho_3+\rho_4+\rho_9+\rho_{10}+\rho_{11}+\rho_{12}) = & \rho_1+\rho_3+\rho_9+\rho_{11} \\
\hline
(\rho_1+\rho_3+\rho_5+\rho_7+\rho_9+\rho_{11}+\rho_{13}+\rho_{15}) & \\ \times (\rho_5+\rho_6+\rho_7+\rho_8+\rho_{13}+\rho_{14}+\rho_{15}+\rho_{16}) = & \rho_5+\rho_7+\rho_{13}+\rho_{15} \\
\hline
(\rho_2+\rho_4+\rho_6+\rho_8+\rho_{10}+\rho_{12}+\rho_{14}+\rho_{16}) & \\ \times (\rho_1+\rho_2+\rho_3+\rho_4+\rho_9+\rho_{10}+\rho_{11}+\rho_{12}) = & \rho_2+\rho_4+\rho_{10}+\rho_{12} \\
\hline
(\rho_2+\rho_4+\rho_6+\rho_8+\rho_{10}+\rho_{12}+\rho_{14}+\rho_{16}) & \\ \times (\rho_5+\rho_6+\rho_7+\rho_8+\rho_{13}+\rho_{14}+\rho_{15}+\rho_{16}) = & \rho_6+\rho_8+\rho_{14}+\rho_{16} \\
\hline
\end{array}
\]
\end{table}
\end{proof}

\subsection{Factorization is equivalent to RCI + ROI}
\label{sec:fcRCIROI}

In the interest of reducing clutter, in this subsection and Appendix~\ref{sec:LCFC}, the \emph{random variables} will be suppressed, and only the corresponding \emph{values} will be shown. For example, \[P(A=s,B=t|M_{A}=u,M_{B}=v) \text{ will be shortened to } P(s,t|u,v)\] for $s,t=\pm1$ and $u,v=1,2$. The following fact was proved in Jarrett~\cite{Jarrett1984}.

\begin{Theorem}
A model instance is factorizable iff it satisfies both the remote context independent (RCI) condition and the remote outcome independence (ROI) condition.
\label{thm:fcRCIROI}
\end{Theorem}

\begin{proof}
($\Rightarrow$) Assume factorizability. Then

\begin{align}
&P(s|u,v)=\sum_{t} P(s,t|u,v)=\sum_{t} P(s|u)P(t|v)=P(s|u) \text{ and } \nonumber \\
&P(t|u,v)=\sum_{s} P(s,t|u,v)=\sum_{s} P(s|u)P(t|v)=P(t|v).
\label{eq:fcToRCI}
\end{align}

In each case, the first equality is a marginal probability calculation, the second equality follows from factorizability, and the third is just the fact that a pmf sums to 1. This establishes that factorizability implies RCI. As for establishing that factorizability implies ROI, note 

\begin{equation}
P(s,t|u,v)=P(s|u)P(t|v)=P(s|u,v)P(t|u,v).
\label{eq:fcToROI}
\end{equation}

The first equality is the definition of factorizability and the second equation follows from RCI, which was proved in Eq.~\ref{eq:fcToRCI}.

($\Leftarrow$) Assume RCI and ROI. Then 
\begin{equation}
P(s,t|u,v) = P(s|u,v)P(t|u,v)=P(s|u)P(t|v).
\end{equation}

The first equality follows from ROI and the second equality follows from RCI.
\end{proof}

\subsection{LC implies factorizability}
\label{sec:LCFC}

Recall that, by the definition of LC, conditional probabilities such as $P(s|t,u,v)$, $P(t|s,u,v)$, must be well-defined, that is, $P(s,u,v)>0$ and $P(t,u,v)>0$ (Def.~\ref{def:LC}). Also $P(u,v)=P(M_{A}=u,M_{B}=v)>0$ is a given -- see Sec.~\ref{sec:measChoice}.

\begin{Theorem}
If a model instance is locally causal (LC), then it is also factorizable.
\label{thm:lcFC}
\end{Theorem}

\begin{proof}
First show that LC implies RCI. Assume LC.

\begin{align}
\label{eq:lcToRCI}
&P(s|u,v)=\sum_{t} P(s,t|u,v)=\sum_{t} P(s|t,u,v)P(t)=\sum_t P(s|u)P(t)=P(s|u) \nonumber \\
& \text{ and }  \\
&P(t|u,v)=\sum_{s} P(s,t|u,v)=\sum_{s} P(t|s,u,v)P(s)=\sum_s P(t|v)P(s)=P(t|v). \nonumber
\end{align}

In each line the third equality invokes LC. The other three equalities follow by ordinary marginal, conditional, and total probability manipulations.

Then factorizability follows from
\begin{align}
\label{eq:lcToROI}
&P(s,t|u,v)=\frac{P(s,t,u,v)}{P(u,v)} = \frac{P(s|t,u,v)P(t,u,v)}{P(u,v)}=\nonumber \\
&\frac{P(s|t,u,u)P(t|u,v)P(u,v)}{P(u,v)}= P(s|u)P(t|u,v) = P(s|u)P(t|v). 
\end{align}
where the first three equalities are conditional probability manipulations and the last two equalities invoke a combination of LC and RCI.
\end{proof}     \qedhere

\subsection{RCI model instances that are not NC}
\label{sec:rciNOTfc}

Consider the generic model instance with 
\[\gamma_k=\frac{1}{2} \text{ for } k=2,3,5,8,9,12,13,16 \text{ and } \gamma_k=0 \text{ otherwise. }\]
Table~\ref{tab:rciNotCHSH} clearly shows that RCI is satisfied, both on Alice and Bob's sides. 

\begin{table}[H]
\caption{Proof that the example satisfies RCI: Each of Alice and Bob's outcomes depends only on her/his own  measurement choice.}
\label{tab:rciNotCHSH}
\[
\begin{array}{|c|c|c|c||c|c|}
\hline
s & t & u & v & P(A=s|M_A=u,M_B=v) & P(B=t|M_A=u,M_B=v)\\
\hline
\hline
-1 & -1 & 1 & 1 & \frac{1}{2} & \frac{1}{2} \\
\hline
-1 & 1 &  1 & 1 & \frac{1}{2} & \frac{1}{2} \\
\hline
1 & -1 & 1 & 1 & \frac{1}{2} & \frac{1}{2} \\
\hline
1 &  1 & 1 & 1 & \frac{1}{2} & \frac{1}{2} \\
\hline

-1 & -1 & 1 & 2 & \frac{1}{2} & \frac{1}{2} \\
\hline
-1 & 1 &  1 & 2 & \frac{1}{2} & \frac{1}{2} \\
\hline
1 & -1 & 1 & 2 & \frac{1}{2} & \frac{1}{2} \\
\hline
1 &  1 & 1 & 2 & \frac{1}{2} & \frac{1}{2} \\
\hline

-1 & -1 & 2 & 1 & \frac{1}{2} & \frac{1}{2} \\
\hline
-1 & 1 &  2 & 1 & \frac{1}{2} & \frac{1}{2} \\
\hline
1 & -1 & 2 & 1 & \frac{1}{2} & \frac{1}{2} \\
\hline
1 &  1 & 2 & 1 & \frac{1}{2} & \frac{1}{2} \\
\hline

-1 & -1 & 2 & 2 & \frac{1}{2} & \frac{1}{2} \\
\hline
-1 & 1 &  2 & 2 & \frac{1}{2} & \frac{1}{2} \\
\hline
1 & -1 & 2 & 2 & \frac{1}{2} & \frac{1}{2} \\
\hline
1 &  1 & 2 & 2 & \frac{1}{2} & \frac{1}{2} \\
\hline
\end{array}
\]
\end{table}

The correlations and $s$-functions for this example are
\begin{align*}
&(w,x,y,z)=(-1,1,1,1) \\
&(s_1,s_2,s_3,s_4)=(4,0,0,0).
\end{align*} 

Hence one of the CHSH inequalities is violated, so this example cannot be NC. Because it is not NC, it also cannot be factorizable or LC by Thm.'s~\ref{thm:fcImpliesPDIntro} and~\ref{thm:lcFC}.

See Thm.~\ref{thm:predetImpliesRCI} in Appendix~\ref{sec:ncRCINotROI} for a proof that every NC model instance satisfies RCI. Together with this example, it explains the positioning of "Remote Context Independence" (box 12) strictly above "NC" (box 8) in Fig.~\ref{fig:modelsGraphPlotNew}. It also explains why box 12 labeled "Remote Context Independence" is outlined in green, since this example has an $s$-function that exceeds a Tsirelson bound.

\begin{figure}
\centering
\includegraphics{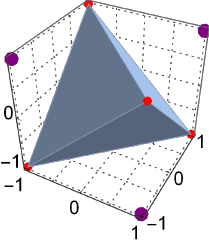}
\caption{Perfect correlations $(x,y,z)$ on (red) and off (purple) the NC region, where $w=-1$ is fixed. Red dots at the corners correspond to correlations that satisfy CHSH. Purple dots off of the tetrahedron correspond to correlations that violate CHSH. (The purple dot corresponding to the point $(-1,1,-1)$ is hidden.)}
\label{fig:bellRCI}
\end{figure}

One can find more examples like these which are RCI but whose correlations have a pattern either of three -1's and one +1 or three +1's and one -1. Consider Fig.~\ref{fig:bellRCI}, which shows the NC region for $w=-1$, a tetrahedron. The vertices of the tetrahedron (red dots) correspond to model instances with correlations consisting of zero, two, or four +1's (or alternatively four, two, or zero -1's), hence have $s$-functions that satisfy CHSH. The purple dots not in the NC region, but on the corners of the correlation cube, correspond to model instances that have correlations consisting of  one -1 and three +1's or three -1's and one +1, hence have an $s$-function that exceeds a Tsirelson bound. The same holds for the NC region for $w=1$, also a tetrahedron, but in which the red dots and the purple dots are switched. See Fig.'s~\ref{fig:models5} and~\ref{fig:models5Combined}.

\subsection{ROI model instance that is not NC}
\label{sec:roiNOTfc}

Table~\ref{tab:roiExampleTab} shows an assignment of generic parameters
$\gamma_{1},\gamma_{2},...,\gamma_{16}$, and the associated correlations and $s$-functions are shown in Table~\ref{tab:roiNotPDCorrSFun}. Note that $s_{3}=-2.6108$ exceeds the lower bound of $-2$ of the CHSH inequalities. Hence the model instance cannot be NC, factorizable, or LC (by Thm.~\ref{thm:lcFC}).
 
\begin{table}[H]
\caption{Example that is ROI but does not satisfy CHSH.}
\label{tab:roiExampleTab}
\[
\begin{array}{|c|c|c|c|c|}
\hline
\text{Context} & \multicolumn{4}{c|}{\text{Outcome }(s,t)} \\
\hline
\hline
(u,v) & (-1,-1)  & (-1,1)   & (1,-1)   & (1,1) \\
\hline
(1,1) & 0.0435 & 0.9540 & 0.0001 & 0.0023 \\
\hline
(1,2) & 0.2198 & 0.7777 & 0.0006 & 0.0020 \\
\hline
(2,1) & 0.7469 & 0.0180 & 0.2296 & 0.0055 \\
\hline
(2,2) & 0.1428 & 0.8142 & 0.0064 & 0.0366 \\
\hline
\end{array}
\]
\end{table}

\begin{table}[H]
\caption{Proof that example does not satisfy CHSH.}
\label{tab:roiNotPDCorrSFun}
\[
\begin{array}{|c||c|}
\hline
w & -0.9083 \\
\hline
x & -0.5565 \\
\hline
y & 0.5048 \\
\hline
z & -0.6412 \\
\hline
\hline
s_1 & 0.2154 \\
\hline
s_2 & -0.4882 \\
\hline
s_3 & -2.6108 \\
\hline
s_4 & -0.3188 \\
\hline
\hline
\text{Min and Max} & -2.6108 \text{ and } 0.2154\\
\hline
\end{array}
\]
\end{table}
However, Table~\ref{tab:roiNotPD} shows that this model instance satisfies ROI. Note that the columns labeled $P(s,t|u,v)$ and $P(s|u,v) \times P(t|u,v)$ are equal, which shows that Alice and Bob's outcomes are conditionally independent, given their joint measurement choices. This example explains the positioning of "Remote Outcome Independence" (box 13) strictly above "Bell local (Factorization)" (box 5) in Fig.~\ref{fig:modelsGraphPlotNew}.

\begin{table}[H]
\caption{Proof that example satisfies ROI.}
\label{tab:roiNotPD}
\[
\begin{array}{|c|c|c|c||c|c|c|c|}
\hline
s & t & u & v & P(s,t|u,v) & P(s|u,v) \times P(t|u,v) \\
\hline
\hline
-1 & -1 & 1 & 1 & 0.0435 & 0.0435 \\
\hline
-1 & 1 &  1 & 1 & 0.9541 & 0.9541 \\
\hline
1 & -1 & 1 & 1 & 0.0001 & 0.0001 \\
\hline
1 &  1 & 1 & 1 & 0.0023 & 0.0023 \\
\hline

-1 & -1 & 1 & 2 & 0.2198 & 0.2198 \\
\hline
-1 & 1 &  1 & 2 & 0.7776 & 0.7776 \\
\hline
1 & -1 & 1 & 2 & 0.0006 & 0.0006 \\
\hline
1 &  1 & 1 & 2 & 0.0020 & 0.0020 \\
\hline

-1 & -1 & 2 & 1 & 0.7469 & 0.7469 \\
\hline
-1 & 1 &  2 & 1 & 0.0180 & 0.0180 \\
\hline
1 & -1 & 2 & 1 & 0.2296 & 0.2296 \\
\hline
1 &  1 & 2 & 1 & 0.0055 & 0.0055 \\
\hline

-1 & -1 & 2 & 2 & 0.1428 & 0.1428 \\
\hline
-1 & 1 &  2 & 2 & 0.8142 & 0.8142 \\
\hline
1 & -1 & 2 & 2 & 0.0064 & 0.0064 \\
\hline
1 &  1 & 2 & 2 & 0.0366 & 0.0366 \\
\hline
\end{array}
\]
\end{table}

\subsection{RCI and ROI are distinct}

In Sec.'s~\ref{sec:rciNOTfc} and~\ref{sec:roiNOTfc} example model instances were shown that are, respectively,
\begin{itemize}
\item
RCI but not NC and
\item
ROI but not NC.
\end{itemize}

These instances are obviously different, demonstrating that neither the set of RCI model instances nor the set of ROI model instances can contain the other. To see this recall that $RCI \bigcap ROI=FC \subset NC$ and both of these instances are outside $NC$. This fact is reflected by the positioning of RCI (box 12) and ROI (box 13) on different paths in Fig.~\ref{fig:modelsGraphPlotNew}.

\section{CHSH inequalities survive convex combinations}
\label{sec:miscellaneous}

The proof of Lemma~\ref{lemma:linearComboCHSH} shows that convex combinations of model instances that satisfy CHSH also satisfy all the CHSH inequalities. The definitions of the correlation matrix $C$ and the $s$-function matrix $S$ can be found in Sec.~\ref{sec:defineCHSH}.

\begin{Lemma}
\label{lemma:linearComboSFun}
Let $\pmb{\gamma}_1$, $\pmb{\gamma}_2$,...,$\pmb{\gamma}_n$ be $n$ generic conditional probability vectors, and let $\pmb{\pi}=(\pi_1,\pi_2,...,\pi_n)$ be a pmf. Define
\[
\pmb{\gamma}=\sum_k \pi_k \pmb{\gamma}_k.
\]
Then 
\[
S C \pmb{\gamma}^T=\sum_k \pi_k S C \pmb{\gamma}_k^T.
\]
That is, the $s$-functions of a convex combination of conditional probability vectors equal the same convex combination of the $s$-functions of the underlying cpvs.
\end{Lemma}

\begin{proof}
This follows immediately from the linearity of matrix multiplication.
\end{proof}

\begin{Lemma}
\label{lemma:linearComboCHSH}
Given the same assumptions as in Lemma~\ref{lemma:linearComboSFun}. Then if the $s$-functions associated with each $\pmb{\gamma}_k$ satisfy all of the CHSH inequalities, so do the $s$-functions associated with $\pmb{\gamma}$.
\end{Lemma}

\begin{proof}
From Eq.~\ref{eq:genericCHSH}, 
\[
-2 \leq S C \pmb{\gamma}_k^T \leq 2
\]
for all $k$ and since the $\pi_k$'s are nonnegative, therefore
\[
-2 \pi_k \leq \pi_k  S C \pmb{\gamma}_k^T \leq 2 \pi_k
\]
for all $k$ as well. By Lemma~\ref{lemma:linearComboSFun} and the fact that $\sum_k \pi_k=1$, adding these $n$ inequalities yields
\[-2 \leq S C \pmb{\gamma}^T \leq 2.\]
\end{proof}


\end{document}